\pdfoutput=1
\documentclass[nofootinbib,twocolumn,showpacs,amssymb,floatfix,aps]{revtex4}

\usepackage [latin1]{inputenc}
\usepackage{graphicx}
\usepackage{dcolumn}
\usepackage{bm}
\usepackage{epsfig}
\usepackage[english]{babel}
\usepackage{amsmath}
\usepackage{amssymb}
\usepackage{verbatim}
\usepackage{alltt}

\def\({\left(}
\def\){\right)}

\begin{document}


\preprint{}
\title{Critical Tests of Leading Gamma Ray Burst Theories II} 

\author{Shlomo Dado${}^{a}$, Arnon Dar${}^{a}$ and A. De R\'ujula${}^{b,c}$}
\affiliation{  \vspace{3mm}
${}^{a}$Physics Dept. Technion, Haifa, Israel\\
${}^b$Instituto de F\'isica Te\'orica (UAM/CSIC), Univ. Aut\'onoma de Madrid, Spain;\\
Theory Division, CERN, CH 1211 Geneva 23, Switzerland
}

\date{\today}

\begin{abstract}

It has been observationally established that supernovae (SNe) of Type Ic produce long duration gamma ray bursts (GRBs) and that neutron star mergers generate short hard GRBs.  SN-Less GRBs presumably originate in a phase transition of a neutron star in a high mass X-ray binary. How these phenomena actually generate GRBs is debated.
The fireball and cannonball models of GRBs and their afterglows have been widely confronted with the huge observational data, with their defenders claiming success. The claims, however, may reflect multiple choices and the use of many adjustable parameters, rather than the validity of the models. Only a confrontation of key falsifiable predictions of the models with solid observational data can test their validity. Such critical tests are reviewed in this report.

\end{abstract}


\maketitle


\section{Introduction}

Gamma-ray bursts (GRBs) are brief flashes of gamma rays lasting between a few
 milliseconds and several hours, message-bearers of extremely energetic astrophysical
phenomena \cite{Fishman 1995}. 
They were first observed on July 2, 1967 by the USA Vela spy satellites,
 which were launched to detect USSR tests of nuclear weapons in the atmosphere, in violation of the 1963 USA-USSR Nuclear Test Ban Treaty. Their
 discovery was first published in 1973 after 15 such events were detected 
 \cite{Klebesadel}, establishing
 their ``natural" character and indicating an extra-solar origin.

During the first 20 years after their discovery hundreds of GRB models were
 published, all assuming that GRBs were Galactic in origin 
 (see, e.g.~\cite{Norris}). An extragalactic
 origin would imply an implausibly large energy release in gamma rays from a very
 small volume in a very short time, if their emission was isotropic, as was generally
 assumed. During that period it was also found that GRBs fall roughly into two classes,
 long duration ones (LGRBs) that last more than $\sim 2$ seconds, and short bursts (SGRBs)
 lasting less than $\sim 2$ seconds \cite{Nemiroff}. 
 Most SGRBs are short hard bursts (SHBs), with a much
 harder spectrum than LGRBs. The origin and production mechanism of GRBs have been
 major astrophysical puzzles until recently.

In 1984, Blinnikov et al.~\cite{Blinnikov} 
suggested that exploding neutron stars in close binaries
 may produce GRBs with an isotropic gamma-ray energy up to $\sim\!10^{46}$ erg. Such GRBs
 could be seen only from relatively nearby galaxies. Per contra, Paczynski 
 maintained \cite{Paczynski86}
 that the sky distribution of GRBs was more consistent with large cosmological distances,
 like those of quasars, with a typical redshift, $z$, between 1 and 2. This would imply a
 supernova-like energy release, $\sim\!10^{51}$ erg, within seconds, making gamma-ray bursters
the brightest known objects, many orders of magnitude brighter than any quasar  \cite{Paczynski86}.

The first plausible physical model of GRBs at large cosmological distances was
proposed by Goodman, Dar and Nussinov in 1987 \cite{Goodman2}.
 They suggested that GRBs were
produced in stripped-envelope SNe and neutron stars mergers (NSMs) by an $e^+ e^- \gamma$
 fireball \cite{Goodman1}
 formed by neutrino-antineutrino annihilation around the newly born compact
 object --a massive neutron star, a quark star or a stellar black hole. Shortly after the launch
 of the Compton Gamma-Ray Observatory (CGRO) in 1991, it became clear that
 neutrino-annihilation fireballs were not powerful enough to produce observable GRBs
 at the cosmological distances indicated by the CGRO observations \cite{Meegan}, 
 unless the
 fireballs were collimated into narrow beams by funneling through surrounding
 matter, as posited by Rees and Meszaros \cite{Meszaros}.

In 1994 Shaviv and Dar suggested  \cite{Shaviv}
that narrowly beamed GRBs can be produced
 by jets of highly relativistic plasmoids of ordinary matter, later called {\it cannonballs} (CBs),
via the inverse Compton scattering (ICS) of light surrounding their launch sites. They
proposed that such jets may be ejected in stripped-envelope core-collapse supernova
explosions, in mergers of compact stars due to the emission of gravitational waves, and
in a phase transitions of neutron stars (NSs)
to an even more compact object, i.e.~a quark star or
a black hole (BH), 
following mass accretion in compact binaries. These hypotheses, unaltered,
constitute the basis of the CB model.

An important prediction of the fireball (FB) model was a transition of the initial short
$\gamma$-ray emission to a longer-lived {\it afterglow} (AG) \cite{Paczynski2}
at longer wavelengths, due to
the slow down of the expansion of the $e^+ e^- \gamma$ fireball by the surrounding medium. 
In 1997 the team behind the satellite BeppoSAX discovered that GRBs are indeed followed
by a longer-lived X-ray AG \cite{Costa}.  
This resulted in an accurate enough sky localization of
GRBs, facilitating the discovery of their AGs at longer wavelengths \cite{van Paradijs}, 
the localization
of their host galaxies \cite{{Sahu}}
and the measurement of their redshifts \cite{Metzger} Also, at least in one early instance, the
association of a GRB with a type Ic SN explosion \cite{Galama}.

During the past twenty years, observers mainly using HETE, Swift, Konus-Wind, Chandra,
Integral, XMM-Newton, Fermi and the Hubble space telescope, plus ground-based
telescopes, measured the spectra of GRBs from $\gamma$-rays to radio. They demonstrated the
association of LGRBs with Type Ic SNe and studied the properties of their host galaxies
and near environments \cite{Fynbo}. 
In particular, this provided clear evidence that LGRBs
take place mainly in star formation regions within the disk of spiral galaxies, where
most Type Ic SNe take place, while SGRBs originate in and outside spiral and elliptical
galaxies and are not associated with SN explosions. These differences led to the wide
spread recent belief \cite{Fong}
that SHBs are produced in mergers of two NSs, or a
NS and a BH, as suggested long before \cite{Goodman2} \cite{Meszaros}.

The Ligo-Virgo gravitational wave (GW) detectors made it possible to test whether
 relatively nearby mergers produce SGRBs. Indeed, SHB170817A,  
 \cite{20a,20b,20c,20d,20e}
 was seen 
$1.74\pm 0.05$ s after the gravitational wave burst GW170817 \cite{Abbott}, 
proving that NSMs produce SHBs.
 Moreover, the universal shape of all the well sampled early AGs of ordinary SHBs and
of SHB170817A  --expected from a pulsar wind nebula (PWN) powered by the spin
 down of a newly born millisecond pulsar-- suggests that most SHBs are produced by
 NSMs yielding a NS remnant rather than a black hole \cite{Dado2018}.

Although LGRBs have been seen in association with type Ic SNe \cite{Galama} 
\cite{Melandri}, no associated 
SN has been detected in several nearby long duration GRBs, despite very deep
searches \cite{Gal-Yam }. The universal behavior of the AG of both SHBs and SN-Less GRBs
\cite{Dado2018} \cite{Dado}
suggests that the latter are also powered by a newly born millisecond pulsar, perhaps
 after a phase transitions of a NS to a quark star \cite{Shaviv} \cite{Dado2017}, 
 following mass accretion
onto the NS in a high mass X-ray binary (HMXRB).

Since 1997 only two theoretical models of GRBs and their afterglows --the standard
 fireball (FB) model \cite{FBM Reviews}
  and the cannonball (CB) model \cite{DD2004}-- have been used extensively
 to interpret the mounting observations. Practitioners of both models have
 claimed to reproduce well the data. But the two models were originally
 and still are quite different in their basic assumptions and predictions. This is despite the
 replacements, to be mentioned, of key assumptions of the standard FB model (but not its
 name) with assumptions underlying the CB model. The claimed success of the models
in reproducing the data, despite their complexity and diversity, may reflect the fact that
most theoretical results depend on free parameters and choices adjusted for each GRB.
 As a result, when successful fits to data were obtained, it was not clear
whether they were due to the validity of the theory or to multiple choices and the use of
many adjustable parameters.

Scientific theories ought to be falsifiable. Hence, only confrontations between solid 
observational data and key predictions
of theories, which do not depend on free adjustable parameters, 
can serve as decisive tests of the theories. 
Such critical tests of the cannonball and fireball models 
of long GRBs and SHBs are summarized in this review.

\section{The GRB Models}

GRBs and SHBs consist of a few $\gamma$-ray pulses with a ``FRED" temporal shape: a fast
rise and an (approximately) exponential decay \cite{Fishman 1995}. 
The number of pulses, their chaotic
time sequence and their relative intensities vary drastically from burst to burst and are
not predicted by the GRB models. The main properties of resolved pulses and of the
AGs of GRBs and SHBs, as well as the correlations between different observables are
what the models ought to predict or understand. Since LGRBs and SHBs have different
progenitors, they will be discussed separately.

\subsection{The cannonball model}

\begin{figure}
\centering
\includegraphics[width=6.5 cm]{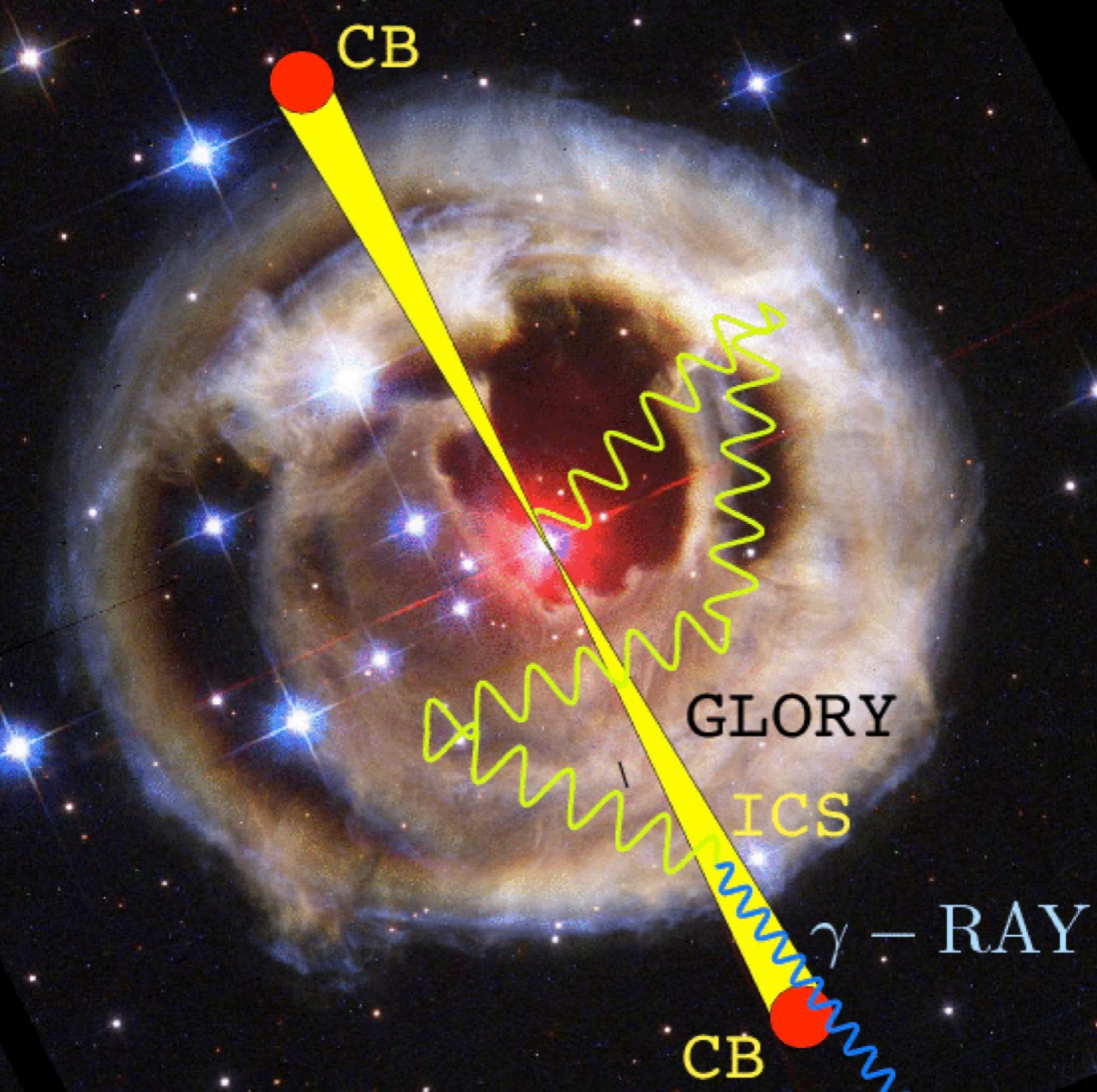}
\caption{Electrons in a cannonball inverse Compton scatter photons in the glory of light
surrounding a newly-born compact object, launching them forward as a narrow beam of 
$\gamma$ rays.}
\label{fig:glory}
\end{figure}

The CB model \cite{DD2004} is illustrated in Figure \ref{fig:glory}. 
In it, bipolar jets of highly
relativistic plasmoids (a.k.a.~CBs) are assumed to be launched by matter falling onto a
newly born compact stellar object \cite{Shaviv}. SNe of Type Ic (the broad-line stripped-envelope
ones) thus generate ``SN-GRBs". Similarly, mergers in NS/NS and NS/BH binaries give rise to SHBs. 
SN-less GRBs are produced in high-mass X-ray binaries, as a NS accreting mass
from a companion suffers a phase transition to a quark star \cite{Shaviv}. 
Finally X-ray flashes (XRFs)
are simply GRBs observed from a relatively large angle relative to the CBs' emission axis.

The prompt $\gamma$-ray pulses are produced by ICS by the electrons enclosed in the CBs
of the radiation surrounding the launch site: the ``glory". In SN-GRBs, this glory is
the light halo formed around the progenitor star by scattered light from pre-supernova
ejections \cite{DD2004}. 
In SN-less GRBs the glory can be light from the massive star companion,
or the radiation emitted from the accretion disk formed around the NS.
In SHBs it can be the X-ray radiation from an accretion disk formed around the NS
remnant by fall back of tidally disrupted material or debris from the final explosion
of the lighter NS after it lost most of its mass \cite{Blinnikov}.

The prompt radiation is more than intense enough to completely ionize the interstellar
medium (ISM) of the host galaxy along the path that the CBs will follow. When a CB enters
this medium, it decelerates by sweeping in the nuclei and electrons in front of it. The
swept-in particles are Fermi accelerated to high energies by the turbulent magnetic fields
present or generated in the CBs by the merging inner and interstellar plasmas. The
accelerated electrons emit synchrotron radiation (SR), the dominant AG of SN-GRBs,
that usually take place in dense stellar regions: the molecular clouds where most SNe
occur.

In SN-less GRBs and SHBs with a millisecond pulsar remnant, which usually take
place in much lower density environments than those of SN-GRBs, the AG appears
to be dominated by the radiation emitted from the pulsar wind nebula (PWN)
 \cite{Dado} \cite{Dado2017}.

\subsection{The fireball model}

\begin{figure}
\centering
\includegraphics[width=8.5 cm]{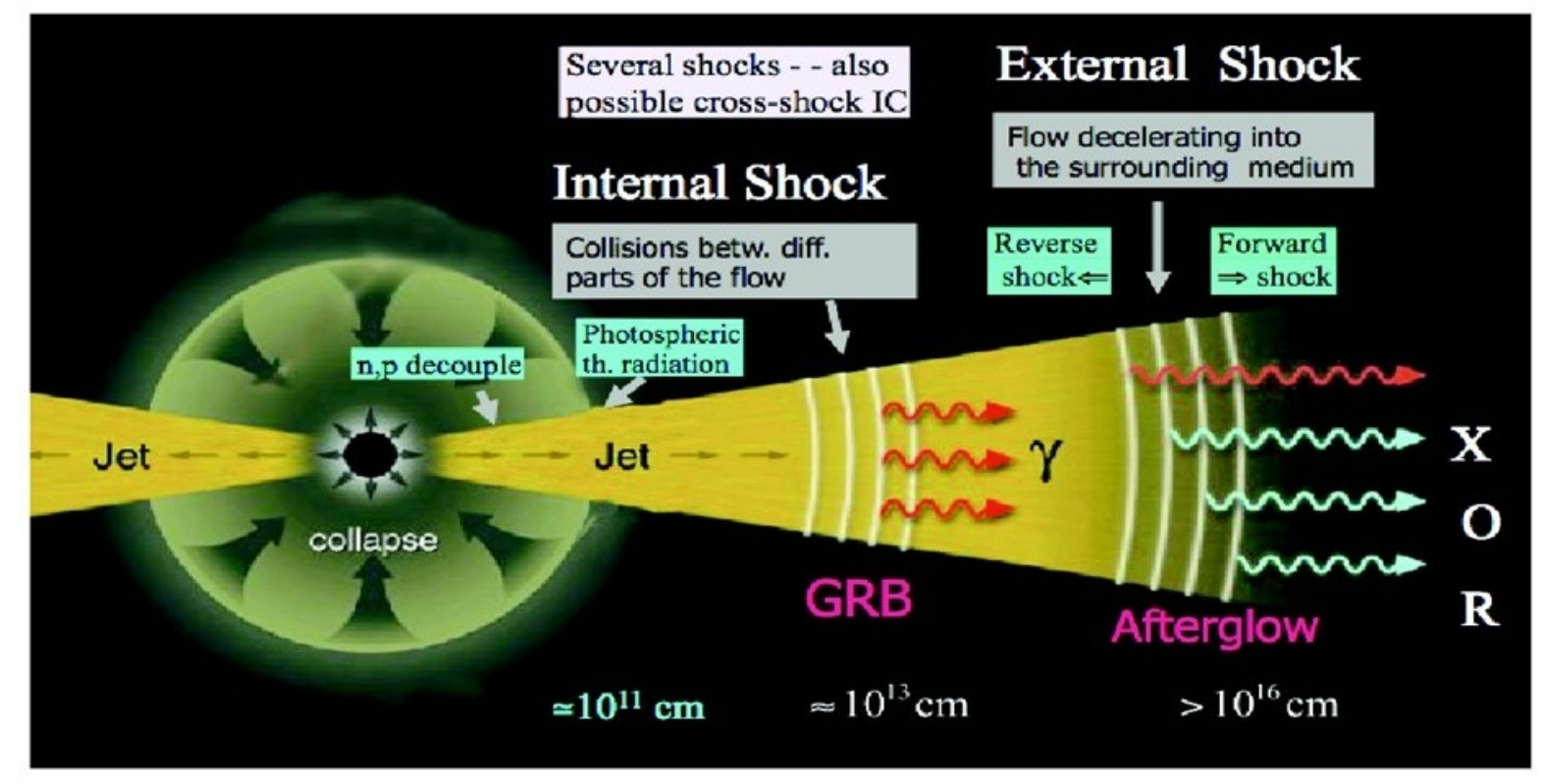}
\caption{Schematic description of the fireball model of GRBs \cite{Meszaros2014}.}
\label{fig:FBmodel}
\end{figure}

The FB models of GRBs evolved a long way from the original spherical $e^+ e^- \gamma$
fireball \cite{Goodman1} 
to the current ``collimated-fireball" models \cite{FBM Reviews}. A very popular version is
illustrated in Figure \ref{fig:FBmodel}, borrowed from \cite{Meszaros2014}. 
Long GRBs are produced by a jet of highly
relativistic conical shells of ordinary matter launched by {\it collapsars} --the collapse of a
massive star to a black hole-- either directly without a supernova ({\it failed} supernova)
\cite{WoosleyA} \cite{WoosleyB}, or indirectly in a  
{\it hypernova}: the delayed collapse of the newly-born compact
object to a BH by accretion of fall back material in a core-collapse SN 
\cite{ADR, MacFadyen }. 
SHBs are assumed to be produced by similar jets, launched in the merger
of a NS with another one, or with a BH.

In the current FB models, the prompt emission pulses are assumed to be produced
by synchrotron radiation (SR) from highly relativistic electrons shock-accelerated in the
collisions between colliding conical shells. The collision of the shells with the circumburst
medium is assumed to drive a forward shock into this medium or the pre-ejected
stellar wind, and a reverse shock towards the inner shells. The electrons produce the
AG \cite{FBM Reviews} on top of the light of a hypernova \cite{MacFadyen } in LGRBs, or a 
{\it macronova} in SHBs \cite{Macronova}.
 The reverse shock produces the optical photons while ICS of the SR in the forward
blast wave produces photons of GeV to TeV energy.

\section{The prompt GRB emission}

 The CB-model's basic assumption is that, in analogy with quasars and micro-quasars, a SNIc event
results in the axial emission of opposite jets of one or more CBs, made of ordinary matter.
The $\gamma$-rays of a GRB are produced by jets of CBs with an initial bulk-motion Lorentz
factor $\gamma_0\!\equiv\!\gamma(t\!=\!0)\!\gg\!1$. Although initially expanding in their 
rest system at a speed of the order of the relativistic sound speed, $c/\sqrt{3}$, 
CBs are effectively point-like for an earthly observer at a cosmological distance.

The electrons in a CB inverse-Compton-scatter the ambient photons they encounter.
This results in a {\it prompt} $\gamma$-ray beam of aperture $\simeq\! 1/\gamma_0\!\ll\! 1$
 around the CB's direction.
Viewed by an observer at an angle $\theta$ relative to the CB's direction, the individual photons
are boosted in energy by a Doppler factor 
$\delta_0\!\equiv\!\delta(t\!=\!0)\!=\!1/[\gamma_0\,(1\!-\!\beta\,\cos\theta)]$
 or, to a good
approximation for $\gamma_0^2\!\gg\!1$ and $\theta^2\!\ll\!1$, 
 $\delta_0\!\simeq\!2\gamma_0/(1\!+\!\gamma_0^2\theta^2)$.
 
 \subsection{The GRB polarization (Test 1)}
 
 For the most probable viewing angles ($\theta\!\sim \!1/\gamma_0$) the polarization of the prompt,
inverse Compton scattered photons is linear and its predicted \cite{jet}
magnitude is:
 \begin{equation}
 \Pi\!=\!2\gamma_0^2\,\theta^2/(1\!+\!\gamma_0^4\theta^4)\!=\!{\cal{O}}(1)
 \label{eq:Pol}
 \end{equation}
Very luminous or very dim GRBs are likely, respectively, to have been observed very
near ($\gamma_0^2\,\theta^2\!\ll\!1$) or very far off-axis ($\gamma_0^2\,\theta^2\!\gg\!1$), 
thus resulting in considerable but not near-to-maximal polarization.

In the standard FB models, both the prompt GRB and the AG are produced by
SR from high energy electrons, shock- and Fermi-accelerated in collisions 
between conical shells, and between conical shells and the ISM, respectively.
Such acceleration requires highly turbulent magnetic fields in the acceleration region,
resulting in a rather small net polarization. Indeed, the AGs are observed to display a
small polarization \cite{PolarizationAG }
as predicted by both the CB and the FB models. But, while the
 polarization of the prompt emission in the FB models is expected to be small, like that of
the AG, a rather large linear polarization of the prompt emission has been observed in
 most GRBs where it was measured \cite{Polarization}.
 In Table \ref{Table1} the available polarization data are reported. Though these measurements
are challenging, the message that the polarization is large is loud and clear.

\begin{table*}
\caption{GRBs with measured $\gamma$-ray polarization during the prompt emission.}
\label{table1}
\centering
\begin{tabular}{l l l l l l}
\hline
\hline
~~~GRB~~~ & Polarization(\%) & ~CL~ &~~~Reference~\cite{Polarization} ~~~ & Polarimetry~~  \\
~930131~~ & \,\,\,$>\!35$,   & 90\% & Willis et al.   2005 & BATSE (Albedo) \\
~960924~~ & \,\,\,$>\!50$    & 90\% & Willis et al.   2005 & BATSE (Albedo) \\
~021206~~ &\,\, $80\pm 20$     & ???  & Coburn \& Boggs 2003 & RHESSI~~~~~~~  \\
041219A~~ & \,\,\,\,$98\pm 33$  & 68\% & Kalemci et al.  2007 & INTEGRAL-SPI~~ \\
100826A~~ &\,\, $27\pm 11$      & 99\% & Yonetoku et al. 2011 & IKARUS-GAP~~~~ \\
110301A~~ & \,\,\,$70\pm 22$      & 68\% & Yonetoku et al. 2012 & IKARUS-GAP~~~~ \\
~110721~~ & 84\,+16/-28      & 68\% & Yonetoku et al. 2012 & IKARUS-GAP~~~~ \\
~061122~~ & \,\,\, $>\!60$   & 68\% & Gotz et al.     2013 & INTEGRAL-IBIS~ \\
140206A~~ & \,\,\, $>\!48$   & 68\% & Gotz et al. ~~~~2014 & INTEGRAL-IBIS~ \\
160821A~~ &\,66\,+27/-26     & 99\% & Sharma et al.   2019 & AstroSat-CZTI~  \\
190530A~~ &\,$55.4\pm 21.3$     & 99\% & Gupta et al. 2022 & AstroSat-CZTI~  \\
\hline
\end{tabular}
\label{Table1}
\end{table*}

\subsection{Prompt-observable correlations (Test 2)}

The ICS of glory photons of energy $\epsilon$ by a CB boosts their energy, as seen by an
observer at redshift $z$, to $E_\gamma\!=\!\gamma_0\,\delta_0\,\epsilon/(1\!+\!z)$. 
Consequently, the peak energy $E_p$ of their time-integrated energy distribution satisfies
\begin{equation}
(1+z)\,E_p\!\approx\! \gamma_0\,\delta_0\, \epsilon_p\,, 
\label{eq:Ep0}
\end{equation}
with $\epsilon_p$ the peak energy of the glory.

In the Thomson regime the nearly isotropic distribution (in the CB rest frame)
of a total number $n_\gamma$
of IC-scattered photons  is beamed into an angular distribution
$dn_\gamma/d\Omega\!\approx\! (n_\gamma/4\,\pi)\,\delta^2$
in the observer's frame. Consequently, the isotropic-equivalent
total energy of the photons satisfies
\begin{equation}
E_{iso}\!\propto\! \gamma_0\, \delta_0^3\, \epsilon_p. 
\label{eq:Eiso}
\end{equation}
Hence, both ordinary LGRBs and SGRBs, which in the CB model are GRBs viewed
mostly from an angle $\theta\!\approx\!1/\gamma$ 
(for which $\delta_0\!\approx\!\gamma_0$), satisfy the correlation
\begin{equation}
(1+z)\,E_p\propto [E_{iso}]^{1/2},
\label{eq:Corr1}
\end{equation}
while far off-axis ones ($\theta^2\! \gg \! 1/\gamma^2$) 
have a much lower $E_{iso}$, and satisfy 
\begin{equation}
(1+z)\,E_p\propto [E_{iso}]^{1/3}.
\label{eq:Corr2}
\end{equation}

\begin{figure}[]
\centering
\includegraphics[width=8.5 cm]{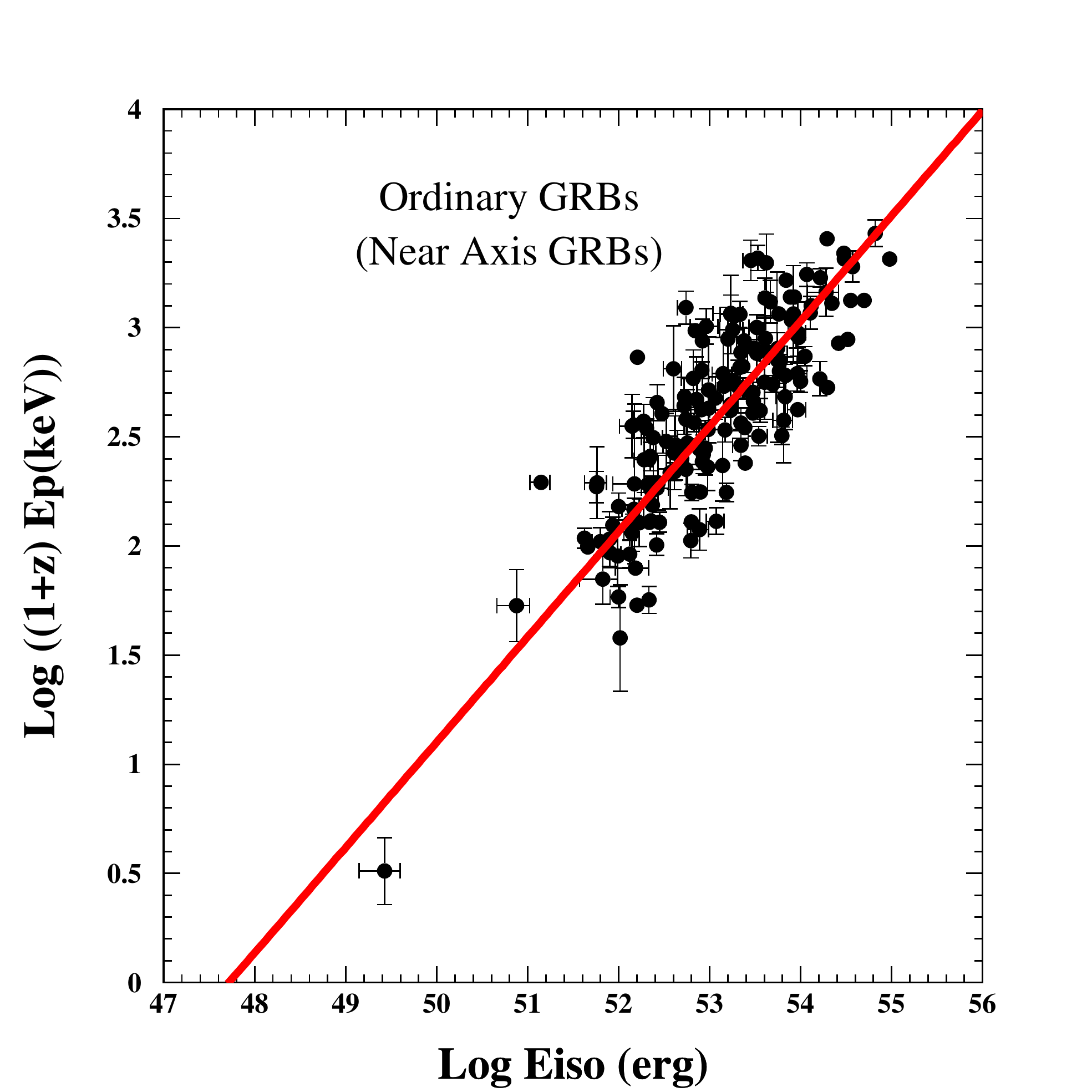}
\caption{The $[E_p, E_{iso}]$ correlation in ordinary LGRBs viewed near axis. 
The line is the best fit, whose slope is $0.48\!\pm\! 0.02$,
consistent with the CB model prediction of 
Equation \ref{eq:Corr1}. 
The lowest $E_{iso}$   GRB  is  020903 at $z\!=\!0.25$  (HETE).}
\label{fig:epeiso17GRBs}
\end{figure}

\begin{figure}[]
\centering
\includegraphics[width=8.5 cm]{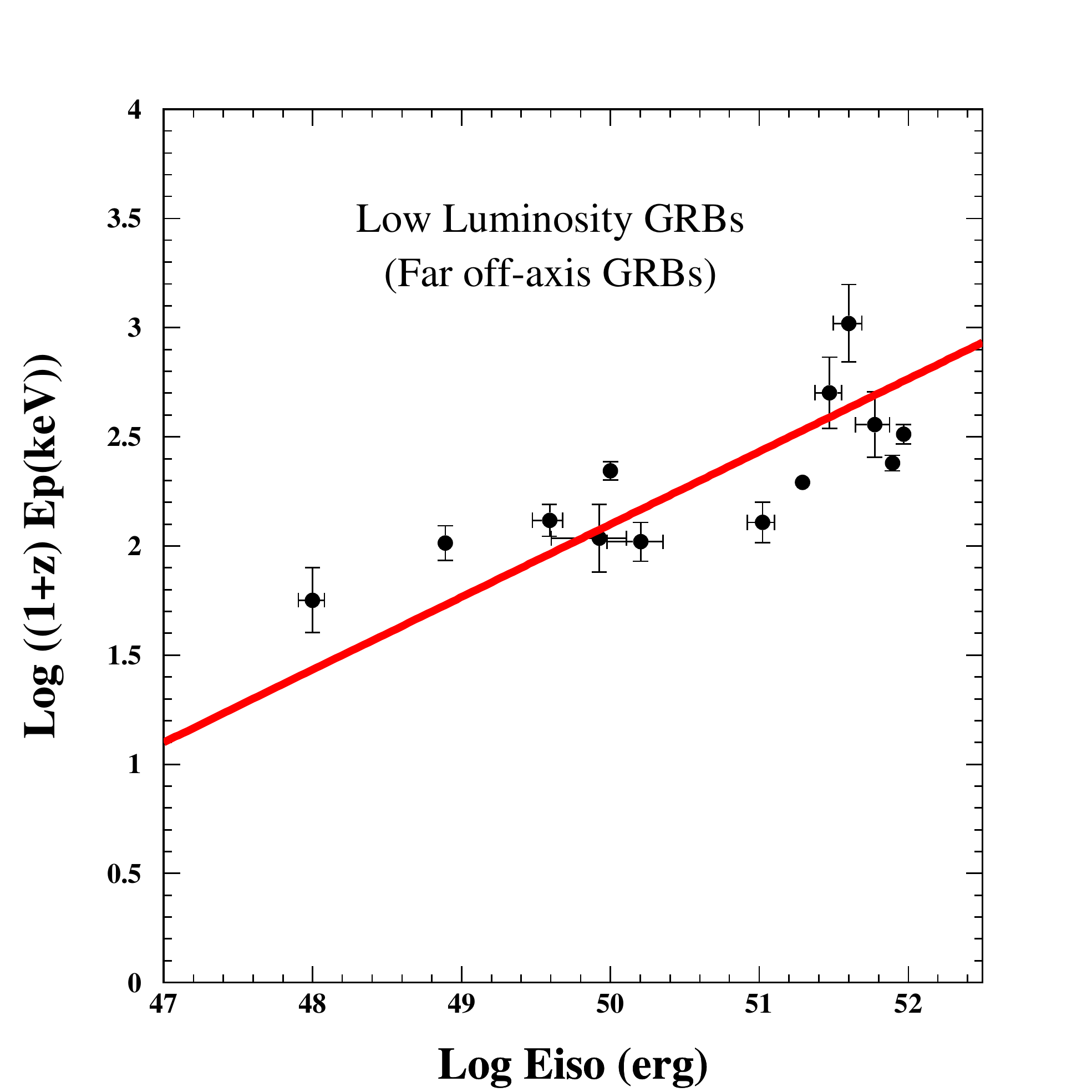}
\caption{The $[E_p,E_{iso}]$ correlation in LGRBs viewed far off axis 
(which include the so-called low-luminosity LGRBs, and XRFs)  \cite{Jonker}. 
The line is the CB model prediction \cite{Correlations} of Equation \ref{eq:Corr2}.}
\label{fig:epeisoLLGRBs}
\end{figure}

\begin{figure}[]
\includegraphics[width=8.5 cm]{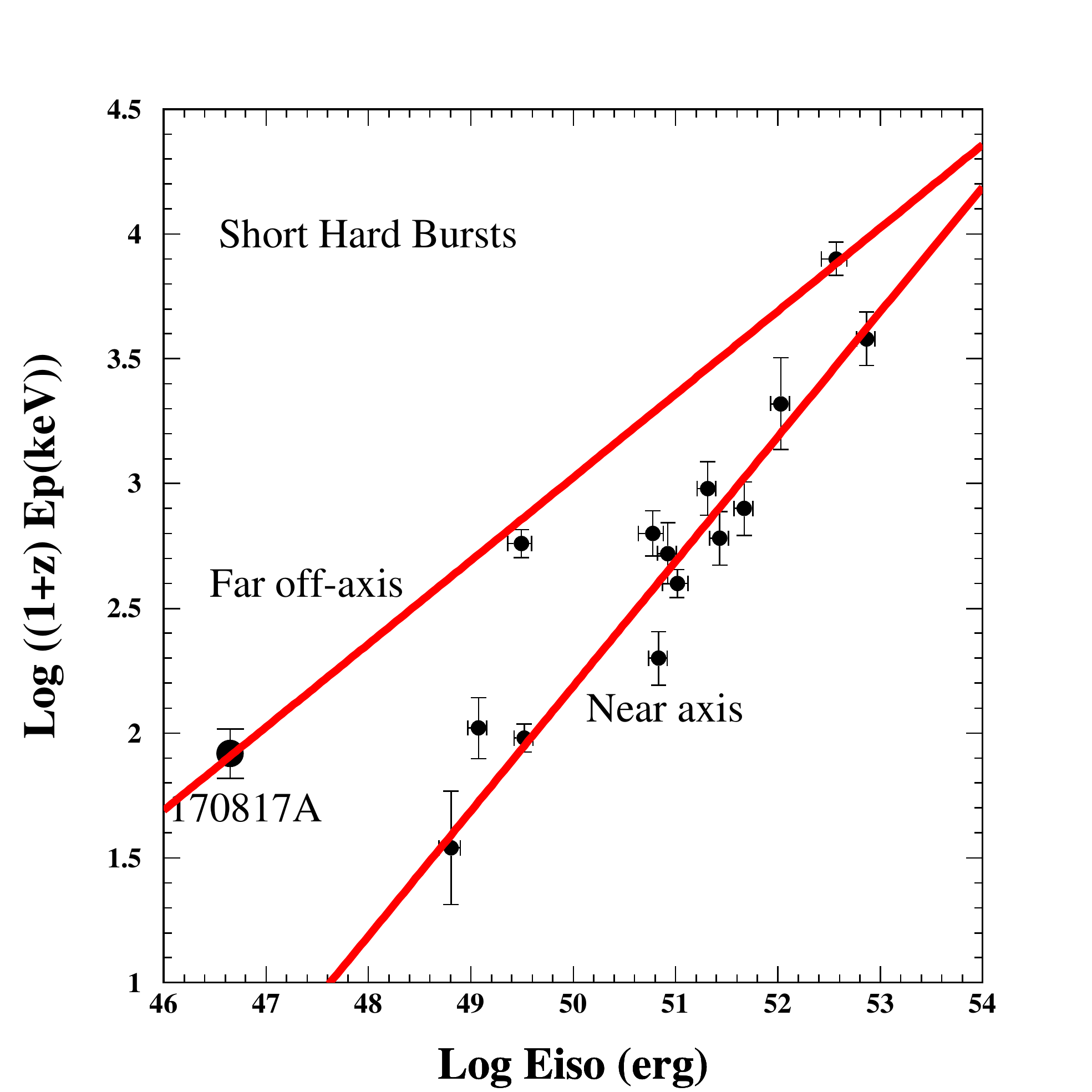}
\caption{The $[E_p,E_{iso}]$ correlations in SHBs.
The lines are the CB model predicted correlations as 
given by Equations \ref{eq:Corr1} and \ref{eq:Corr2}.}
\label{fig:epeisoallshbs}
\end{figure}

These $[E_p, E_{iso}]$ correlations
predicted by the CB model  \cite{Correlations}, were 
later empirically discovered  \cite{Amati2002} in ordinary LGRBs. 
They are shown in Figures \ref{fig:epeiso17GRBs}
 and \ref{fig:epeisoLLGRBs} for
GRBs of known redshift. The $[E_p, E_{iso}]$ correlation predicted by the CB model for low luminosity SGRBs is presented in Figure \ref{fig:epeisoallshbs}. The figure includes SHB170817A, a very 
soft burst known for its association with a gravitational-wave signal  \cite{20a,20b,20c,20d,20e}.
This SHB will deserve its own chapter.

An apparent weakness of Equations \ref{eq:Ep0} and \ref{eq:Eiso} is that, strictly 
speaking, they refer to single-pulse GRBs: they are written for just one value 
of the Lorentz factor and of the viewing angle. Based on Fermi data, it
has been shown that the $[E_p,E_{iso}]$ 
correlation is very well fit by 
$(1+z)E_p\! \propto\! E_{iso}^b$ for single or first pulses ($b\!=\!0.465\!\pm\!0.044$;
$\chi^2 / {\rm dof}\!=\!1.10$),
for the rest of the pulses ($b\!=\!0.503 \pm 0.050$;  $\chi^2/{\rm dof}\!=\!1.09$)
and even for the entire GRB ($b\!=\!0.499\!\pm\!0.035$; $\chi^2/{\rm dof}\!=\!1.04$) \cite{BR}.
These results are in excellent agreement with the CB-model prediction, 
Equation \ref{eq:Corr1}.

The correlations just discussed, snuggly satisfied by the data and extending over
many orders of magnitude, strongly support the contention that the prompt photons of
high and low luminosity GRBs --as well as SHBs and XRFs-- are emitted by an effectively
point-like highly relativistic source --such as a CB-- viewed at different angles. They are
not a consequence of Fireball models.

\subsection{Temporal shape of prompt pulses (Test 3)}

As stated before,
GRBs consist of fast rising, roughly exponentially decaying individual short pulses
\cite{Fishman 1995}, 
dubbed ``FRED". Their number, time sequence, and relative intensities vary drastically
between bursts and are not predictable by the current GRB models.

\subsubsection{\it Pulse shapes in the CB model}

The typical
FRED shape of individual pulses is predicted in the CB model, and their durations 
are understood.
The glory's light has a thin thermal bremsstrahlung spectrum 
$dn_g/d\epsilon\!\propto\! \epsilon^{-\alpha}\,\exp [-\epsilon/\epsilon_p]$ \cite{ThBrem}.
Inverse Compton Scattering of this light by the electrons in a CB produces a GRB pulse
with a time and energy dependence well described \cite{Dado2009a} by:
\begin{equation}
E\,{d^2N_\gamma\over dE\,dt}\!\propto\!{t^2\over(t^2\!+\!\Delta^2)^2}\,
 E^{1-\alpha}\,\exp[-E/\mathcal{E}_p(t)]
 \label{eq:pulseShape}
\end{equation} 
where $\mathcal{E}_p(t)$ is the pulse's peak energy at time $t$, 
$E_p\!\equiv\!\mathcal{E}_p(t\!=\!0)$ is its maximum,
$\Delta$ is approximately the peak time of the pulse in the observer's frame --originating
at the time when the CB becomes transparent to its internal radiation.

In Equation \ref{eq:pulseShape}, the early quadratic temporal rise is due to the increasing cross
section, $\pi\, R_{CB}^2\!\propto\! t^2$
of the fast-expanding CB when it is still opaque to radiation. When
it becomes transparent its effective cross section for ICS becomes a
 constant: the Thomson cross section times the number of electrons in the CB. That and
 the density $n_g$ of the ambient photons --which for a distance 
$r\!=\!\gamma\,\delta\,c\,t/(1\!+\!z)\!>\! R_g$ (the
 radius of the glory) decreases like $n_g(r)\!\propto\! 1/r^2\!\propto 1/t^2$--
contributes a factor $t^{-2}$ to the late temporal decline.

If CBs are launched along the axis of the glory of a torus-like pulsar wind nebula,
or of an accretion disk with a radius $R_g$, the glory photons at a distance $r$ from the
center intercept the CB at an angle $\arccos[r/(r^2+R_g^2)^{1/2}]$
resulting in a $t$-dependent peak energy 
\begin{equation}
\mathcal{E}_p(t)\approx\mathcal{E}_p(0)[1\!-\!t/(t^2\!+\!\tau^2)^{1/ 2}]
\label{eq:Epoft}
\end{equation}
with $\tau\! =\! R_g (1 \!+\! z)/\gamma\, \delta\ c$, valid also at late times for an
approximately spherical glory.

\begin{figure}[]
\centering
\includegraphics[width=8.5 cm]{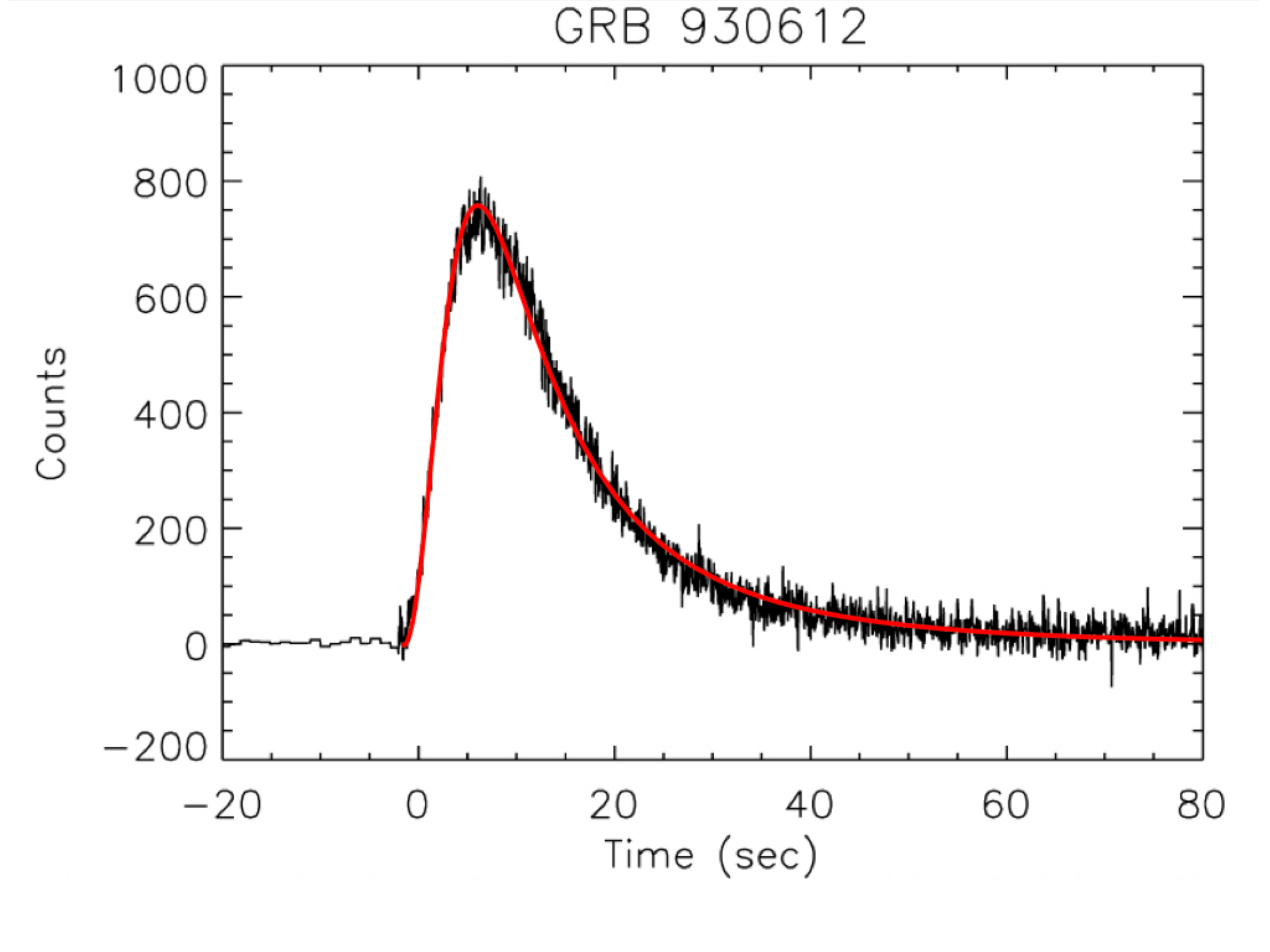}
\caption{The pulse shape of GRB930612 measured with BATSE 
 aboard CGRO and the shape given by 
Equation \ref{eq:pulseShape} for
$\Delta\!=\!7.1$ s and $\tau\!=\!70.1$ s, the best CB-model fit to re-bined data.}
\label{fig:Pulse930612}
\end{figure}

For LGRBs with $\tau\!\gg\!\Delta$, Equations \ref{eq:pulseShape} and \ref{eq:Epoft}
yield half maximum values at 
$t\!\simeq\!0.41\Delta$ and $t\!\simeq\!2.41\Delta$, 
resulting in a full width at half maximum
${\rm FWHM}\!\simeq\!2\Delta$, a rise time from half maximum to peak value
${\rm RT}\!\simeq\!0.59\Delta$ and a decay time from peak count to half peak, 
${\rm DT}\!\simeq\!1.41\Delta$. Consequently
$\rm {RT/DT}\!\simeq\!0.42$ and $\rm{RT\!\simeq\!0.30 \,FWHM}$.
The pulse shape given by Equations \ref{eq:pulseShape} and \ref{eq:Epoft},
with $\Delta\!=\!7.1$ s and $\tau\!=\!70.1$ s,
 is shown in Figure \ref{fig:Pulse930612} for the particularly well measured single-pulse of
GRB930612 \cite{Kocevski}.

\begin{figure}[]
\centering
\includegraphics[width=8.5 cm,angle=0.6]{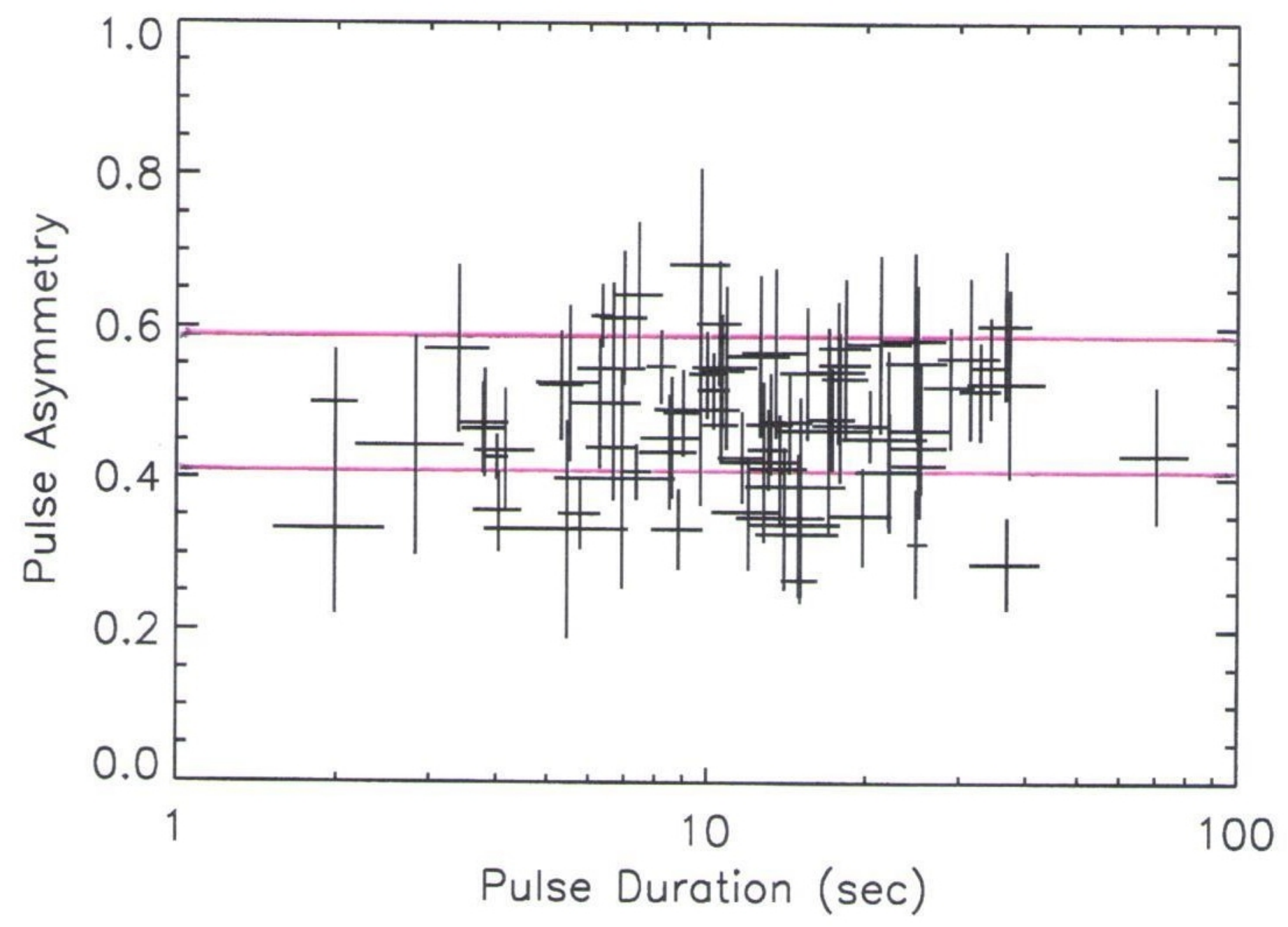}
\caption{Comparison between the observed asymmetry ratio RT/DT 
as function of pulse duration reported in \cite{Kocevski}
for a sample of 77 resolved LGRB pulses measured with BATSE 
aboard CGRO (with a  mean value ${\rm RT/DT}\!=\!0.47\pm 0.08$), 
and the CB model predicted asymmetry $0.41\!<\!{\rm RT/DT}\!<\!0.58$
for $\Delta\!<\!\tau\!<\!\infty$ (solid lines).}
\label{fig:Asymmetry_LGRBs}
\end{figure}

\begin{figure}[]
\centering
\includegraphics[width=8.5 cm,angle=0.6]{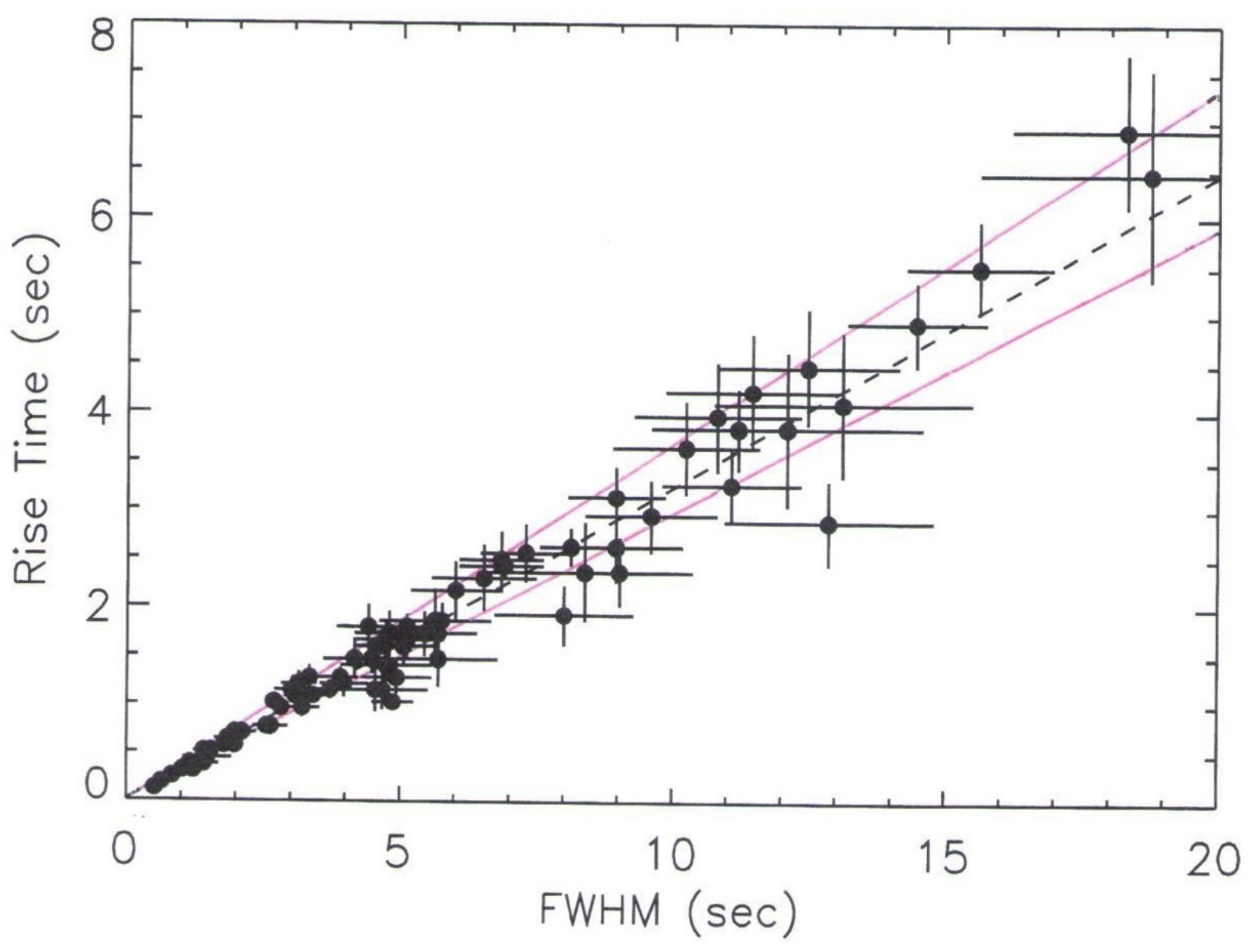}
\caption{Comparison between the rise time RT versus the FWHM reported  
in \cite{Kocevski} for a sample of 77 resolved pulses measured with BATSE aboard
CGRO. The dotted line is best fit ratio ${\rm RT/FWHM}\!=\!0.32$ and the solid lines
are CB model expected boundaries  $0.29\!<\! {\rm RT/FWHM} \!<\!0.36$ for LGRBs.}
\label{fig:RTVFWHM_LGRBs}
\end{figure}

In most LGRBs $\tau\!\gg\!\Delta$ and the CB model predicts ${\rm RT/DT}\!\approx\!0.42$,
 and ${\rm RT/FWHM}\!\approx\!0.29$, changing very little with $\tau$ if $\tau\!\gg\!\Delta$.
Even in the very rare cases where
 $\tau/\Delta\!\approx\!1$, ${\rm RT/DT}\!\approx\! 0.57$ and ${\rm RT/FWHM}\!\approx\! 0.36$. 
 In Figures 
 \ref{fig:Asymmetry_LGRBs} and \ref{fig:RTVFWHM_LGRBs} the predicted ratios
 RT/DT and RT/FWHM for $\Delta\!<\!\tau\!<\!\infty$ are compared to their best fit values in the 77
 resolved pulses of BATSE/CGRO LGRBs reported in \cite{Kocevski}. 
 As shown in
 these figures their best fit values lie well within the narrow area between the predicted 
 CB-model boundaries. 
The mean observed values, ${\rm RT/DT}\!=\!0.47\pm 0.08$ and 
${\rm RT/FWHM}\!=\!0.31\pm 0.03$ 
 \cite{Kocevski},
are very close to the CB-model's expected values 
${\rm RT/DT}\!=\!0.44$ and ${\rm RT/FWHM}\!=\!0.31$ for $\tau\!=\! 10\,\Delta$.

\begin{figure}[]
\centering
\includegraphics[width=8.5 cm]{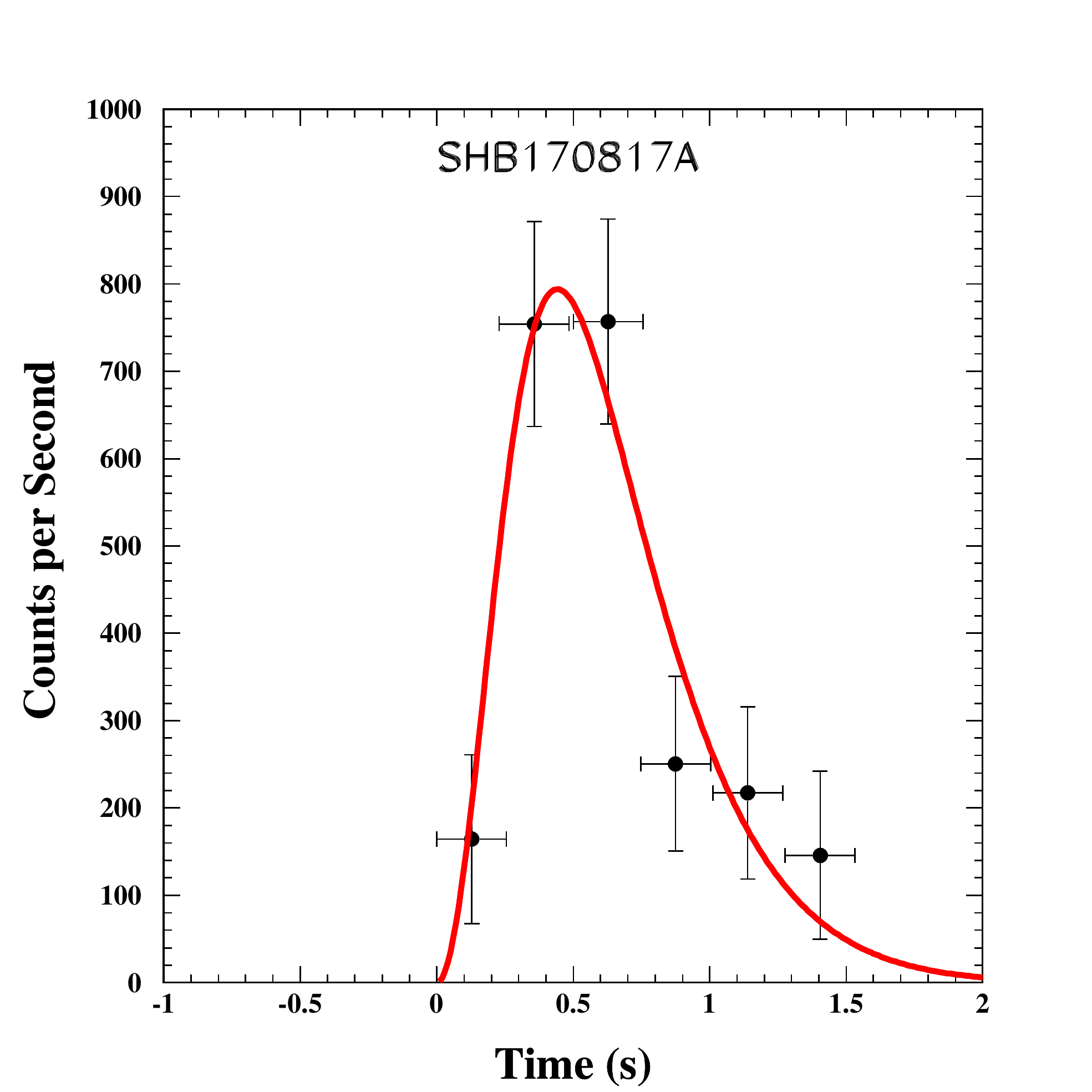}
\caption{The pulse shape of SHB170817A measured with the Fermi-GBM 
 \cite{20a,20b,20c,20d,20e}
and the best fit pulse shape given by Equations \ref{eq:pulseShape} 
and \ref{eq:Epoft} with 
$\Delta\!=\!0.62$ s and $\tau\!=\!0.57$ s, $\chi^2/\rm {dof}\!=\!0.95$. }
\label{fig:fig02}
\end{figure}

In Figure \ref{fig:fig02} the measured and CB-model  pulse shapes of
SHB170817A are compared.
The best-fit
light curve has a maximum at $t\!=\!0.43$ s,  half maxima at $t\!=\!0.215$ s 
and $t\!=\!0.855$ s, 
${\rm RT/DT}\!=\!0.50$ and ${\rm RT/FWHM}\!=\!0.34$.

\subsubsection{\it Pulse shapes in fireball models}

In current standard FB models \cite{FBM Reviews} the GRB prompt pulses are produced by
synchrotron radiation from shock-accelerated electrons in collisions between overtaking
thin shells ejected by the central engine, or by internal shocks in the ejected conical jet.
Only for the fast decline phase of the prompt emission, and only in the limits of very
thin shells and fast cooling, falsifiable predictions have been derived. In these limits the
fast decline phase of a pulse was obtained from the relativistic curvature effect 
\cite{Curvature},\cite{Kobayashi},\cite{NorrisHakkila}.
It yielded a power law decay
$F_\nu(t)\!\propto\!(t\!-\!t_i)^{-(\beta + 2)}\nu^{-\beta}$, where $t_i$ is the beginning time
of the decay phase, and $\beta$ is the spectral index of prompt emission.

The observed decay of the SHB170817A pulse, accompanied by a fast spectral
softening before the afterglow took over, could be roughly reproduced by adjusting a
 beginning time of the decay and replacing the constant spectral index of the FB model
by the observed time-dependent one \cite{Curvature}.

\section{The afterglow of GRBs}

In the CB model, the afterglow (AG) of SN-associated GRBs (SN-GRBs) is mainly
synchrotron radiation (SR) from the relativistic jets of CBs launched in core collapse SNe
of type Ic into the dense interstellar medium --such as a molecular cloud-- where most
SNeIc take place. The AG of SN-less GRBs and SHBs is dominated by a pulsar wind
nebula (PWN) emission powered by the spin down of a newly-born millisecond pulsar 
\cite{Dado,Dado2017}.

The colossal energy of the collimated $\gamma$ rays of a GRB is more than sufficient to
fully ionize the interstellar medium (ISM) it travels through. In SN-GRBs, this ionized
medium is swept into the CBs and generates within them a turbulent magnetic field. Its
magnetic energy density is assumed to be in approximate equipartition with that of
the swept in particles, as indicated by simulations \cite{PlasmaMerge}
and by the equilibrium between
the energy densities of galactic cosmic rays and magnetic fields.

A CB's Lorentz factor $\gamma(t)$ decreases as it collides with the ISM. The electrons that
enter a CB are Fermi accelerated there and cool by emission of SR, isotropic in the CB's
rest frame. In the observer's frame, the radiation is forward-beamed into a cone of
opening angle $\theta\!\sim\!1/\gamma(t)$ and is aberrated, Doppler boosted in energy and redshifted in
the usual way \cite{DD2004}.

The observed spectral energy density (SED) flux of the {\it unabsorbed} synchrotron
X-rays, $F_\nu(t)\!=\!\nu\,dN_\nu/d\nu$, has the form (see, e.g.~Eqs.~(28)-(30) in \cite{Dado2002}),
\begin{equation}
F_{\nu} \propto n(t)^{(\beta_x+1)/2}\,[\gamma(t)]^{3\,\beta_x-1}\,
[\delta(t)]^{\beta_x+3}\, \nu^{-\beta_x}\, ,
\label{eq:Fnu}
\end{equation}
where $n$ is the baryon density of the external medium encountered 
by the CB at a time $t$ and  $\beta_x$ is the spectral index 
of the emitted X-rays, $E\,dn_x/dE\propto E^{-\beta_x}$. 

The CBs are decelerated by the swept-in ionized material. Energy-momentum conservation
for such a plastic collision\footnote{The original assumption in the CB model was that the
interactions between a CB and the ISM were elastic. It was later realized, in view of the
shape of AGs at late times, that a plastic collision was  a better approximation
in the AG phase.}
between a CB of baryon number $N_B$,  radius $R$ and
initial Lorentz factor $\gamma_0\!\gg\! 1$, propagating in a constant-density ISM,
decelerates according to \cite{Dado2009a}:
\begin{equation}
\gamma(t) = {\gamma_0\over \left[\sqrt{(1+\theta^2\,\gamma_0^2)^2 +t/t_d}
          - \theta^2\,\gamma_0^2\right]^{1/2}}\,,
\label{eq:decelerate}          
\end{equation}
where $t$ is the time in the observer frame since the beginning of the AG emission
by a CB, and $t_d$ is its deceleration time-scale
\begin{equation}
t_d\!=\!{(1\!+\!z)\, N_{_B}/ 8\,c\, n\,\pi\, R^2\,\gamma_0^3}.
\label{eq:td}
\end{equation}

In the case of SN-less LGRBs the AGs --to be discussed in detail below--
 are compatible with the radiation
emitted by the pulsar's wind nebula, powered by the rotational energy loss of the newly
born quark star through magnetic dipole radiation, relativistic wind and high energy
charged-particle emission along open magnetic field lines \cite{Dado2017}.

\subsection{``Canonical" behavior of the AG of LGRBs (Test 4)}

\begin{figure}[]
\centering
\includegraphics[width=8.5 cm]{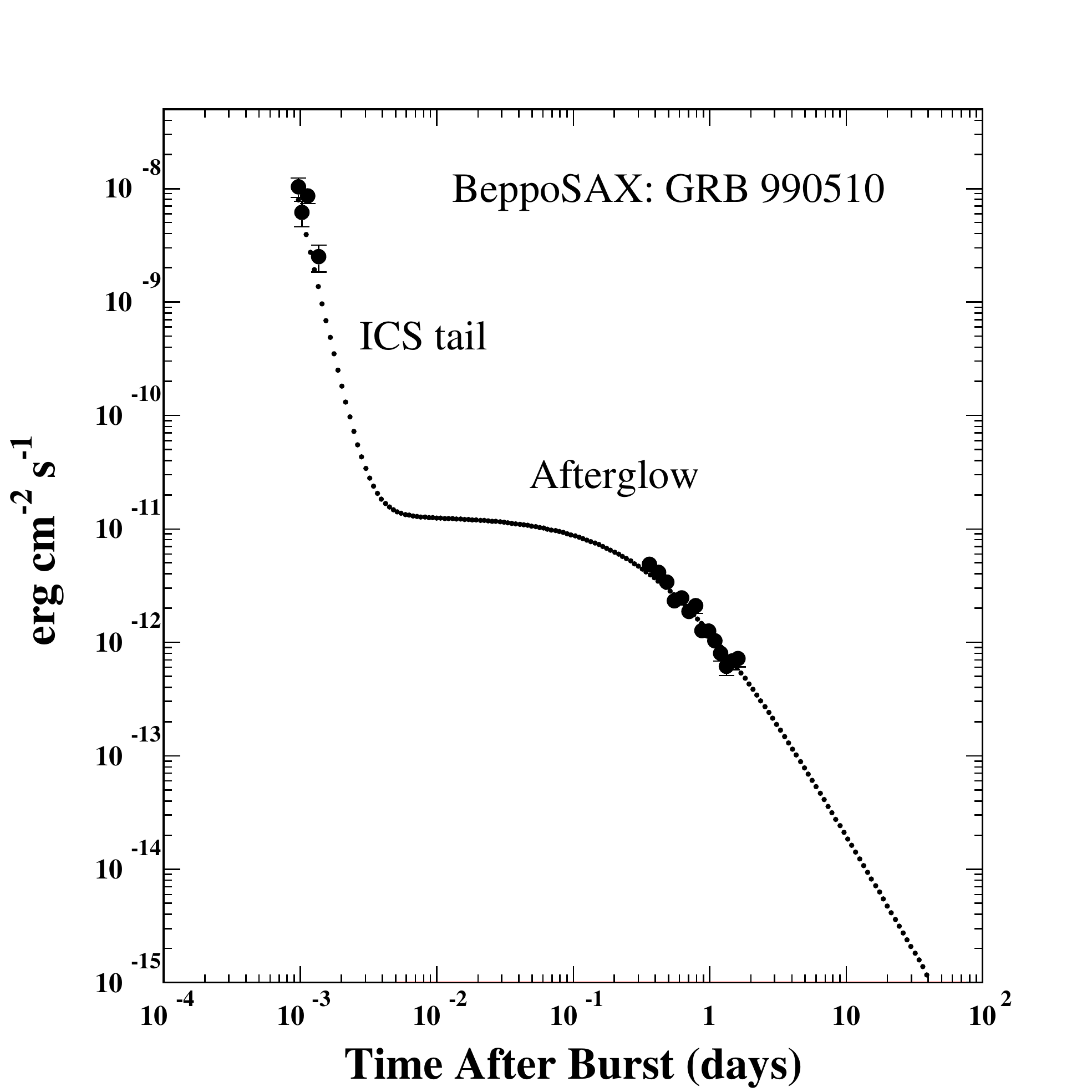}
\caption{The X-ray afterglow of GRB 990510 
measured with the telescopes aboard the BeppoSAX satellite
compared to the canonical X-ray afterglow 
predicted by the CB model \cite{Dado2002} of SN-GRBs.}
\label{fig:XAG_GRB990510}
\end{figure}

\begin{figure}[]
\centering
\includegraphics[width=8.5 cm]{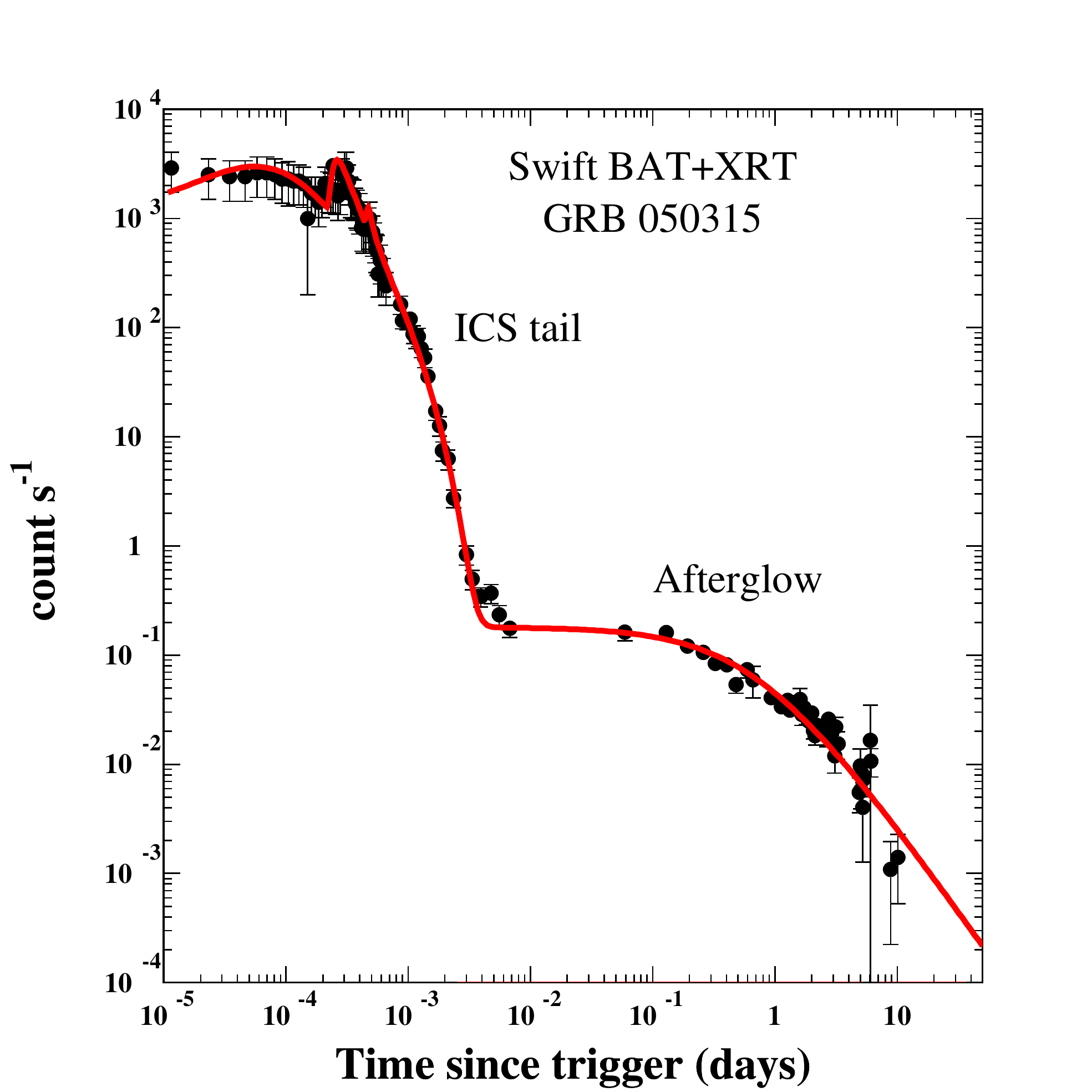}
\caption{The X-ray afterglow of GRB050315 
measured with the telescopes aboard  Swift 
compared to its best fit canonical X-ray afterglow
predicted by the CB model \cite{Dado2002} for SN-GRBs.}
\label{fig:XAG_GRB050315}
\end{figure}

In the CB model the prompt $\gamma$-ray emission was predicted to end with the exponential
temporal decay and fast spectral softening of Equation \ref{eq:pulseShape}, subsequently taken over
by a {\it canonical} X-ray AG, i.e.~an initial shallow decay phase (the {\it plateau}) that
breaks smoothly into a power-law decline.
The shallow decay phase lasts until
the time when the initial rest mass of the CB  is approximately doubled by
the swept-in relativistic mass. In Figures \ref{fig:XAG_GRB990510} 
and \ref{fig:XAG_GRB050315} this behaviour of the X-ray
AG of two LGRBs is shown to agree
 with the CB model prediction (see, e.g.~Figures 27-33)  in
\cite{Dado2002}). This AG-shape prediction was
 made long before it was first observed with the Swift X-ray 
Telescope in the AGs
of GRB050315 \cite{Vaughan} and GRB050319 \cite{Cusumano}
and subsequently categorized as ``canonical".

\subsection{Break time correlations (Test 5)}

In the CB model the Lorentz factor $\gamma(t)$ of a CB, decelerating as per Equation 
\ref{eq:decelerate}, changes slowly until $t_b$, a break time approximately given by 
\cite{Dado2009a}
\begin{equation}
t_b\!\approx\! (1\!+\!\gamma_0^2\theta^2)^2 t_d\,
\label{eq:tb}
\end{equation} 
where $t_d$ as in Equation \ref{eq:td}. This slow change is responsible for the plateau 
phase of SN-GRBs.

\begin{figure}[] 
\centering 
\includegraphics[width=8.5 cm]{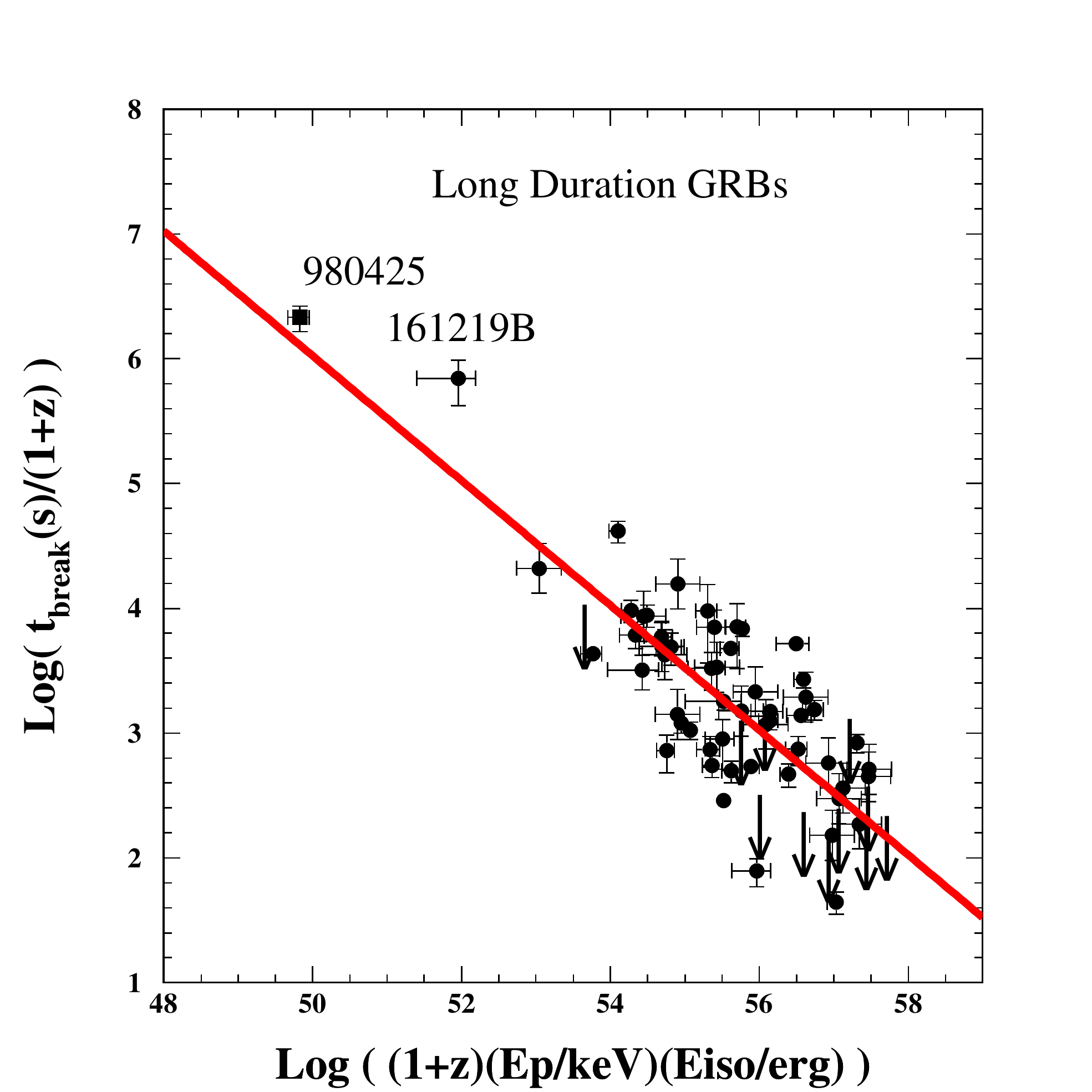}
\caption{The break time $t_b/(1+z)$ of the X-ray AG of SN-GRBs 
measured with the Swift XRT \cite{Swift}, as a function of $[(1\!+\!z)\,E_p\,E_{iso}]$. 
The line is the CB model correlation of Equation \ref{eq:pulseShape},  expected 
in SN-GRBs. SN-Less GRBs, to be discussed in Section \ref{subsec:SN-less},
are not included.} 
\label{fig:tbSNGRBs}
\end{figure}

The dependence of $E_p$ and $E_{iso}$ on $\gamma_0$ and $\delta_0$
 can be used to obtain from Equation \ref{eq:td} the correlation \cite{Dado 2013}
 \begin{equation} 
t_b/(1+z)\!\propto\![(1\!+\!z)\,E_p\,E_{iso}]^{-1/2}, 
\label{eq:3Corr}
\end{equation} 
The observed break time of the X-ray afterglow of LGRBs measured with the Swift satellite
XRT (X Ray Telescope) \cite{Swift} for LGRBs with known $z$, $E_p$ and $E_{iso}$, 
is compared in Figure \ref{fig:tbSNGRBs} to that
predicted by Equation \ref{eq:tb}, with satisfactory results.

\subsection{Post-break closure relations (Test 6)}

\begin{figure}[]
\centering
\includegraphics[width=8.5 cm]{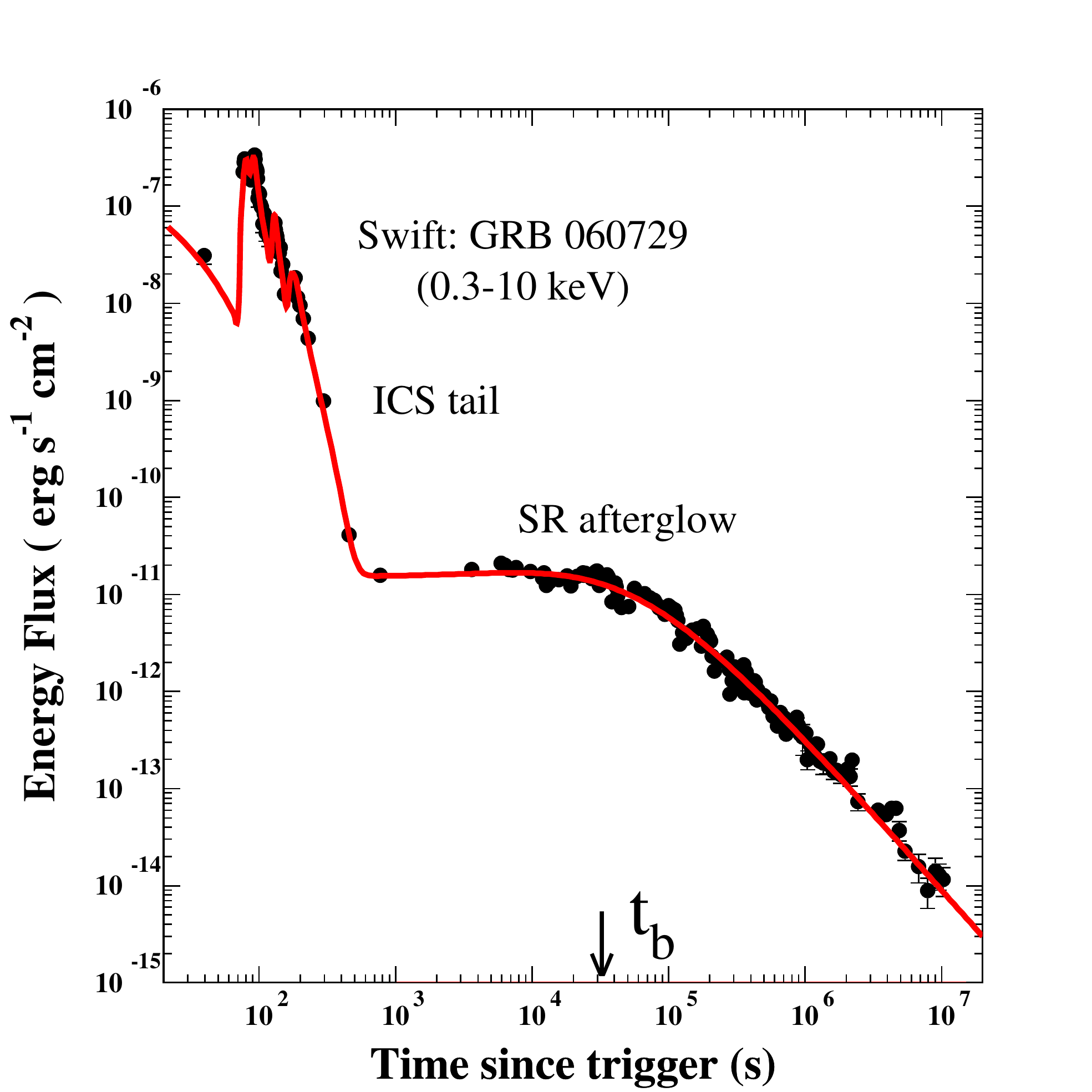}
\caption{The canonical light curve of the X-ray AG of
the SN-GRB 060729 measured with Swift XRT \cite{Swift}, 
and its best fit CB-model AG \cite{DD2004}, given by Equation \ref{eq:Fnu}.
The data also satisfy 
the CB model prediction $\alpha_x\!=\!\beta_x\!+\!1/2$.}
\label{fig:AGX_GRB060729}
\end{figure}

Well after the break, Equation \ref{eq:decelerate}  
yields $\delta(t)\!\approx\! 2\gamma(t)\!\propto\! t^{-1/4}$, 
which, when substituted in Equation \ref{eq:Fnu}, results in the late-time behavior
\begin{equation}
F_\nu(t\!\gg\!t_b)\!\propto\! t^{-\alpha_\nu}E^{-\beta_\nu},  
\label{eq:Fnulate}
\end{equation}
and the {\it closure} relation
\begin{equation}
\alpha_\nu\!=\!\beta_\nu\!+\!1/2.
\label{eq:closure}
\end{equation}
This relation is well satisfied by the X-ray AG of SN-GRBs \cite{Dado 2013} 
--shown in Figure \ref{fig:AGX_GRB060729} for the canonical case of
GRB060729 \cite{Swift}--
as long as the
CBs move in an ISM of roughly constant density. Indeed, this long followed up and well
X-ray AG is best fit with a temporal index $\alpha_x\!=\! 1.46\!\pm\!0.03$,
which agrees well with the predicted
$\alpha_x\!=\!\beta_x\!+\!1/2=\!1.49\!\pm\!0.07$ for an observed \cite{Swift}
$\beta_x\!=\!0.99\!\pm\!0.07$.

\begin{figure}[]
\centering
\includegraphics[width=8.5 cm]{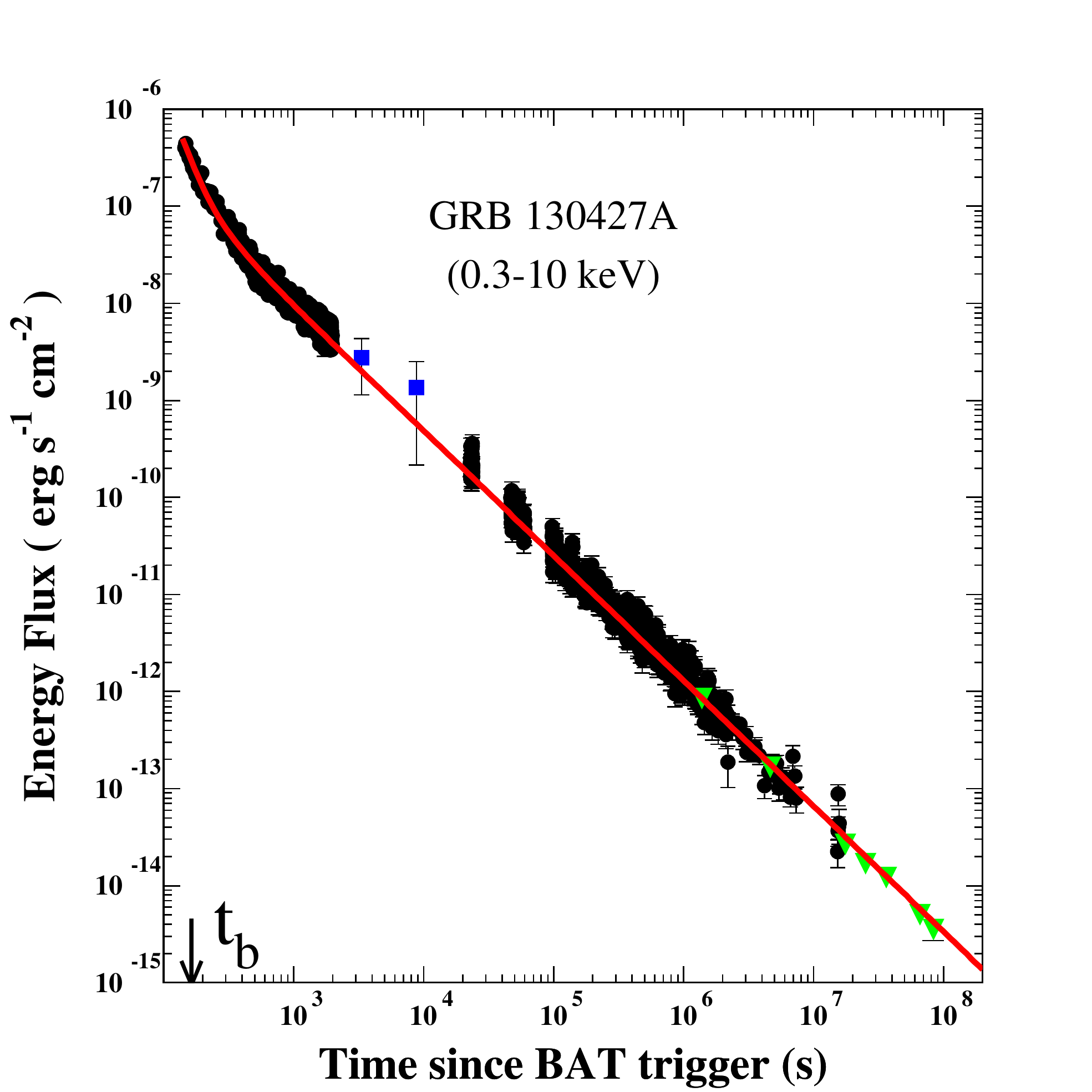}
\caption{The X-ray light-curve of the intense GRB 130427A, 
measured with Swift XRT \cite{Swift} (circles)
and with  XMM Newton  and  Chandra \cite{Pasquale} (triangles) up
to 83 Ms after burst, and its CB model best-fit with a start time and an
early break hidden under the prompt emission phase.
Also shown are the two MAXI data points \cite{Maselli} (squares)
at $t\!=\! 3257$ s and $t\!=\! 8821$ s. The best-fit 
power-law decline has an index $\alpha_x\!=\!1.29$. The temporal decay index 
predicted  by the CB model, Equation \ref{eq:closure}, for the measured spectral index 
\cite{Pasquale}
$\beta_x\!=\!0.79\!\pm\!0.03$  is  $\alpha_x\! =\!1.29\!\pm\! 0.03$.} 
\label{fig:Xar130427A}
\end{figure}

The most accurate test of the CB model prediction of Equation \ref{eq:closure} for a single
 SN-GRB was provided by the measurements of the X-ray AG of GRB 130427A, the
most intense GRB ever detected by Swift and followed with the Swift XRT and X-ray
telescopes aboard XMM Newton and CXO up to a record time of 83 Ms after burst
\cite{Pasquale}. 
The measured light-curve has a single power-law decline with $\alpha_x\!=\!1.309\!\pm\!0.007$
in the time interval 47 ks to 83 Ms. The best single power-law fit to the combined
measurements of the X-ray light-curve of GRB 130427A with the Swift-XRT \cite{Galama}, XMM
Newton, CXO \cite{Pasquale}, and MAXI \cite{Maselli}, 
shown in Figure \ref{fig:Xar130427A} yields $\alpha_x\!=\!1.294\!\pm\!0.03$. The CB
model prediction, as given by Equation \ref{eq:closure} for the measured 
spectral index $\beta_x\!=\!0.79\!\pm\! 0.03$ \cite{Pasquale}, is $\alpha_x\!=\!1.29\!\pm\!0.03$, in remarkable agreement with its best fit value.

No doubt, the assumptions of a constant density circumburst medium is an over
simplification: SN-LGRBs are produced by SN explosions of type Ic, which
generally take place in super-bubble remnants of star formation. These 
environments may have a bumpy density, which deviates significantly from an assumed
constant-density ISM. This explains, in the CB model, the observed deviations from the
predicted smooth light-curves \cite{ManyBumpAG}. 

In a constant-density ISM, the late-time distance $x$ of a CB from its launch site is
\begin{equation}
x(t\!\gg\!t_b)\!=\!{2 \, c \int^t \gamma\,\delta\, dt\over 1+z}\!\approx\! 
{4\,c\,\gamma_0^2\,\sqrt{t_b\,t}\over {1+z}}\,.    
\label{eq:xlate}
\end{equation}
This distance can exceed the size of the super-bubble and even the scale-height of the
disk of a GRB's host galaxy. In such cases, the transition of a CB from the super-bubble
into the Galactic ISM or into the Galactic halo in face-on disk galaxies, will bend the
late-time single power-law decline into a more rapid decline, depending on the density
profile above the disk. For instance, when the CB exits the disk into the halo, its Lorentz
and Doppler factors tend to constant values while, given Equation \ref{eq:Fnu}, its AG decays like
\begin{equation}
F_\nu(t)\!\propto\![n(r)]^{(1\!+\!\beta_\nu)/2}
\label{eq:Fouter}
\end{equation}
with $r\!\propto \!t$. This behavior may have been observed with Swift's XRT \cite{Swift}
 in several GRBs,
such as 080319B and 110918A, at $t\!>\! 3\times 10^6$ s and by CXO in GRB 060729 at $t\!>\!3\times 10^7$ s 
\cite{Grupe2010}.

\subsection{Missing breaks (Test 7)}

The CB model's Equation \ref{eq:Fnu} implies a single power-law for the temporal decline of
the light curve of an AG well beyond its break time $t_b$, as long as the CB moves in a
constant density interstellar medium. Consequently very energetic LGRBs, i.e.~those
 with a large product $(1+z)E_pE_{iso}$, may have a break time $t_b$ smaller than the time at
 which the AG takes over the prompt emission, or before the AG observations
 began \cite{DDD2008}. 
 In such cases the observed light curve has a single power-law behavior with
 a temporal decay index $\alpha_\nu\!=\!\beta_\nu\!+\!1/2$ and the break is missing.

\begin{figure}[]
\includegraphics[width=8.5 cm]{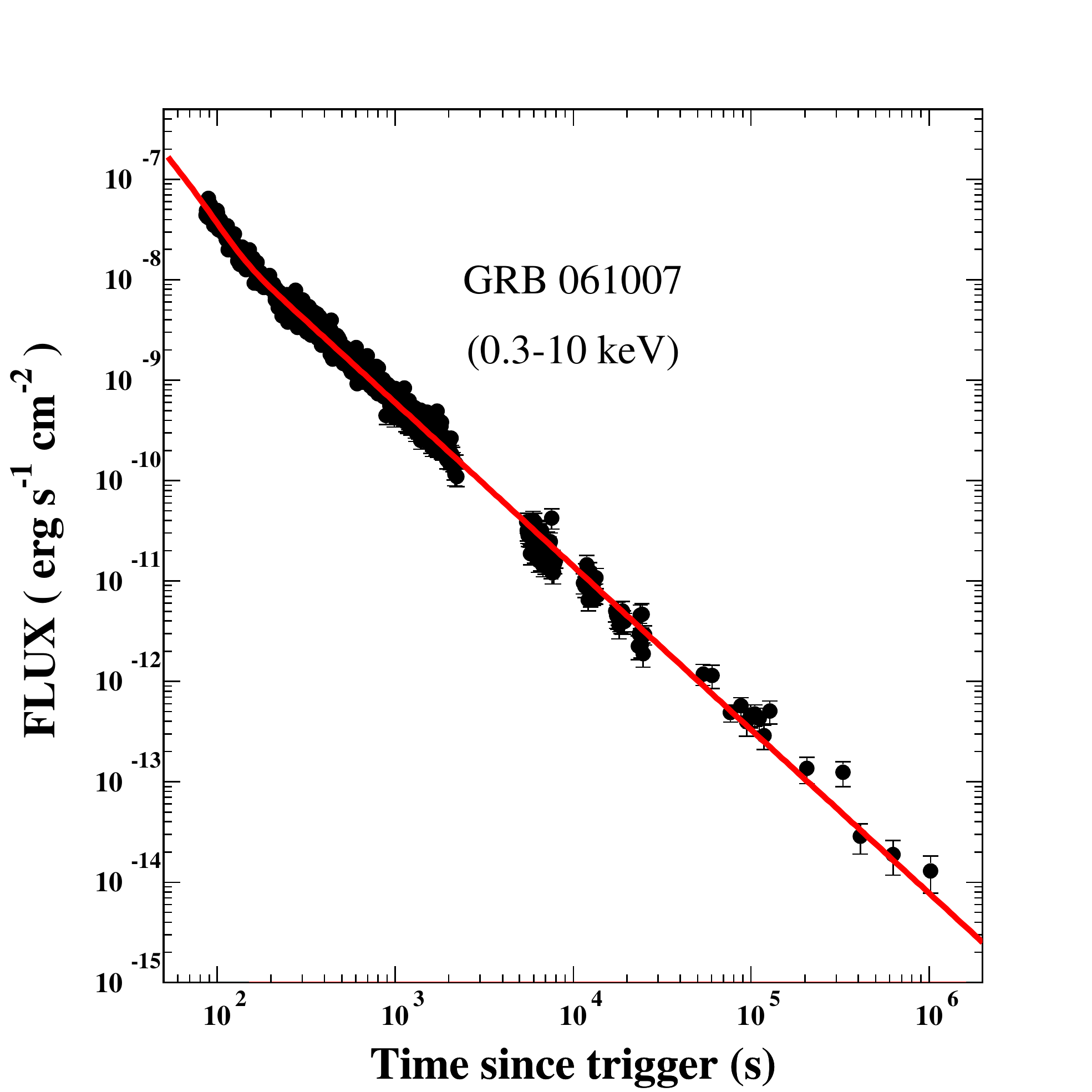}
\caption{The single power-law best fit to the afterglow of GRB061007 with  
a ``missing jet break" measured with Swift XRT \cite{Schady}. The best fit temporal 
index $\alpha_x\!=\!1.65\!\pm\!0.01$ satisfies the CB model prediction 
$\alpha_x\!=\!\beta_x\!+\!1/2\!=\!1.60\!\pm\! 0.11$.}
\label{fig:Xag061007}
\end{figure}

\begin{figure}[]
\includegraphics[width=8.5 cm]{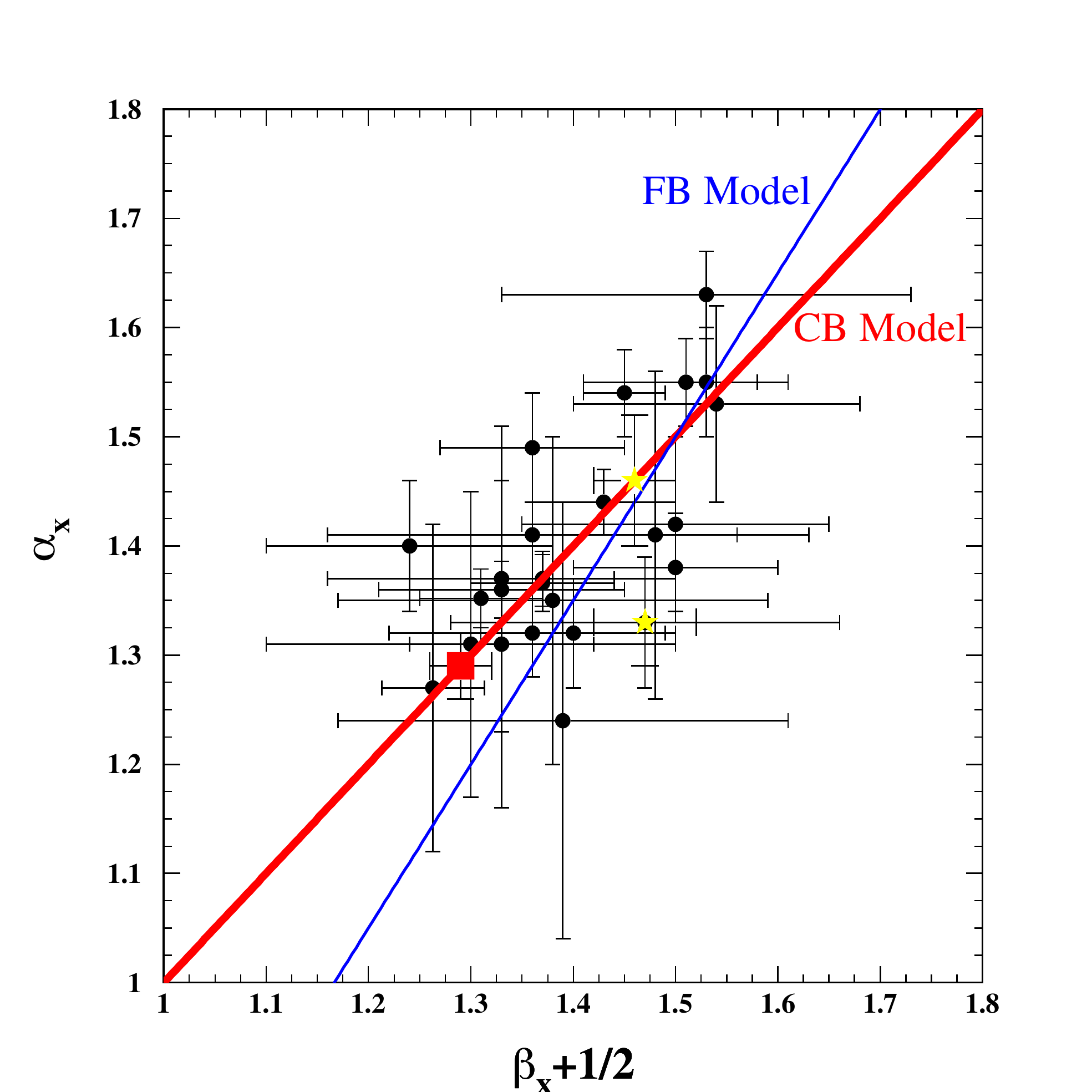}
\caption{
The values of the post-break temporal index $\alpha_x$ as a function
 of the spectral index  $\beta_x+1/2$
for the 28 most intense GRBs with known redshift \cite{Dado2016} obtained from the follow-up
measurements of their 0.3-10 keV X-ray AG with Swift's XRT \cite{Swift}. The square indicates
the result for GRB 130427A. The thick line is the CB model prediction 
Equation \ref{eq:closure}, the thin one is the FB's model.
The models have similar $\chi^2/$dof but 3/4 of the points lie above the FB-model prediction.}
\label{fig:alfavsbeta}
\end{figure}

The first {\it missing break}, shown in Figure \ref{fig:Xag061007}, 
was observed in GRB061007 
 \cite{Schady} with
Swift's XRT. The $\alpha_x$ values of the most energetic Swift XRT LGRBs with known
 redshift are plotted in Figure \ref{fig:alfavsbeta} along with their $\beta_x\!+\!1/2$ 
 values. Also plotted is the
best fit linear relation $\alpha_x\!= \!a (\beta_x\!+\!1/2)$, which yields $a\!=\!1.007$, 
in agreement with $a\!=\!1$, predicted by the CB model.

\section{GRB afterglows in Fireball models}

As illustrated in Figure \ref{fig:FBmodel} the prompt GRB emission is emitted by
shells colliding with each other while moving in the same direction, unlike in a particle collider.
The AG is produced by the ensemble of the shells stopping against the ISM, 
like in a beam dump.
The center-of mass energy in this last process is much bigger than in the first mentioned 
one. This implies that the prompt emission's energy ought to be orders of magnitude 
smaller that the AG's one, exactly contrary to observation.

During the first two years after the discovery of afterglows, the observed light
curves were claimed in FB models to be well fitted by a single power-law \cite{Weijer1997}, as
predicted for spherical fireballs \cite{Meszaros1997} \cite{Piran99}. 
Shortly after, when the observations clearly
showed a smoothly broken power-law, the spherical $e^+e^-\gamma$ fireball was replaced first
with conical flows or thin conical shells of $e^+e^-\gamma$ plasma, later to be 
{\it baryon-loaded} and
eventually to consist in an ordinary-matter plasma. These flows were called {\it collimated
fireballs}, retaining the fire{\it ball} denomination. The collimated fireballs could
neither explain nor correctly reproduce the observed behavior of the AG of SN-GRBs:
 they failed tests 4 to 7:
 
\subsection{The canonical AG shape (Test 4)}

In the FB models a plateau phase in the AG of GRBs was not expected \cite{FBM Reviews}. 
An additional power supply by a newly born millisecond pulsar was later adopted as the
origin of such a phase \cite{ZFD2006}.

\subsection{Break-time correlations (Test 5)}

In the standard FB models, the opening angle of the conical jet satisfies $\theta_j\!\gg \! 1/\gamma_0$.
Because of relativistic beaming, only a fraction $\sim\!1/\gamma^2\, \theta_j^2 $
of the front surface of the jet
is initially visible to a distant near axis observer. This fraction increases with time like
$[\gamma(t)]^{-2}$, due to the deceleration of the jet in the ISM, until the entire front surface of
the jet becomes visible, i.e.~until 
$t\!\approx \! t_b$, with $\gamma(t_b\!)\!=\!1/ \theta_j$. If the total 
 $\gamma$-ray energy $E_{\gamma}$  is assumed to be a constant 
fraction $\eta$ of the kinetic energy
$E_k$ of the jet, which decelerates in an ISM of constant density $n_b$ via a 
{\it plastic collision}, it follows that \cite{Rhoads99}:
\begin{equation}
t_b/(1\!+\!z)\approx {1\over 16\,c}\,\left[{3\,E_{iso}\over \eta\,\pi\, n_b\, 
m_p\,c^2}\right]^{1/3}\,[\theta_j]^{8/3}.
\label{eq:tboftheta}
\end{equation}

Though in the FB model the AG is SR by the shocked ISM
--through {\it elastic scattering} of the ISM particles in front of the jet, not by a {\it plastic collision,}
see Table \ref{Table2}-- Equation \ref{eq:tboftheta} has been widely used, with no rationale, to estimate 
$\theta_j$.

If $E_\gamma\!\approx\! \eta\, E_k\!\approx\!E_{iso}\theta_j^2/4$
is a {\it standard candle}, as argued in \cite{Frail}, $E_{iso}\,\theta_j^2$
is roughly constant (the ``Frail relation") and
\begin{equation}
t_b/(1\!+\!z)\!\propto\! [E_{iso}]^{-1}.
\label{eq:tbEiso}
\end{equation}
The same $[t_b,E_{iso}]$ correlation is obtained for the deceleration of a conical jet in a wind-like
circumburst density \cite{Chevalier2000}. Equation \ref{eq:tbEiso} is inconsistent with 
$t_b/(1\!+\!z)\propto E_{iso}^{-0.69\!\pm\!0.06}$,
the best fit to the data, but it is consistent with the correlation 
$t_b/(1\!+\!z)\propto E_{iso}^{-3/4}$
expected in the CB model \cite{hey Scioscia}.

\begin{table*}
\caption{The late time t-dependence of the bulk motion Lorentz factor 
of highly relativistic conical jets which decelerate by
collision with the surrounding medium.} 
\centering
\begin{tabular}{c c c c c c c}
\hline \hline
Collision:~~~ & Plastic & Plastic  & & Elastic & Elastic\\
\hline 
Density:~~~~~ & ISM &  Wind  &~~~~~ & ISM & Wind\\
\hline\hline
$\gamma(t)\!\propto$& $t^{-3/8}$ & $t^{-1/4}$&~~~~~ &
$t^{-3/7}$ & $t^{-1/3}$ \\
\hline 
\end{tabular}
\label{Table2}
\end{table*}

\subsection{Closure relations (Test 6)}

 In the conical fireball model the increase of the visible area of the jet until the break
 time results in an achromatic break in the AG. If the spectral energy density flux is
 parametrized as $F_\nu(t)\!\propto t^{-\alpha}\nu^{-\beta}$, 
the predicted achromatic change in $\alpha$ across the break
 satisfies $\Delta(\alpha)\!=\! \alpha(t\!>\!t_b)\!-\!\alpha(t\!<\!t_b)\!=\! 3/4$ 
for a constant ISM density. For a wind-like
 density,  $\Delta(\alpha)\!=\!1/2$. None of these closure relations is satisfied by most GRB breaks,
 e.g.~the ones of Figures
 \ref{fig:XAG_GRB990510}, 
\ref{fig:XAG_GRB050315},  
  \ref{fig:AGX_GRB060729}, \ref{fig:Xar130427A} and \ref{fig:Xag061007}.

Liang, et al.~\cite{Liang2008} analyzed the AG of 179 GRBs detected by Swift between January
2005 and January 2007 and the optical AG of 57 pre-Swift GRBs. They did not find any
afterglow with a break satisfying tests 5 or 6 of the FB model.

\subsection{Missing breaks (Test 7)}

 It was hypothesized that the missing break in the X-ray AG of several
GRBs with long follow up measurements took place after the 
observations ended \cite{Racusin16}.
But Equation \ref{eq:tboftheta} implies that late-time breaks are present only in GRBs with a small $E_{iso}$.
This contradicts the fact that missing breaks in GRBs extending to late times are limited
to events with very large, rather than small, $E_{iso}$. This is demonstrated in Figures 
   \ref{fig:Xag061007} and \ref{fig:Xar130427A}
 by the unbroken power-law X-ray AGs of GRBs 061007 and 130427A, for which
$E_{iso}\!=\!1.0\times 10^{54}$ erg \cite{Schady}
and $E_{iso}\!=\! 8.5 \times 10^{53}$ erg \cite{Pasquale} respectively.
 These AGs satisfy the CB model post-break closure relation given by Equation \ref{eq:3Corr}.
 
 \section{Further afterglow tests}
\subsection{Chromatic jet breaks (Test 8)}
 
 In the CB model the jet deceleration break in the afterglow of jetted SN-GRBs is
 dynamic in origin and usually chromatic \cite{DD2004}, while in
the FB model they are basically
geometrical in origin and are therefore predicted to be dominantly achromatic 
\cite{FBM Reviews}, in
 conflict with observations.
 
\subsection{The Universal afterglow of SN-less GRBs (Test 9)}
\label{subsec:SN-less}
 
\begin{figure}[]
 \centering
\includegraphics[width=8.5 cm]{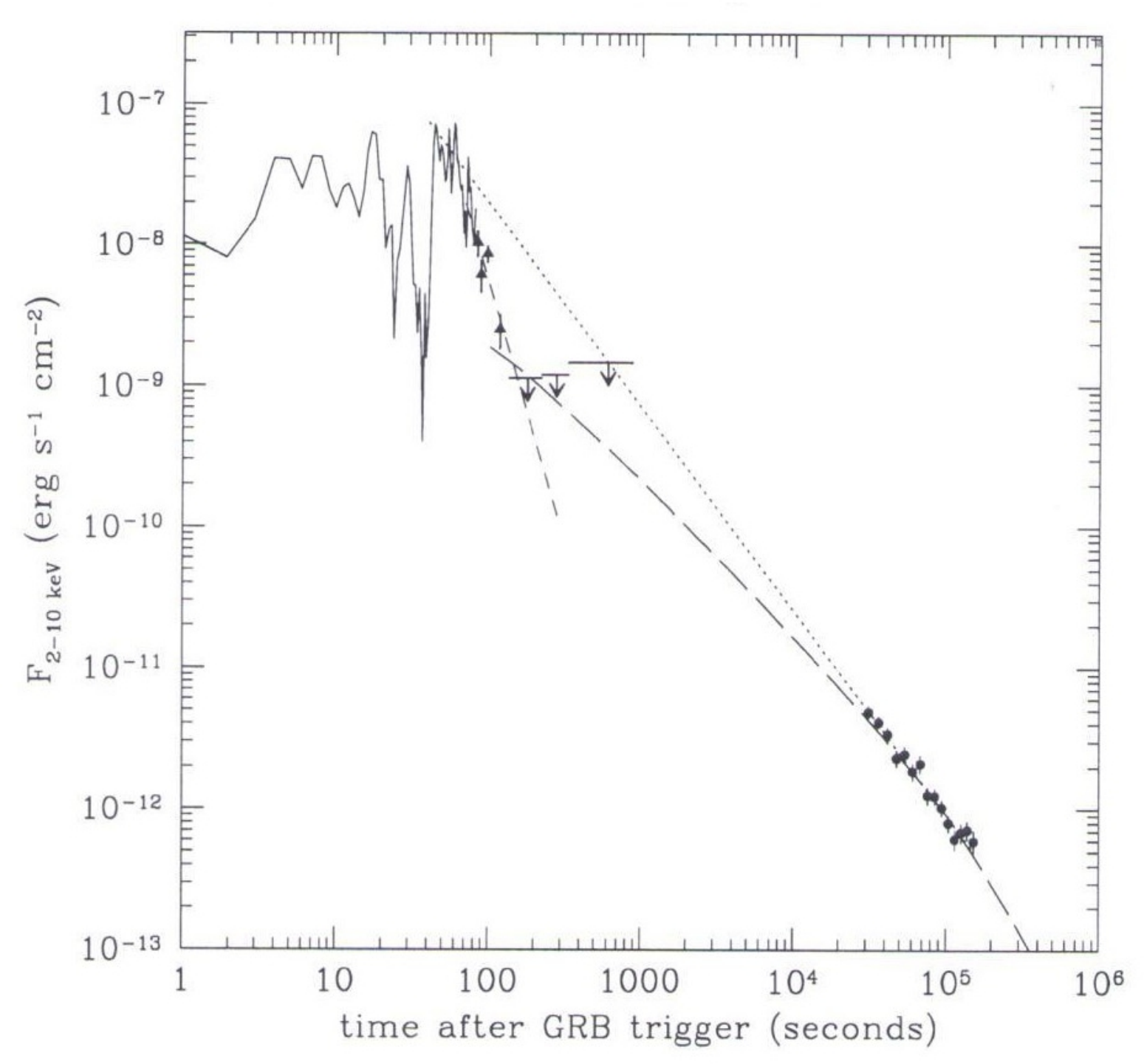}
\caption{The X-ray light curve of GRB 990510 and its AG, 
measured with BeppoSAX 
together with a single power-law fit and a smoothly broken single  power-law fit 
to its X-ray AG \cite{Pian2001}.}
\label{fig:990510pian}
\end{figure}

\begin{figure}[]
\centering
\includegraphics[width=9 cm]{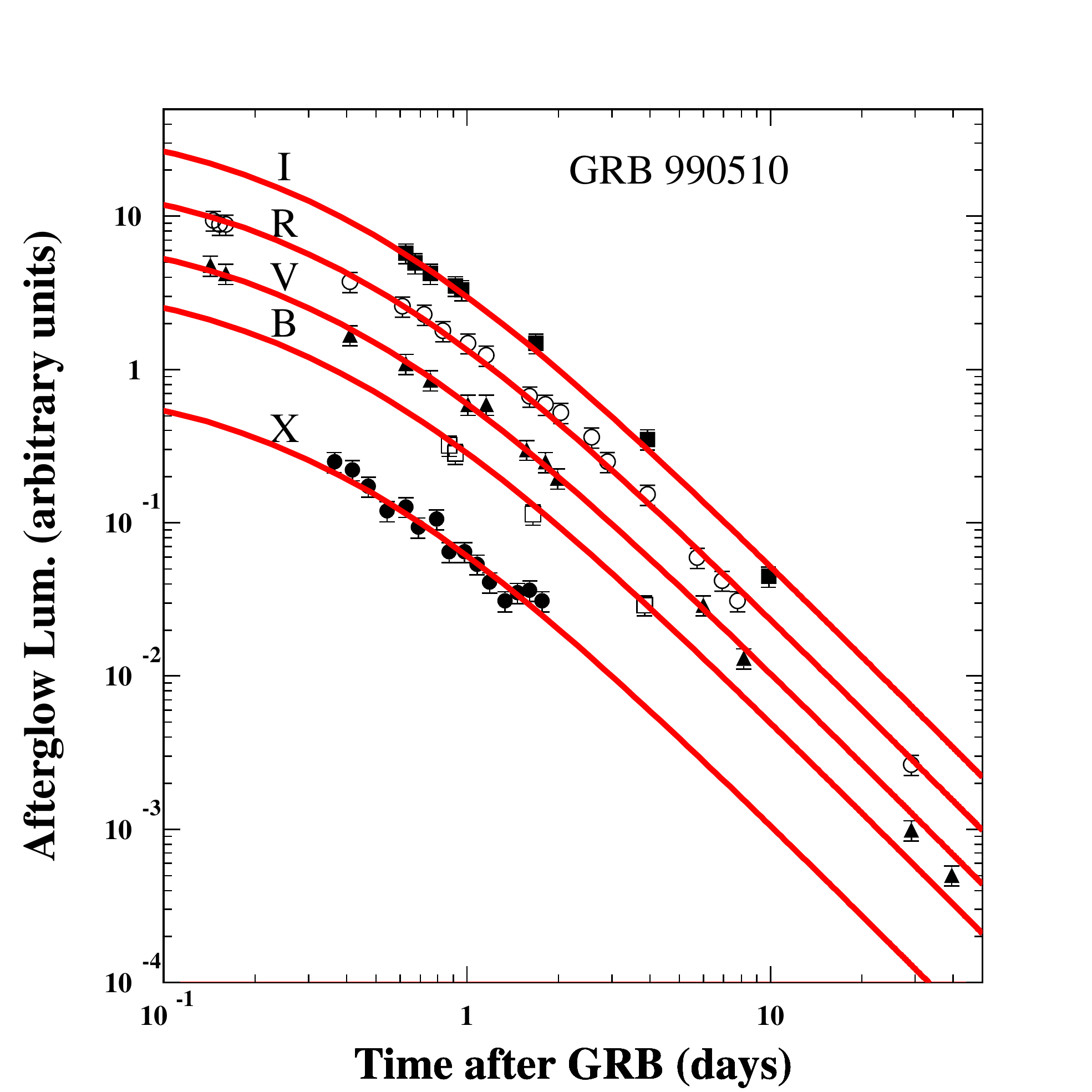}
\caption{Comparison of Equation \ref{eq:PWN}, the predicted temporal behavior 
of the light curves of the X-ray and optical AGs of GRB990510,
with the observations. The X-ray data at 5 keV (filled circles) 
is from \cite{Pian2001}. The data in the bands 
I (filled squares), R (empty circles), V (filled triangles)
and B (open squares) are the ones compiled in \cite{Pian2001} from \cite{Stanek1999}. 
The flux normalization is arbitrary.}
\label{fig:XOAG990510MSP}
\end{figure}

Figure \ref{fig:990510pian}, adapted from \cite{Pian2001}, 
shows the X-ray AG of GRB 990510 measured with
BeppoSAX. It could not be fit by the single power-law predicted by spherical FB models
 (e.g.~\cite{Pian2001}). But it could be fit well by an achromatic ``smoothly broken power law"
parametrization \cite{Pian2001}, as shown in Figure \ref{fig:XOAG990510MSP}. That --and the observed optical and X-ray
AGs of a couple of other GRBs, which could be fit by such a parametrization-- is
what led to the replacement of the spherical $e^+e^-\gamma$ fireball 
 by a conical one, later replaced by a conical jet of ordinary 
 matter which became the current FB model of GRBs.
 
The AG of GRB 990510 and others which were fit by smoothly broken
power laws are not conclusive evidence of conical jets. In fact, an isotropic radiation
from a pulsar wind nebula (PWN), powered by a newly born millisecond pulsar, has an
expected luminosity \cite{Dado5} satisfying
\begin{equation}
L(t,t_b)/L(t=0)\!=\!(1\!+\!t/t_b)^{-2},
\label{eq:PWN}
\end{equation}
where $t_b\!= \! P(0)/2\, \dot P(0)$, and $P(0)$ and 
$\dot P(0)$ are the pulsar's initial period and its time derivative. 
This is shown in Figure \ref{fig:XOAG990510MSP} for GRB 990510 which, even though
it is not a ``certified" SN-less GRB. 

\begin{figure}[]
\centering
\includegraphics[width=8.5 cm]{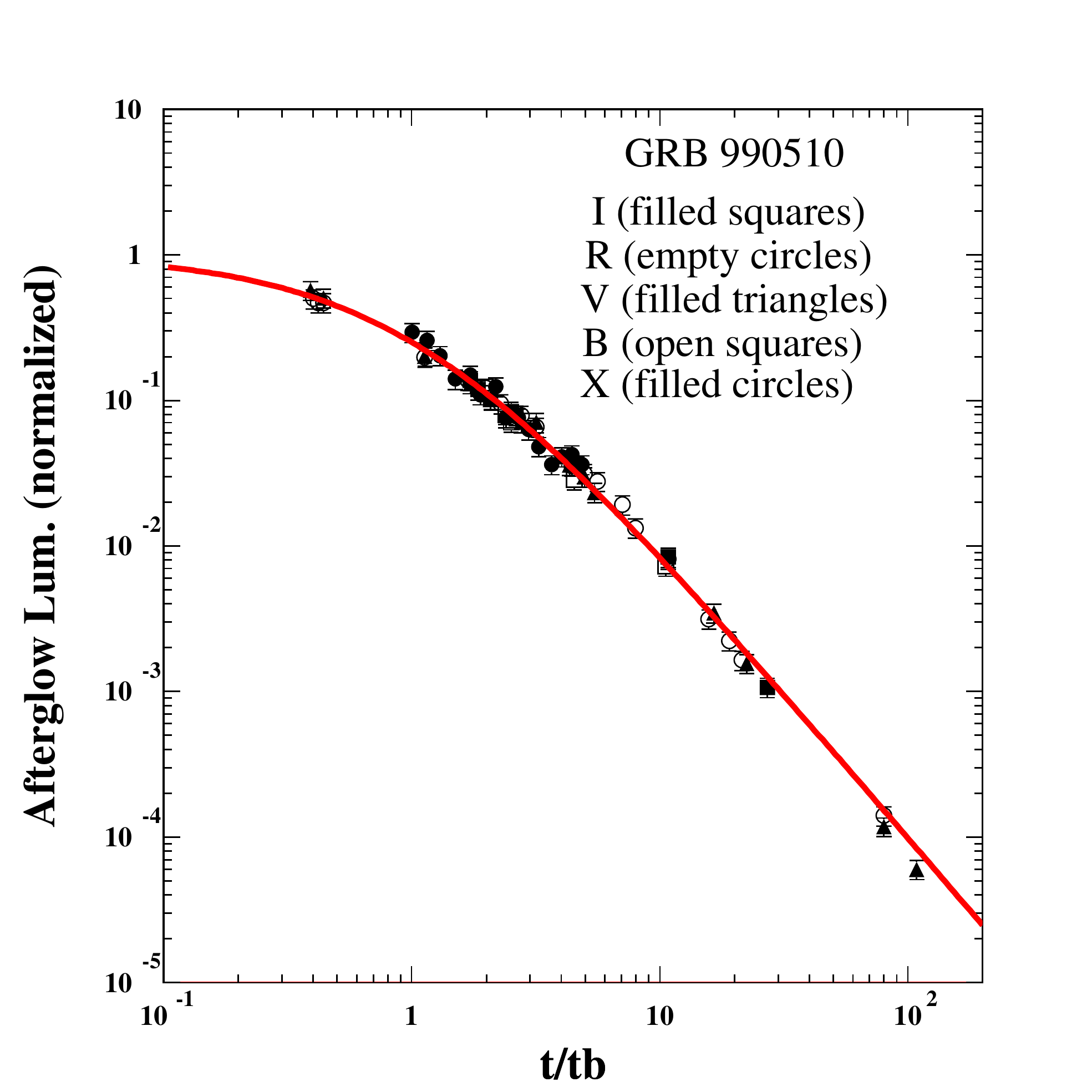}
\caption{Comparison between the normalized  light curves of the X-ray 
and optical AGs of GRB990510
and their universal shape: Equation \ref{eq:PWN}. 
The data are the same as in Figure \ref{fig:XOAG990510MSP}.}
\label{fig:AG990510MSP5B}
\end{figure}

\begin{figure}[]
\centering
\includegraphics[width=8.5 cm]{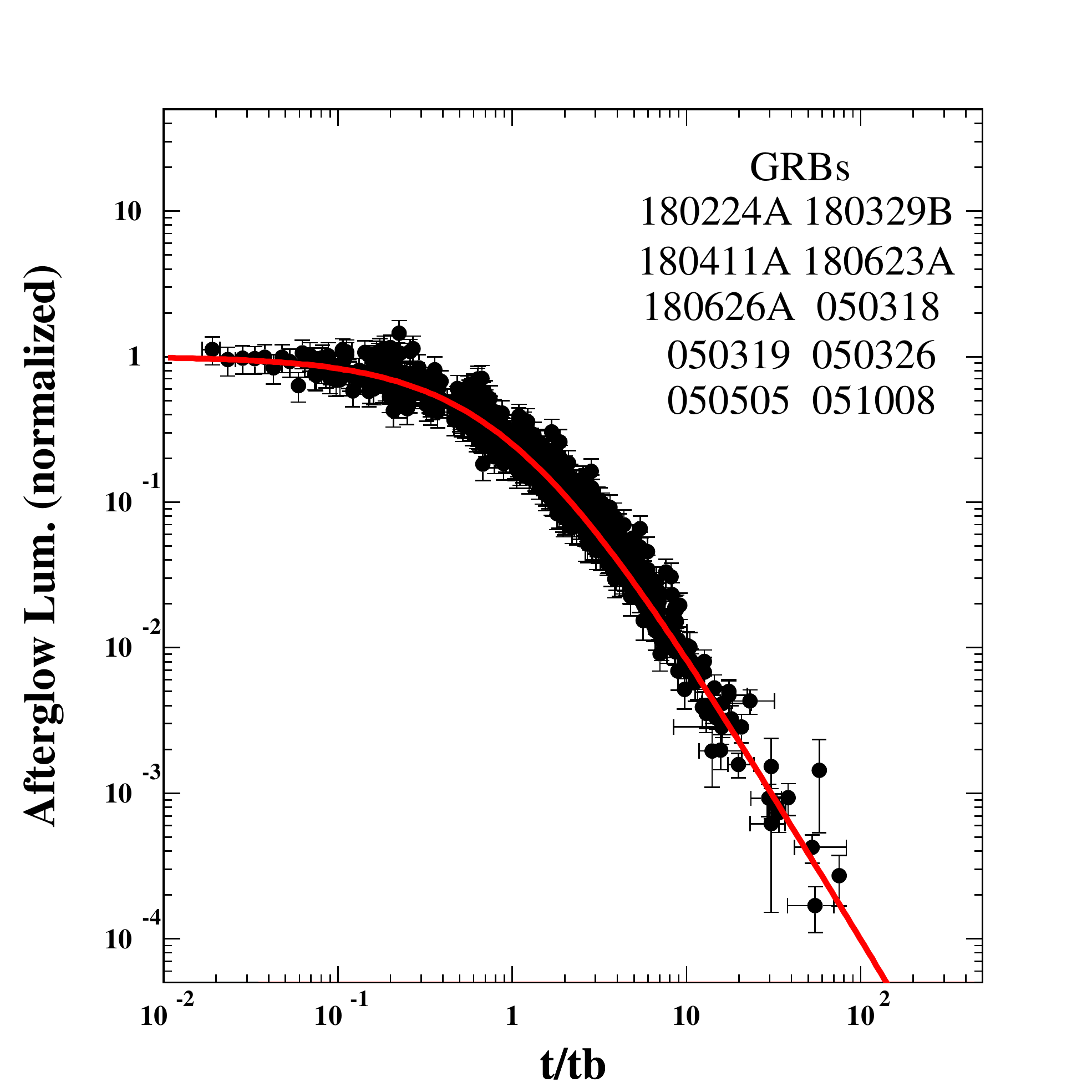}
\caption{Comparison between the normalized light curve
of the X-ray afterglow measured with Swift XRT \cite{Swift} 
of 12 SN-less GRBs with a well sampled AG
in the first couple of  days after burst
and their predicted universal behavior, Equation \ref{eq:PWN}.}
\label{fig:XAG10GRBMSP}
\end{figure}

\begin{figure}[]
\centering
\includegraphics[width=8.5 cm]{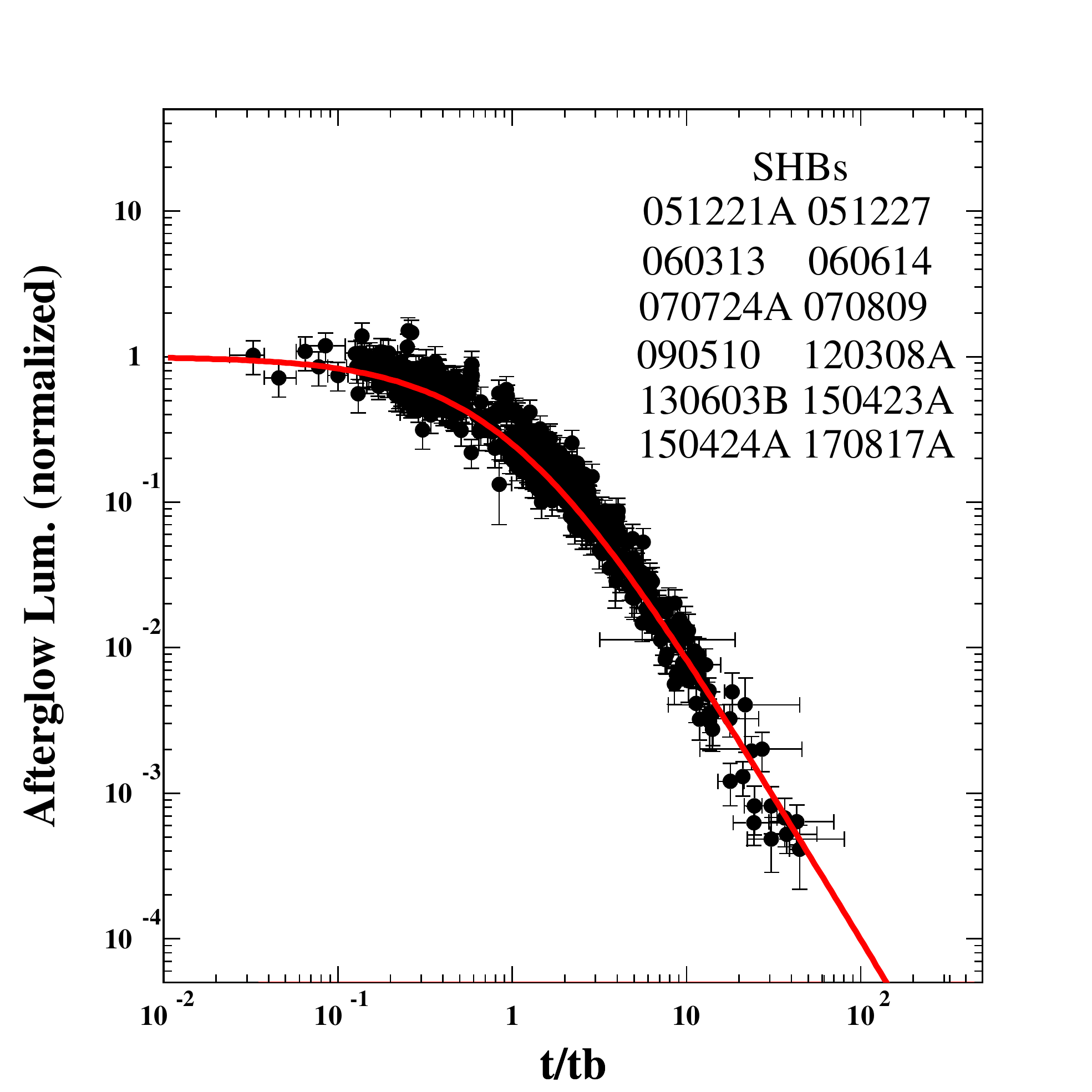}
\caption{Comparison between the normalized light curve
of the X-ray AG of 11 SHBs 
with a well sampled AG measured with Swift's XRT \cite{Swift}
during the first couple of days after burst 
and the predicted universal behavior of Equation \ref{eq:PWN}.
The bolometric light curve of SHB170817A \cite{Drout2017}
is included, and shown separately in Figure \ref{fig:epvt_shb}.} 
\label{fig:XAGS12SHBMSP}
\end{figure}

The {\it universal behaviour} \cite{Dado2017} 
Equation \ref{eq:PWN} describes well the X-ray and optical AG light curves 
of GRB 990510, as shown in Figures \ref{fig:XOAG990510MSP} and \ref{fig:AG990510MSP5B} and
of the AG of all the SN-less GRBs and SHBs with a well
sampled AG during the first few days after burst. This is demonstrated in 
Figure \ref{fig:XAG10GRBMSP} for
the X-ray AG of twelve SN-less GRBs, and in Figure \ref{fig:XAGS12SHBMSP}
for the twelve SHBs \cite{Dado2018}
from the Swift XRT light curve repository \cite{Swift}, 
that were well sampled in the mentioned period.

\section{The progenitors of GRBs}
\subsection{Long GRBs}
\label{GRBProg}

The possibility that GRBs be produced in SN explosions was suggested very long
ago \cite{SNGRB} \cite{Goodman2}.
This hypothesis was incorporated in the CB model as the daring bet that there
ought to be a 1:1 correspondence between LGRBs and Type Ic SNe, for a long time a
tenable opinion \cite{DD2004}. 
The first direct evidence for a SN-GRB association was provided by
the discovery \cite{Galama} --within the Beppo-SAX error 
circle around the position of GRB980425 \cite{Soffita, Pian99}--
of SN1998bw, at $z\!=\! 0.0085$. The SN light curve indicated that the time of the explosion
was within - 2 to + 7 days of the GRB  \cite{Galama,Iwamoto98}.
A physical association between GRB980425 and
SN1998bw was consistent with the GRB being entirely normal, produced by a relativistic
jet in a SN explosion and viewed at a relatively large angle \cite{Shaviv, WW}. 
Our explicit CB model
analysis was entirely consistent with this, as can also be seen in Figures 
\ref{fig:epeisoLLGRBs}  and \ref{fig:XAG980425}.

During the first few years after GRB980425, the optical afterglow of several relatively
nearby ($z \!<\! 0.5$) GRBs provided clear evidence for a GRB-SNIc association, independent of  GRB luminosity. The most convincing evidence was provided by the fairly nearby 
($z\!=\! 0.1685$) GRB030329 \cite{VSB},
one of the brightest GRBs detected by that time.
The CB-model's fit to its early AG data was extrapolated, 
with the addition of a superimposed 
SN1998bw light curve (transported in distance) to predict the exact date
when a SN would be discovered as a bump in the AG and a dramatic change in the
spectrum \cite{DDD030329}. 
SN2003dh duly obeyed the CB model's forecast. 
Admittedly, there was an
element of timing luck in all this: SNLess GRBs were discovered later and they proved
wrong the original hypothetical 1:1 SN/GRB association.

An early standard view of the SN/GRB association was that of 
Woosley \cite{WoosleyA,WoosleyB}, who
 argued that GRBs were not be produced by SNe, but by ``failed SNe", or ``collapsars",
direct collapses of massive stars into black holes without an associated 
supernova.
Concerning SN1998bw and GRB980425, the FB model defenders concluded that
this pair represented a new subclass of rare events 
\cite{Galama},
\cite{Bloom98,WES99,Hurley2002}.
These would be associated with ``hyper-novae" \cite{Macronova, Iwamoto98},
super-energetic explosions with kinetic energy exceeding $10^{52}$ erg, as
was inferred for SN1998bw from its high expansion velocity and 
luminosity \cite{Patat2001} 
and from the very strong radio emission from its direction \cite{Kulkarni}.

\section{The progenitors of SHBs}
\label{sec:Progenitors}
After the discovery of the gravitational wave GW170817 and its electromagnetic
sibling SHB170817A, a consensus was reached that SHBs originate in the merger of
neutron star pairs. Two days before the discovery date, a paper appeared on arXiv 
\cite{DDlargeangle}, not
only reiterating that earlier view, but predicting that a SHB found in this ``multi-messenger"
way would be seen far of axis. This fact, also currently generally accepted, is based on
the greater red-shift reach of GRB or SHB observations relative to the GW ones. In the
CB model, within the volume reach of GW observations, it would be most unlikely for a
SHB to point close to the observer.

\section{Further tests}
\subsection{Redshift distribution of LGRBs (Test 10)}

In the CB model long duration GRBs belong to two classes, SN-GRBs produced by
Type Ic SNe and SN-less GRBs presumably produced in phase transitions of neutron
stars to quark stars following mass accretion in high mass X-ray binaries \cite{Shaviv}
\cite{Dado2018}. In both
cases, the progenitors involve a short-lived high mass star. Hence the production rate of
these GRBs is proportional to the star formation rate \cite{Hogg} 
modified by beaming \cite{Dado2014}.

\begin{figure}[]
\centering
\includegraphics[width=8.5 cm]{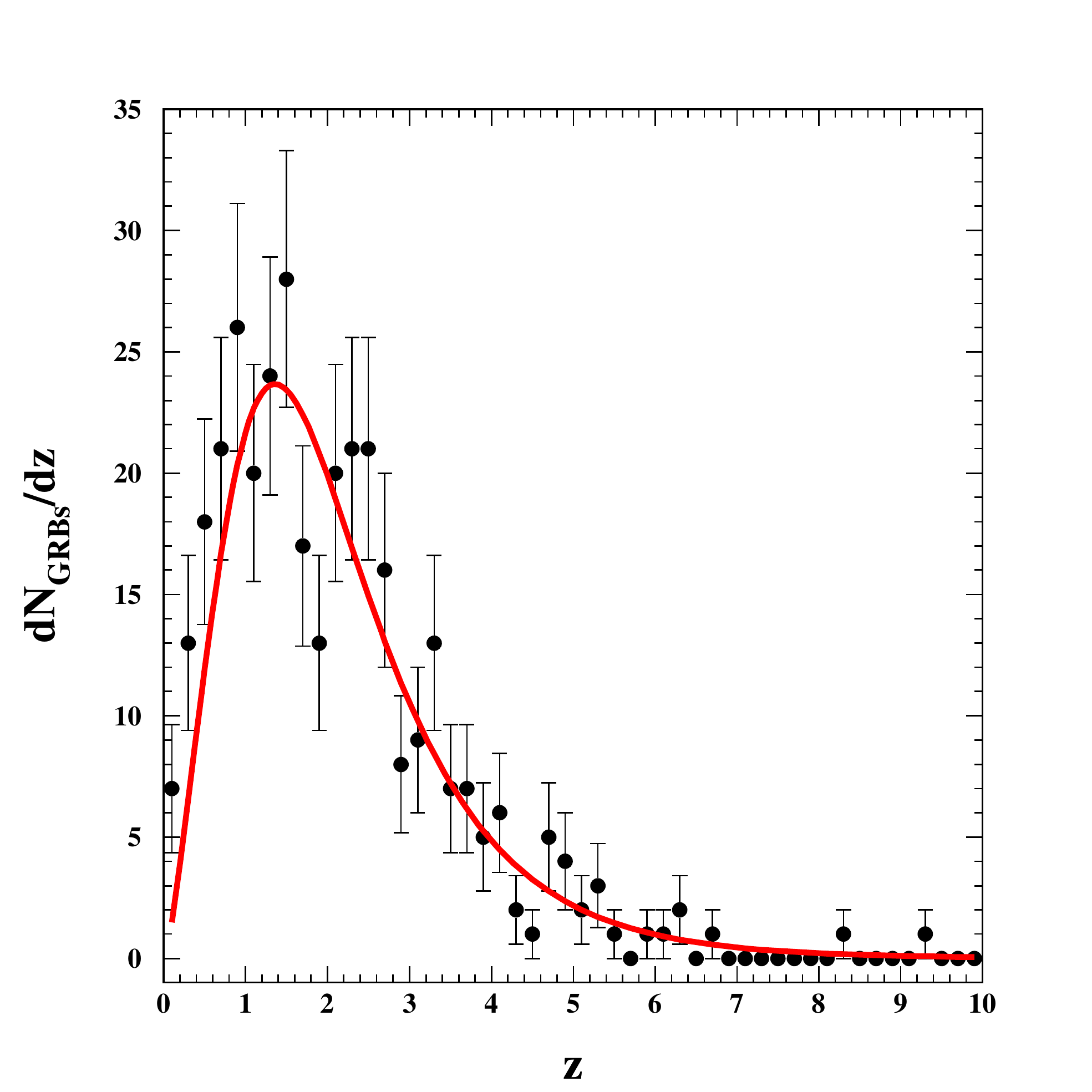}
\caption{Comparison between the redshift distribution
of 356 long GRBs with known redshift observed before June 2018
and their CB-model's expected distribution if the production rate of 
GRBs is proportional to the SFR; ($\chi^2/{\rm dof}\!=\!37.57/49\!=\!0.77$).}
\label{fig:zlgrbs2018}
\end{figure}

The CB-model's beaming fraction, $f_b(z)$, depends on the standard $\Lambda$CDM cosmological
luminosity distance, $f_b(z)\!=\! N/\sqrt{D_L(z)}$.
The proportionality factor $N$ is a function of
the minimal luminosity for GRB detection. It can be estimated by use of the properties
of XRF060218 at a redshift $z_{min}\!=\!0.0331$ \cite{MirHal}
with its record lowest prompt-emission peak
energy for a Swift GRB, $E_p^{min}\!=\! 4.5$ keV. 
The result is $N\!\simeq\!\sqrt{D_L(z_{min})}\,\epsilon_g/E_p^{min}$, with $\epsilon_g\!=\! 3$ eV
the estimated peak spectral energy of the glory surrounding the progenitor (a
Wolf-Rayet star). In Figure \ref{fig:zlgrbs2018} this prediction is tested 
by a comparison with the data.

\begin{figure}[]
\centering
\includegraphics[width=8.5 cm]{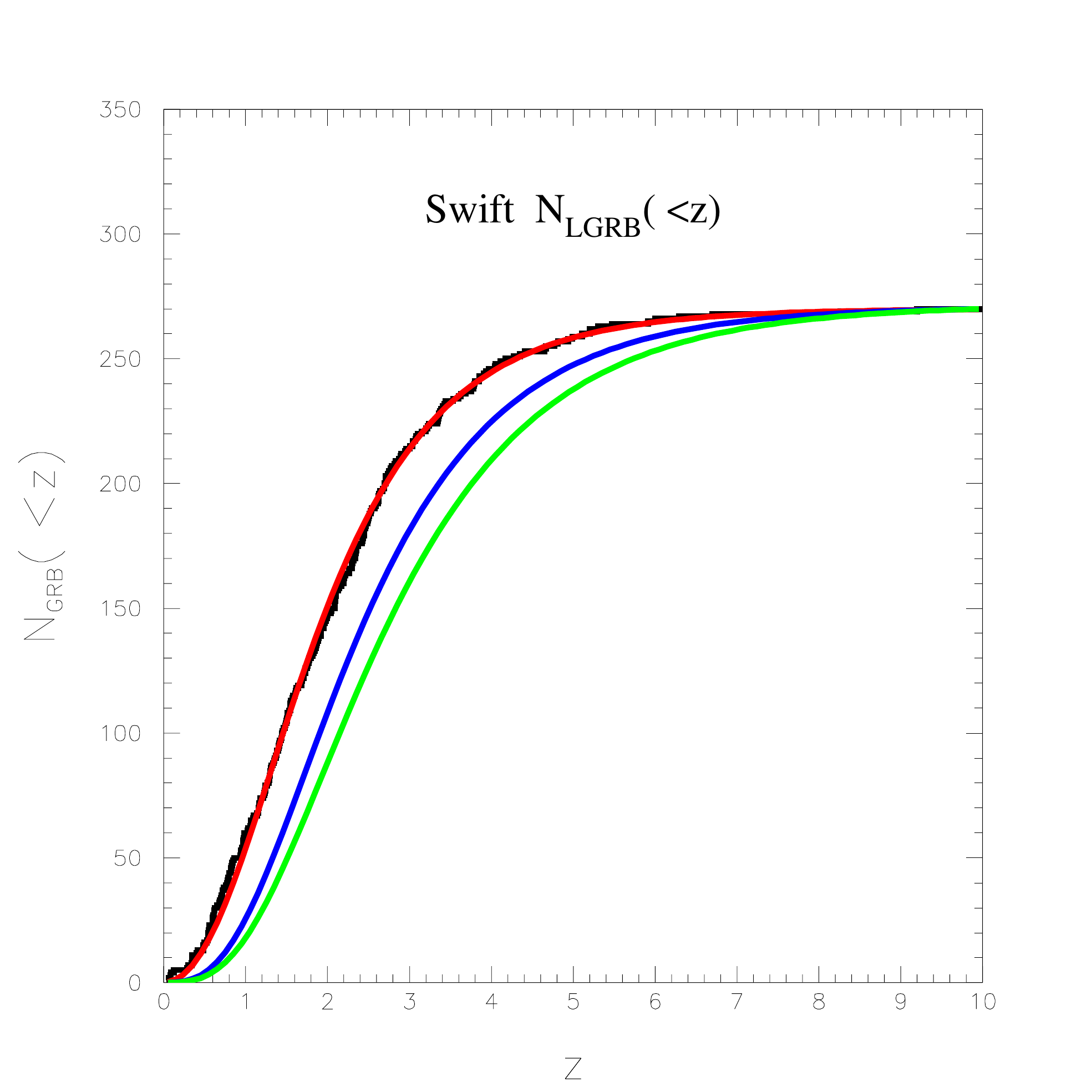}
\caption{Comparison between the cumulative distribution function, $N(\!<\! z)$,
of the 262 LGRBs with known redshift (histogram) detected by
Swift before 2014 and the $N(\!<\!z)$ expected in the CB model (left curve)
for long GRBs, whose rate is assumed to trace the SFR. Also shown are
the distributions expected in FB models with no evolution (rightmost
curve) and with it (middle curve) \cite{GRBSFR}.}
\label{fig:zlgrbscumstep}
\end{figure}

In the FB models the jet's aperture obeys $\alpha_j\!\gg\! 1/\gamma_0$ 
so that the rate of GRBs would 
be expected to be proportional to the 
star formation rate (SFR) up to very large redshifts beyond those
accessible to optical measurements, in contradiction with the data. Unlike the SFR in
a comoving unit volume, which first increases with redshift \cite{Dado2014}, the observed rate of
LGRBs decreases with increasing $z$ in the range $z\!\le\!0.1$, even after correcting for detector
flux threshold \cite{Guetta2006}. 
At larger $z$, it increases faster than the SFR \cite{Daignr2006}. The discrepancy at
small $z$ was interpreted as evidence that ordinary LGRBs and low-luminosity LGRBs
and XRFs belong to physically distinct classes \cite{LLGRBs}. 
The discrepancy at $z\!\gg\!1$ was claimed
to be due to a different evolution \cite{GRBSFR}. In Figure \ref{fig:zlgrbscumstep} 
the cumulative distribution, $N(\!<\!z)$, of GRBs in CB and FB models are shown.
In the latter model, with and without the evolution of LGRBs relative to the SFR
assumed in \cite{GRBSFR}. 
As can be seen in the figure, the data does not support the FB model's
assumptions.

\subsection{Low Luminosity GRBs (Test 11)}
In the CB model the observed properties of GRBs depend strongly on their viewing
angle relative to the CBs' direction of motion. Ordinary (O) GRBs are viewed from
angles $\theta\!\sim\!1/\gamma_0$, so that their Doppler factor is $\delta_0\!\sim\!\gamma_0$. 
In the CB model, low luminosity
(LL) GRBs are ordinary GRBs with similar intrinsic properties viewed from far off-axis,
i.e.~$\delta_0\gamma_0\!=\!1/(1\!-\!\beta_0\cos\theta)\!\simeq 2/\theta^2$.
Consequently, in the rough approximation that GRBs are
standard candles, ordinary GRBs and LL GRBs satisfy the relations:
\begin{equation}
E_{iso}{\rm (LL\,GRB)}\!\simeq\!E_{iso}{\rm (O\,GRB)}/[\gamma_0^2\,\theta^2]^3
\label{eq:LLOEiso}
\end{equation}
\begin{equation}
L_p{\rm (LL\,GRB)}\!\simeq\!L_p{\rm (O\,GRB)}/[\gamma_0^2\,\theta^2]^4
\label{eq:LLOL}
\end{equation}
between their isotropic-equivalent energy $E_{iso}\!\propto\!\gamma_0\,\delta_0^3$
and their peak luminosity $L_p\!\propto\gamma_0^2\,\delta_0^4/(1\!+\!z)^2$
 \cite{DD2004}. The best fit CB model light curve to the X-ray AG of GRB980425
\cite{Pian2004,Kulkarni}, shown in Figure \ref{fig:XAG980425}, resulted
 in $\gamma\,\theta\!\approx\! 8.7$. 
 The mean $E_{iso}$ of ordinary GRBs is 
 $\langle E_{iso}{\rm(O\,GRB)}\rangle\!\simeq 7\times 10^{52}$ erg.
 Thus, Equation \ref{eq:LLOEiso} yields 
 $E_{iso}{\rm (GRB980425)}\!\simeq\!
1.84\times 10^{-5}
\langle E_{iso}{\rm(O\,GRB)}\rangle
\!\approx 1.3 \times 10^{48}$ erg,
agreeing with the observed value \cite{Amati76}
$E_{iso}{\rm (GRB980425)}\!\simeq\!(1.0\pm 0.2)\times 10^{48}$ erg.

Equations \ref{eq:LLOEiso} and \ref{eq:LLOL}, as well as the correlations 
between other properties of LL GRBs
(e.g.~the ones tested in Figures \ref{fig:epeisoLLGRBs} 
and \ref{fig:tbSNGRBs}) establish that they are ordinary GRBs viewed
from far off axis. So does the fact that O and LL GRBs have the same proportionality
factor in the relation between their birth rates and the SFR, as in \cite{Dado2014} and 
Figure \ref{fig:zlgrbscumstep}.

One of the best evidences for low-luminosity and ordinary SN-GRBs belonging
to the same class is that they are both produced by very similar SNeIc \cite{Melandri}, 
akin to
SN1998bw. For instance, SN2013cq and its GRB130427A at $z\!=\!0.34$, with a
record high GRB fluence and $E_{iso}\!\sim\!10^{54}$ erg \cite{Melandri}, 
was very similar to SN1998bw,
which produced the LL GRB980425 with a record low $E_{iso}\!\sim\!10^{48}$ erg \cite{Amati76}, 
six orders of magnitude smaller.

In the FB model low luminosity GRBs were claimed to be intrinsically different
from ordinary SN-GRBs and to belong to a different class \cite{LLGRBs}. 
A frequent and inescapable conclusion. In this case, given that:

a) The model could not explain the $\sim 6$ orders of magnitude difference between 
the $E_{iso}$
of LL SN-GRBs, such as that of GRB980425, and that of very high luminosity SN-GRBs,
such as GRB130427A, though they are produced by very similar SNeIc \cite{Melandri}.

 b) As previously discussed, the FB model cannot describe the GRB distribution as a
function of redshift, recall e.g., Figure \ref{fig:zlgrbscumstep}.

The CB model's resolution of these two apparent problems is simple: the interplay
between a narrow radiation beam and the observer's direction, occasionally far off-axis.
The far off-axis observation of SHB170817A has now been accepted by the FB community,
following its measured viewing angle --unexpected in a FB model-- relative to the axis of
the neutron star binary which produced GW170817 \cite{Abbott} --extracted from the behavior of
its late time AG \cite{MooleyJET} -- and from the superluminal motion --unexpected in a FB model-- of
its point-like radio source \cite{MooleySL},
 items to be discussed anon (tests 12, 13 and 16). Consistently,
GW170817 had $E_{iso}\!=\!(5.4\pm 1.3)\times 10^{46}$ erg, five orders of magnitude smaller
than that of ordinary SHBs.

\subsection{The CB's superluminal velocity in SN-GRBs (Test 12)}
The first observation of an apparent superluminal velocity of a source in the plane of
the sky was reported \cite{Kapteyn}
 in 1902, and since 1977 in many high resolution observations of
highly relativistic jets launched by quasars, blazars, and micro-quasars. The interpretation
of such observations within the framework of special relativity was provided in \cite{Rees1996}.

A source with a velocity $\beta\,c$ at redshift $z$, viewed from an angle $\theta$ relative to its
direction of motion and timed by the local arrival times of its emitted photons has an
apparent velocity in the plane of the sky:
\begin{equation}
V_{app}\!=\!{\beta\,c \sin\theta \over (1\!+\!z)(1\!-\!\beta \cos\theta)}\,\approx\,
{\beta\,c\,\gamma\,\delta \sin\theta \over (1\!+\!z)}
\label{eq:Vapp}
\end{equation}
For $\gamma\!\gg\!1$, $V_{app}$ has a maximum value $2\,\gamma\,c/(1\!+\!z)$ at $\sin\theta\!=\!1/\gamma$.

The predicted superluminal velocity of the jetted CBs cannot be verified during
the prompt emission phase, because of its short duration and the large cosmological
distances of GRBs. But the superluminal velocity of the jet in far off-axis, i.e.~nearby
low-luminosity GRBs, can be obtained from high resolution follow-up measurements of
their AGs \cite{Dar2000b}. Below,  two cases are treated in detail.

\subsubsection{\bf GRB980425}
The radio and X-ray afterglow of GRB980425, the nearest observed
SN-GRB with a known redshift, $z\!=\!0.0085$, has so far offered the best opportunity to
look for the superluminal signature of the highly relativistic jets which produce GRBs
\cite{Dar2000b}. 
 For some reason this has been totally overlooked in the late-time X-ray \cite{Kulkarni} and
radio observations \cite{Soder} of SN1998bw/GRB980425. But if the transient sources observed
on days 1281 and 2049.19 from the direction of SN1998bw are the CB which produced
GRB980425, this source moved by a viewing angle $\theta_s\,\simeq\, 2\, c/V_\perp\!\simeq\! 1/170$ 
rad with an average $V_{app}\,\simeq\!340\,c$.
 This interpretation implies that these sources are not present there
anymore, and were not to be seen before SN1998bw/GRB980425 was observed.

\begin{figure}[]
 \centering
\includegraphics[width=8.5 cm]{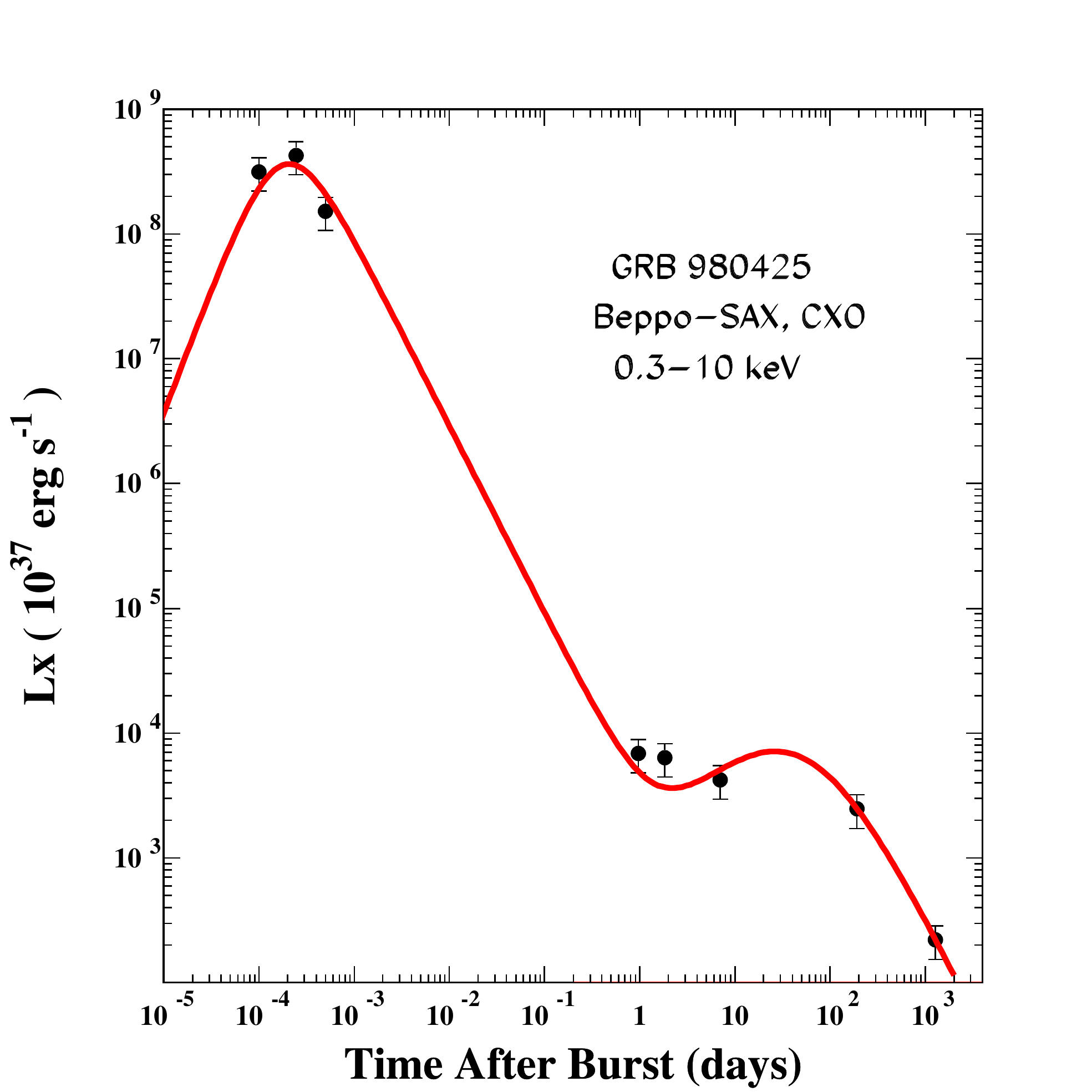}
\caption{The 0.3-10 keV X-ray light-curve of GRB980425 measured by Beppo-SAX 
\cite{Pian2004} (first 7 points). The last point at 1281 days is 
due to the source S1b resolved by CXO \cite{Kulkarni}.
The line is the CB model best fit light-curve to the prompt emission 
pulse and the AG of GRB980425, yielding $\gamma\,\theta\!\approx\! 8.7$.}
\label{fig:XAG980425}
\end{figure}

Supportive CB-model evidence for the above values of $\theta_s$ and $V_{app}$ in GRB980425 is
provided by other observations:\\
\indent (i) The expected value of $E_p$, as given by Equation \ref{eq:Ep0}
with a typical $\epsilon_p\!=\!1\,{\rm eV}$ \cite{DD2004},
is 58 keV. This is in good agreement with the observed \cite{Amati76}.
$E_p\!=\!55\!\pm\!21$ keV.\\
\indent (ii) Recall that $R_g$ stands for the radius characterizing the extention of the glory
surrounding a SN. A highly relativistic CB crosses it in a (local) time $R_g/c$. For an
observer this corresponds to a GRB pulse of
 ${\rm FWHM}\!\simeq\!2\,(1\!+\!z)R_g/\gamma_0\,\delta_0\,c$. The averages of
these widths and of their redshifts for ordinary GRBs are 0.89 s \cite{Soder} and 
$\langle 1\!+\!z\rangle\!\simeq\!3$.
 The ratio of these normalizing values to the 
 ${\rm FWHM}\!\simeq\!R_g\theta^2/c\!\simeq\!2$ s
 duration of GRB980425 results in $\gamma\,\theta\!\simeq\!9$. Upon substitution in Equation
\ref{eq:LLOEiso} one obtains $E_{iso}\rm (GRB980425)\!\simeq\!1.1\!\times\!10^{48}$ erg, 
 in good agreement with its measured  $(1.0\!\pm\!0.2)\!\times\!10^{48}$ erg 
  \cite{Amati76}.\\
 \indent (iii) The 0.3-10 keV X-ray light-curve of the AG of GRB980425 
\cite{Pian2004,Kulkarni} can be well
 fit by the CB model with $\gamma\,\theta\!\approx\!8.7$, 
 as shown in Figure  \ref{fig:XAG980425}.

\subsubsection{\bf GRB030329}

 As mentioned in Section \ref{GRBProg}, the CB model was used to fit the 
 early AG of GRB030329 and to predict the discovery date of its associated
 SN, SN2003dh. This being a two-pulse GRB, the fits to its $\gamma$ rays and to its
 AG consequently involved two cannonballs. The prediction of the amount of
 their superluminal motion, based on the approximation of a constant ISM density,
  turned out to be wrong \cite{Taylor2003}. 
 Subsequent observations of the AG showed a series of 
very clear re-brightenings, interpreted in the CB model as encounters of the CBs with ISM 
over-densities. Corrected by the consequent faster slow-down of the CBs' motion, 
the new CB-model results, were not a prediction, but
were not wrong (see \cite{ManyBumpAG}  and its Figure 2 for 
controversial details not mentioned here).

The relative proximity ($z\!=\!0.1685$) of GRB030329 
and its record-bright
radio AG made possible its record long, high-resolution follow-up observations with the
Very Long Baseline Array (VLBA) and Very Long Baseline Interferometry (VLBI), until
3018.2 days post GRB \cite{Bhat}. 
The earlier observers \cite{Taylor2004} adopted a circular Gaussian fit of the location of 
the radio source(s). They could not resolve separate images, except on day 51
after burst, when two sources $0.28\!\pm\!0.05$ mas ($\sim\!0.80$ pc) away from each other were seen,
with a statistical significance of $20\,\sigma$. By then the two sources 
(two CBs or a CB and the supernova?) had moved away from one another at 
an average relative apparent superluminal velocity $\langle V_{app}\rangle\!\approx\! 19\,c$.
The ``second component" was discarded: {\it 
``Since it is only seen at a single frequency, it is remotely possible that
this image is an artifact of the calibration.''} \cite{Taylor2003,Taylor2004}
This second component, should it not be an artifact, would be, in the CB model,
initially {\it hyperluminal}. An approximation to a detailed analysis would
be the following.
For an initial CB with $\gamma^2\!\gg\!1/\theta^2\!\gg\!1$
the apparent velocity of Equation \ref{eq:Vapp} is $\approx\!2\,c/\theta$.
At later times, $t\!\approx\!t_b\!\approx\! 0.05$ days, 
when the CB has decelerated to the point that 
$[\gamma(t)\theta]^2\! < \! 1$, $V_{app}\!\propto\! t^{-1/2}$ in a (rough!)
constant-density approximation. The early {\it local} approximate apparent velocity is
 $V_{app}(t_b)\!\sim \! (1+z)\langle V_{app}\rangle
 (51\,{\rm d} /t_b)^{1/2}\!\approx\!710\,c$.

\begin{figure}[]
\centering
\includegraphics[width=8.5 cm]{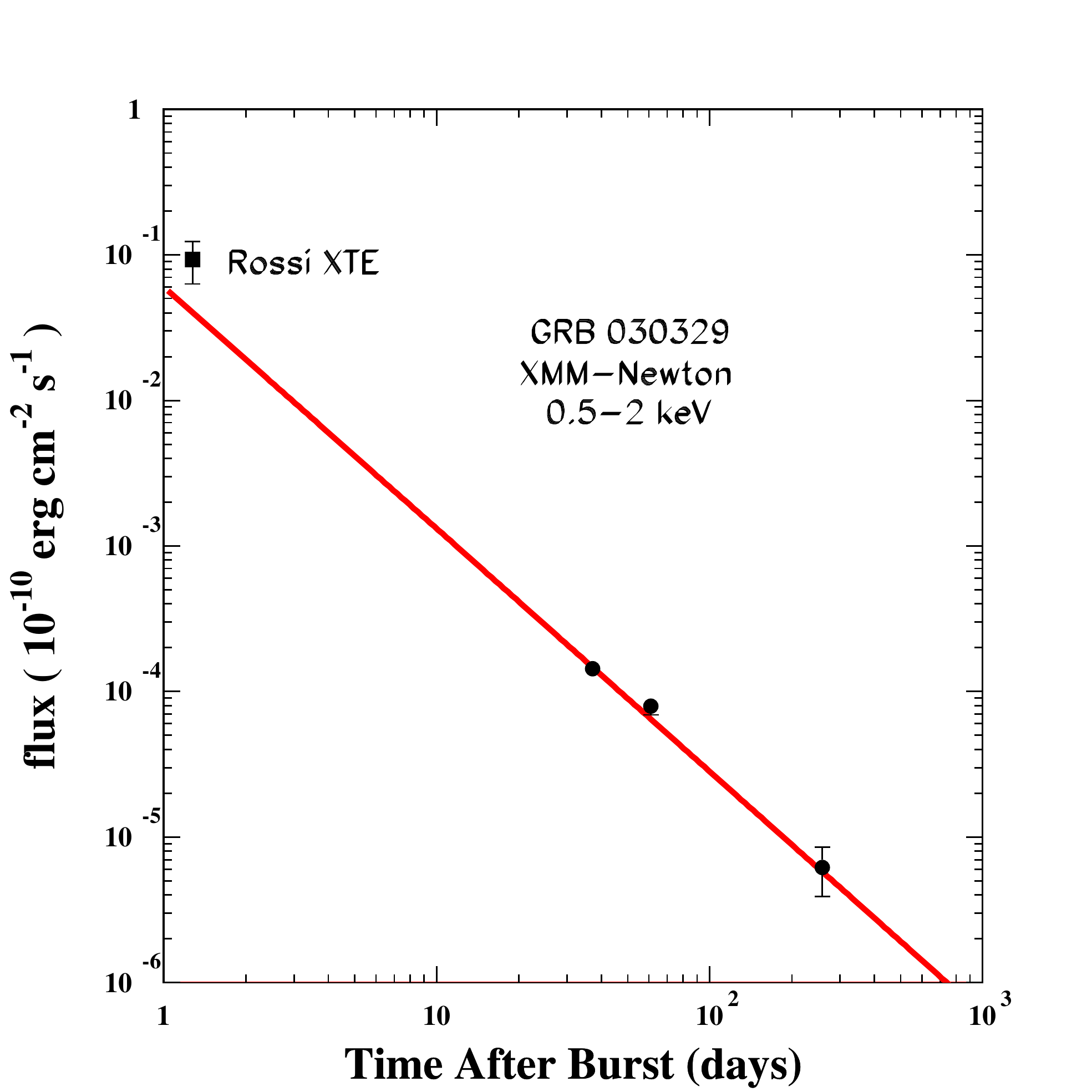}
\caption{The late-time 0.5-2 keV X-ray AG of the joint source 
GRB030329/SN2003dh as measured by Rossi-XTE and
XMM-Newton \cite{Tiengo}. The line is the CB 
model expectation (for the predicted
$\alpha_x\!=\!\beta_x+1/2$ ) and, in this case, $\beta_x\!=\!1.17\pm 0.04$.}
\label{fig:LTXAG030329}
\end{figure}

A way to test whether a constant-density approximation 
for the bumpy ISM traversed by the CBs of GRB030329 is good --on average--
is the following. At late times 
 $\delta(t)\!=\!2\,\gamma(t)\!\propto\! t^{-1/4}$.
Consequently, $F_\nu\!\propto\! t^{-\alpha_\nu}\,\nu^{-\beta_\nu}$ with 
$\alpha_\nu\!=\!\beta_\nu\!+\!1/2$, 
which are all well satisfied 
by the late-time X-ray afterglow \cite{Vanderspek} and radio observations  
of GRB030329 \cite{Taylor2004}. For example, the measured spectral index 
$\beta_X\!=\!1.17\!\pm\! 0.04$ 
in the 0.2-10 keV X-ray band \cite{Bhat}, results in a late-time temporal decay index 
$\alpha_x=\beta_x\!+\!1/2\!=\!1.67\!\pm\!0.04$, in good agreement with 
the observed $\alpha_x\!=\!1.67$ \cite{Vanderspek} 
as shown in Figure \ref{fig:LTXAG030329}. 

\begin{figure}[]
\centering
\includegraphics[width=8.5 cm]{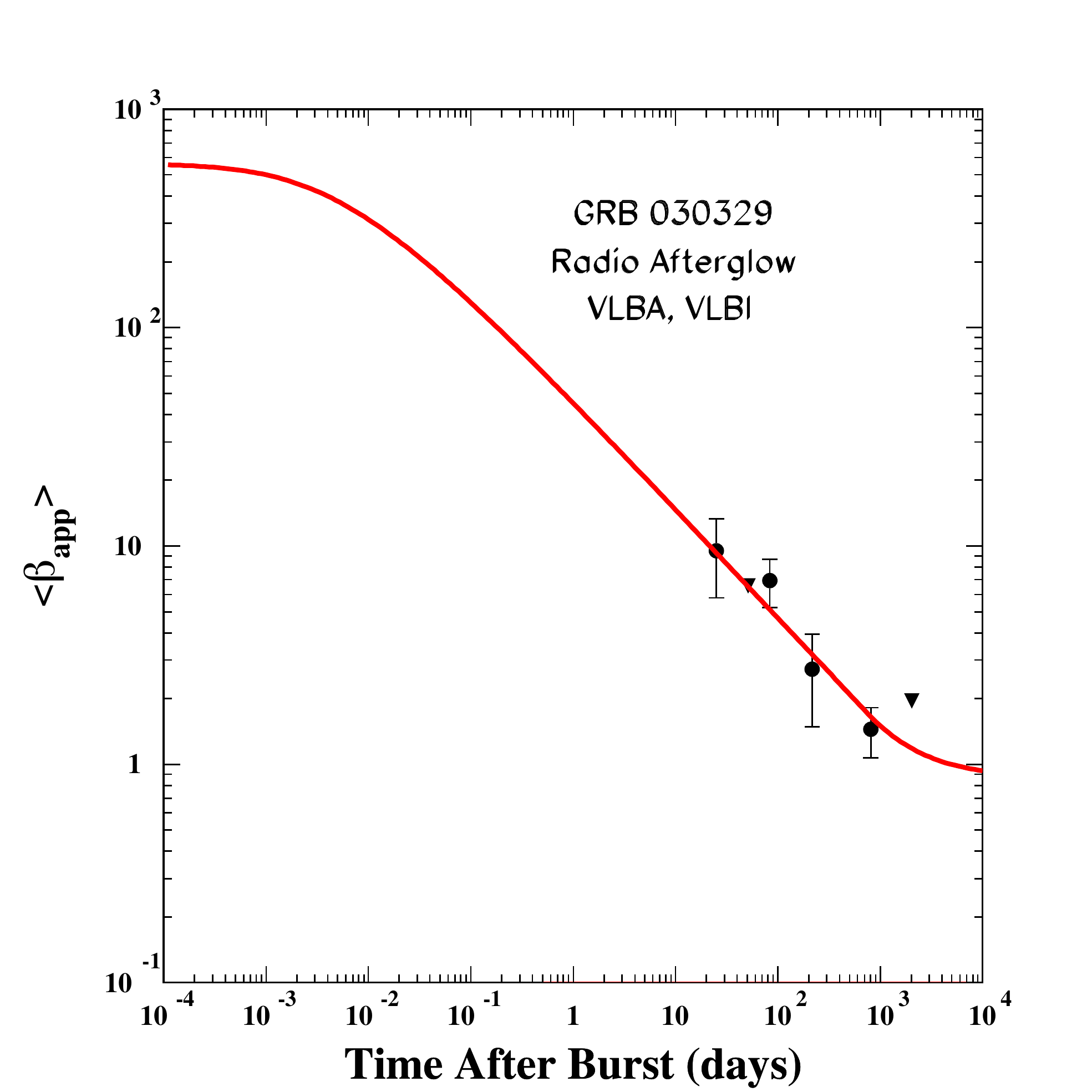}
\caption{The time-averaged expansion rate of the radio image of 
GRB030329/SN2003dh \cite{Bhat}. The line is the predicted 
$\langle \beta_{app}\rangle$ of the CB, which produced GRB030329 
and its AG, assuming that $2R_\perp$ is its distance  
from SN2003dh.}
\label{fig:SLMbeta030329}
\end{figure}

 The earlier VLBA  measurements \cite{Taylor2003}
 could not  resolve  the separate images of
SN2003dh and the CBs which produced GRB030329 and its AG.
Assuming a disk shape, the radio image was fit with a circular Gaussian 
of diameter $2\,R_\perp(t)$. If this distance is a rough measure of the displacement of the second
CB fired by SN2003dh, it corresponds to a mean superluminal velocity  
declining like $t^{-1/2}$, in agreement with the data,
shown in Figure \ref{fig:SLMbeta030329}.
The prediction is for $\gamma_0\,\theta\!=1.76$, obtained
for the observed \cite{Taylor2004} $E_{iso}\rm(GRB030329)\!=\!(1.86\pm 0.08)\!\times\!10^{52}$ erg
and $\langle E_{iso}\rm{(O\,\, GRB)}\rangle \!\approx\! 7\!\times\!10^{52}$
erg for ordinary GRBs, assuming a standard candle approximation for GRBs 
and $\epsilon_p\!\simeq\!1$ eV \cite{DD2004} in Equation \ref{eq:Ep0}.  
The observed late-time behavior, shown in Figure \ref{fig:SLMbeta030329}, 
suggests deceleration in a constant density ISM of a CB with $\gamma_0\!\approx\!400$.

Initially, the radius of a CB in its rest frame increases at the speed of sound in a relativistic
plasma, $c_s\!=\!c/\sqrt{3}$.
At an observer's time $T$
the radius of a CB has increased to: 
\begin{equation}
R(T,\theta)\!\approx\! {c_s\over (1+z)}\int_0^T\delta(t,\theta)\,dt,
\label{eq:CBradius}
\end{equation}
where use has been
made of the relation between $T$ and the time in the CB's rest system. 
When the first radio observations took place, $T\!=\!2.7$ days after burst,
the result for this GRB --for the parameters of the CB-model description of its AG and the
deceleration law of Equation \ref{eq:decelerate}-- is $R(T)\!\approx\!5.7\!\times\! 10^{17}$ cm.
This is more than an order of magnitude larger than the largest source size 
that could have resulted in diffractive scintillations.

The FB model has been used to provide successive a posteriori interpretations of.
the observed superluminal expansion \cite{Bhat} of the image size of the source of the radio
AG of the GRB030329-SN2003dh pair, as the observations progressed. All of them
were parametrizations with many adjustable parameters. 
A re-brightening of the source was expected in the FB model as the counter jet became non-relativistic \cite{Granot&al}. But no re-brightening was detected up to 10 years after the burst 
\cite{Tiengo}. As stated by the discoverers of the ``second source" \cite{Taylor2003}:
{\it This component requires a high average velocity of 19c 
and cannot be readily explained by any of the standard models.}

\section{Fast Extragalactic X-ray Transients, Tests 14 \& 15}

In the CB model GRBs produce two main types of extragalactic X-ray transients:
X-ray flashes , which are narrowly beamed LGRBs viewed  far off axis \cite{DDD5}, 
and fast X-ray transients (XRTs) which are emissions from a wind nebula
powered by a newly-born millisecond pulsar in a 
binary neutron-star merger and in SN-less LGRBs \cite{{DDD5}}.

As stated earlier, GRB pulses produced by highly relativistic 
CBs and observed from a far off-axis angle $\theta\! \gg 
1/\gamma$, appear as XRFs  \cite{{DDD5}} with $L_p$, $E_{iso}$ 
and $E_p$ reduced by a factor $\approx[1/\gamma^2 (1\!-\!\cos\theta)]^{4,3,1}$, 
respectively, and a duration $\rm T(FWHM)$ larger  by a factor 
$\approx \gamma^2(1\!-\!\cos\theta)$, all of it
in comparison with ordinary GRBs viewed 
from angles $\theta\!\sim\!1/\gamma$. 
The strong reduction of the peak luminosity $L_p$ of far 
off-axis GRBs makes them visible only from relatively small $z$. 
The orphan isotropic X-ray AGs of far off-axis (unobserved SGRBs \cite{{DDD5}})
are much longer and are visible up to very  large cosmological distances.

\begin{figure}[]
\centering
\includegraphics[width=8.5 cm]{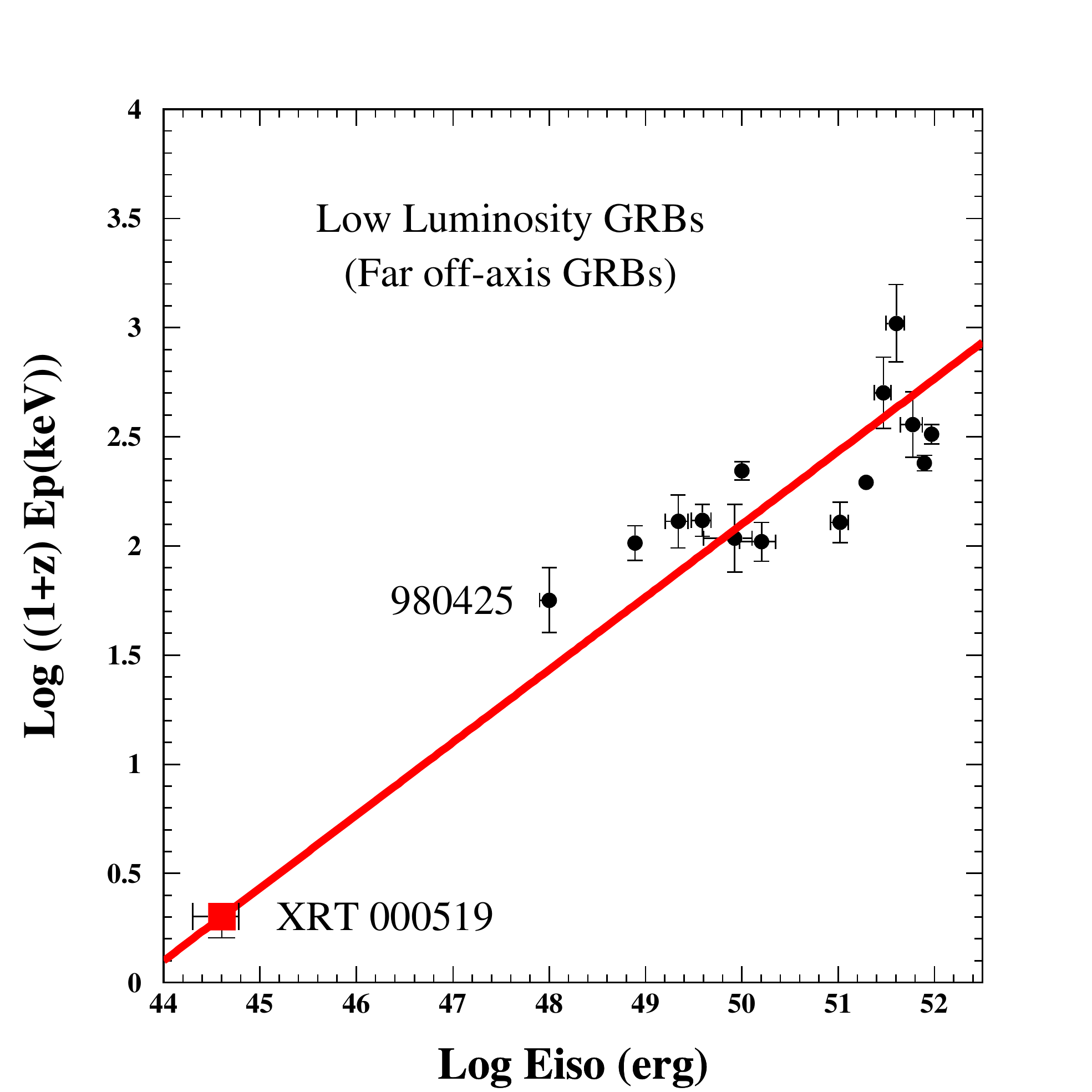} 
\caption{The $[E_p,E_{iso}]$ correlation 
in LGRBs viewed far off axis (which include  low-luminosity 
LGRBs and XRFs) of Figure \ref{fig:AGSHBXT},  with the addition of XRT 000519 \cite{Jonker}  
(full red square).
The line is the CB model predicted correlation \cite{Correlations},  Equation \ref{eq:Corr2}.} 
\label{fig:EPEISOXRTLLGRB}
\end{figure}

The X-ray transients XRT 000519 \cite{Jonker} and XRT 110103 \cite{Glennie}
could have been classified as XRFs. The correlation of
Equation \ref{eq:Corr2}
for XRFs (far off axis GRBs \cite{{DDD5}}), shown in Figure \ref{eq:Corr2}, is well 
satisfied by $E_p\approx 1.5\pm 0.5$ keV and 
$E_{iso}\!\approx\!(4\pm2)\times 10^{44}$ erg, obtained \cite{Jonker} for XRT 000519, as 
shown in Figure \ref{fig:EPEISOXRTLLGRB}. This is quite impressive, since
the the $E_p$ of XRT 000519 is more than three orders of magnitude lower than that
of GRB 980425, the next low-$E_p$ record holder in the figure.

\begin{figure}[]
\centering
\includegraphics[width=8.5 cm]{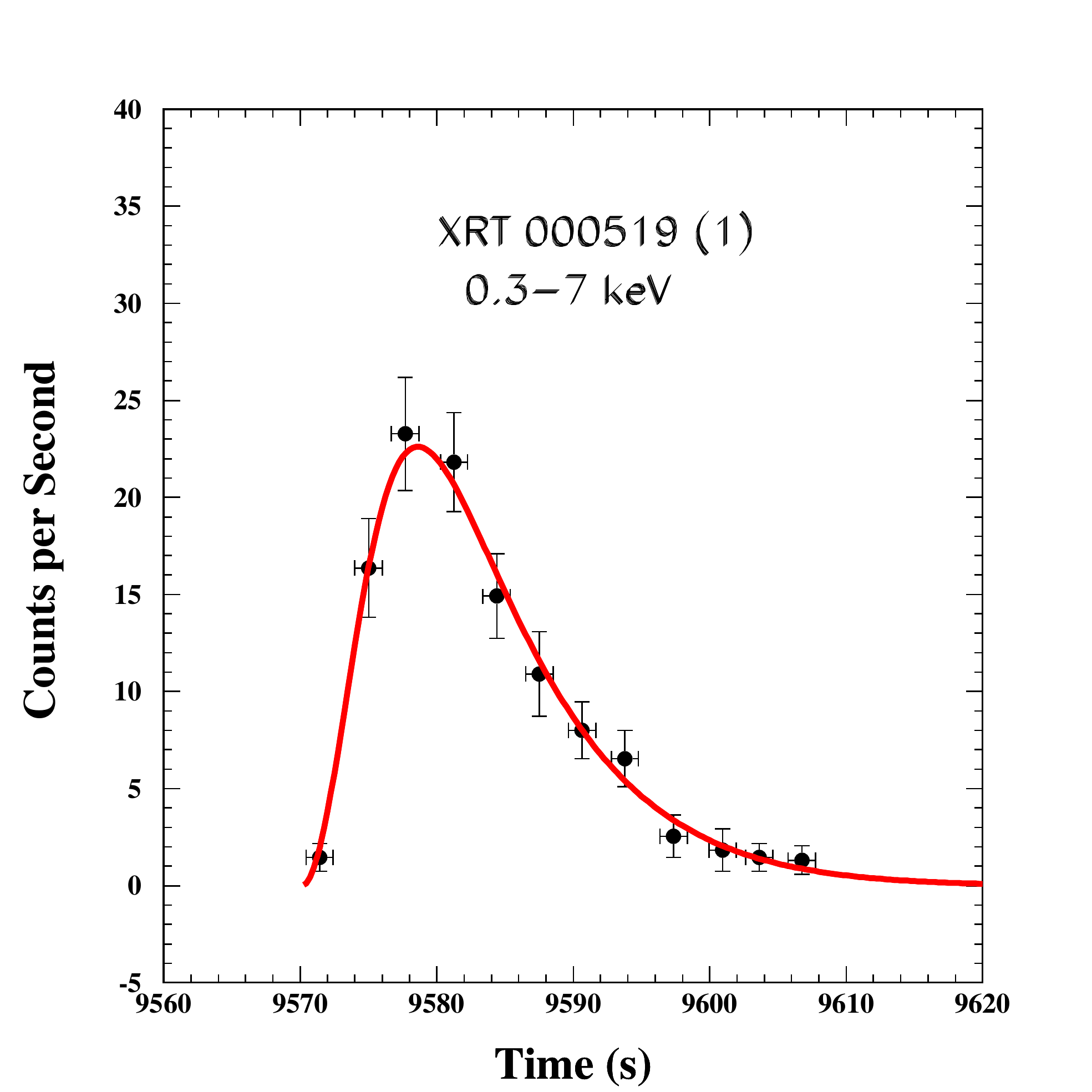} 
\caption{The best fit pulse shape given by Equation \ref{eq:OffAxisPulseShape}
for the first pulse of XRT 000519, reported in \cite{Jonker}.
The fit parameters are $\Delta\!=\!9.65 $ s, $\tau\!=\!17.50$ s,  
$E_p(0)\!=\!13.70$ keV 
(for time $t$ since 9570.10 s); $\chi^2/{\rm dof}\!=\!0.39$.}
\label{fig:XRT000519P1}
\end{figure}

\begin{figure}[]
\centering
\includegraphics[width=8.5 cm]{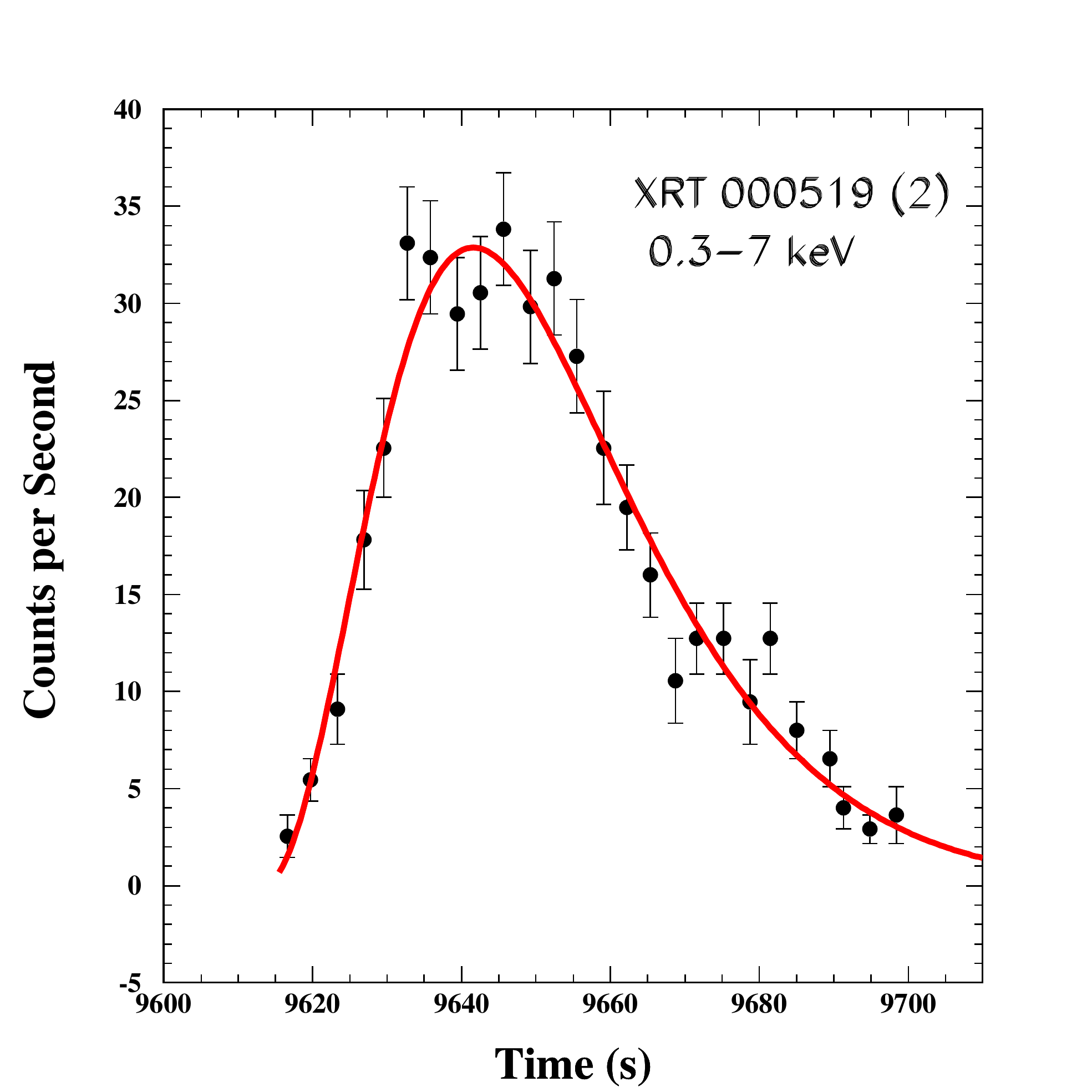} 
\caption{The light curve of the second pulse 
of XRT 000519 \cite{Jonker} and the CB-model's predicted shape 
of Equation \ref{eq:OffAxisPulseShape} with the best-fit parameters 
$\Delta\!=\!35.54$ s, $\tau\!=\!28.58$ s, $E_p(0)\!=\!27.11$ keV
(for time $t$ since 9613.28 s); $\chi^2/{\rm dof}\!=\!1.28$.}
\label{XRT000519P2}
\end{figure}

\begin{figure}[]
\centering
\includegraphics[width=8.5 cm]{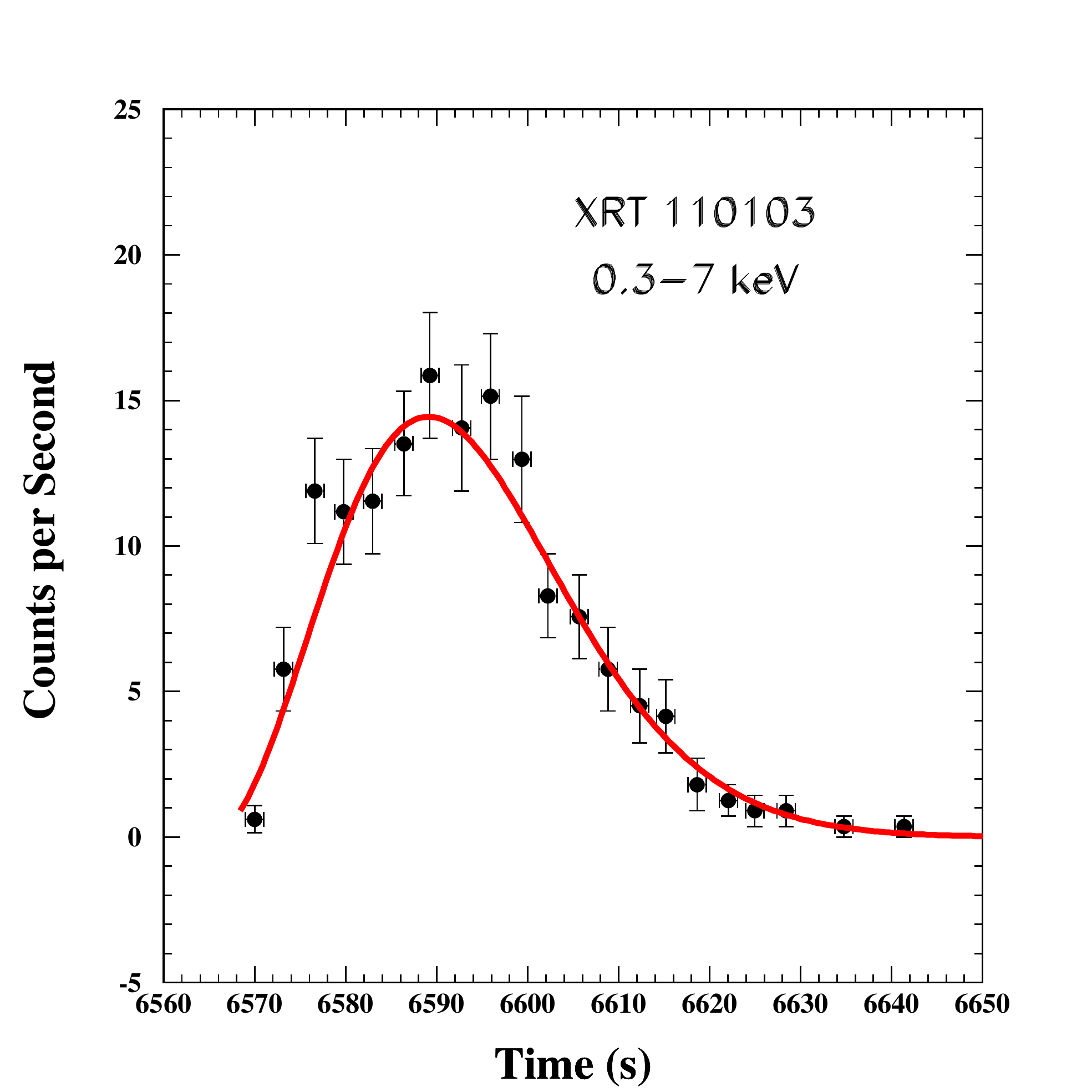} 
\caption{The pulse shape of  XRT 110103  \cite{Glennie}
and the CB-model shape  
of Equation \ref{eq:OffAxisPulseShape} with the best fit parameters, 
$\Delta\!=\!45.03$ s, $\tau\!=\!23.16$ s, $E_p(0)\!=\!12.18$ keV
(for time $t$ since  6564.90 s); $\chi^2/{\rm dof}\!=\!1.20$.}
\label{fig:XRT110103P3}
\end{figure}

To test further  the contention that
XRT 000519 \cite{Jonker} and  XRT 110103 \cite{Glennie} 
can be interpreted as far off-axis LGRBs, 
one may fit
their pulse shapes (count rate as function of time) by the approximate 
CB-model shape of  far off-axis GRBs following from Equation \ref{eq:pulseShape}
\begin{equation}
{dN_\gamma(E\!>\!E_m)\over dt}\!\propto\!{t^2\over(t^2\!+\!\Delta^2)^2}\,
e^{-E_m/E_p(0)}\left[1\!-\!{t\over\sqrt{t^2\!+\!\tau^2}}\right],
\label{eq:OffAxisPulseShape}
\end{equation}
where $E_m$ is a minimum energy of detection, $E_m \!=\! 0.3$ keV for the
two pulses of XRT000519 shown and fit in Figures \ref{fig:XRT000519P1}
and \ref{XRT000519P2}. The single pulse of XRT110103 is fit with Equation 
\ref{eq:OffAxisPulseShape} in Figure
\ref{fig:XRT110103P3}.

\begin{figure}[]
\centering
\includegraphics[width=8.5 cm]{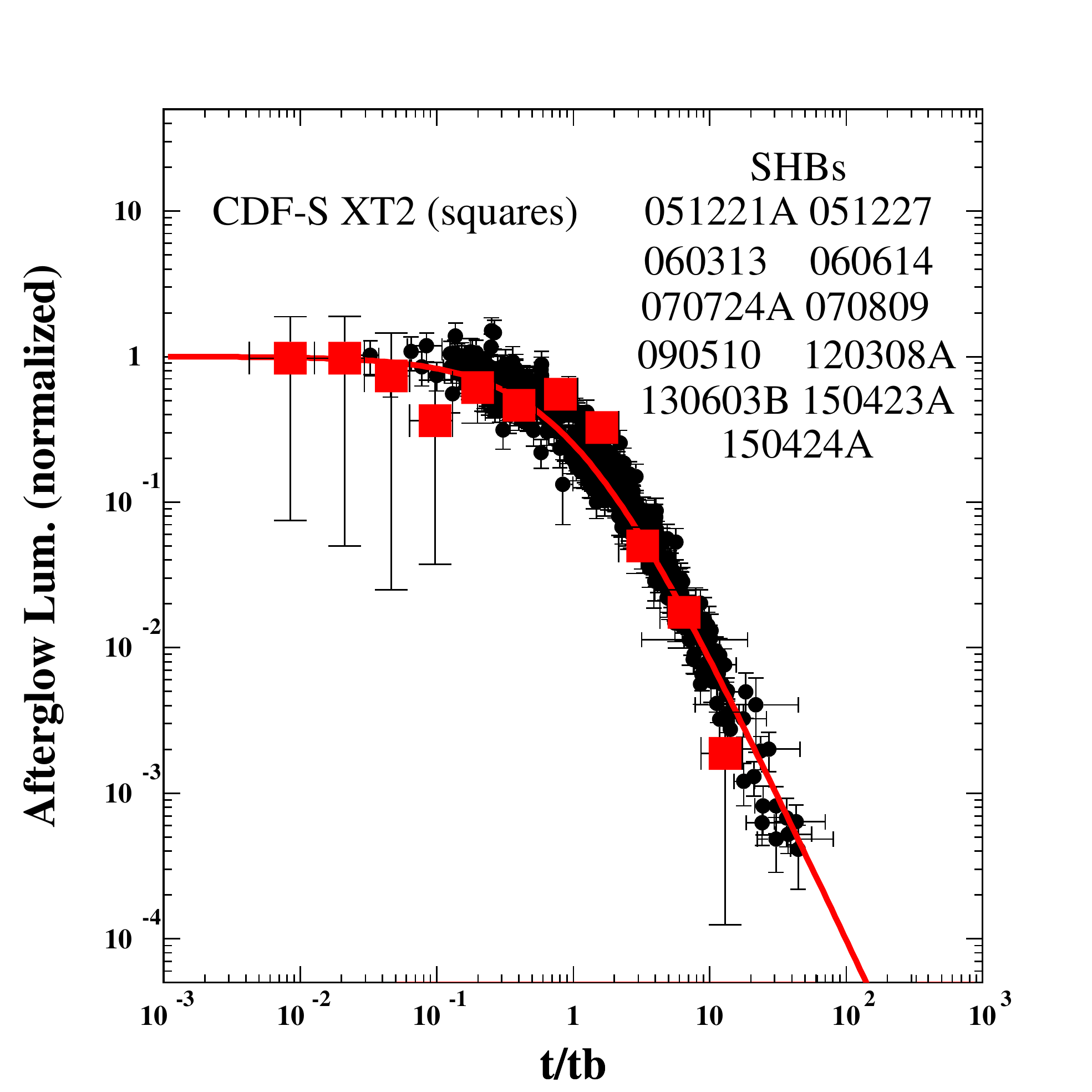} 
\caption{Comparison between the scaled 0.3-10 keV
light curves of the well sampled X-ray AGs 
of SGRBs during the first couple of days after 
burst measured with the Swift XRT  
and the 0.3 to 10 keV light curve of CDF-S XT2 \cite{Xue}. 
The line is the  universal behavior 
of Equation \ref{eq:PWN} for a PWN AG powered by a 
newly-born millisecond pulsar with a braking index $n\!=\!3$.} 
\label{fig:AGSHBXT} 
\end{figure}

In the CB model the early time X-ray  AGs  of  both 
near or far off-axis SGRBs are hypothesized to be powered by a 
newly born millisecond pulsar with a braking index 
$n\approx 3$. Their energy flux satisfies Equation \ref{eq:PWN}.
In Figure \ref{fig:AGSHBXT}  the  X-ray light curve of the fast 
extragalactic transient CDF-S XT2 
\cite{Xue} is shown and
compared to the early-time observations of the 
X-ray AGs of all SGRBs well sampled at early times and
reported in the Swift-XRT Light Curves Repository \cite{Swift}.

\subsection {XRFs in the FB model}
In fireball models XRFs are assumed to belong to a different 
class of GRBs. But their consequently estimated yearly rate over the whole sky is 
negligible compared to the rate of XRFs extracted from observations:
 $\sim\!1.4\!\times\! 10^5$ per year over the whole sky \cite{Glennie}.

\section{GRB Theories confront SHB170817A (Test 16)}

This event is an optimal case to
close the discussion of the comparisons between different models of GRBs. 

GW170817 was the first binary neutron-star merger detected with Ligo-Virgo \cite{Abbott} in
gravitational waves (GWs). It was followed by SHB170817A, $1.74\!\pm\! 0.05$ s  after the end
of the GW's detection, with an afterglow across the electromagnetic spectrum, which
was used to localize it \cite{Hjorthetal2017} to the galaxy NGC 4993. The
GW170817/SHB170817A association was the first indisputable confirmation that pairs of
neutron stars merging due to GW emission produce GRBs. Earlier, other mechanisms
of GRB emission in such mergers had been suggested: in 1984 
the explosion of the
lighter NS after tidal mass loss \cite{Blinnikov}, later called a ``Macronova" event 
 \cite{Paczynski86} and, in
1987, a neutrino annihilation fireball \cite{Goodman2}
around the remnant neutron star. In 1994, the
fireball mechanism was replaced \cite{Shaviv} by ICS of external light by a highly relativistic jet
of ordinary plasma launched by fall back ejecta on the remnant star after the merger,
which became the basis of the CB model of SGRBs and their afterglows.

The relative proximity of SHB170817A provided many critical tests of
SGRB theories. As mentioned in Section \ref{sec:Progenitors}, two days
before the GW170817/SHB170817A 
event, the CB model of GRBs was used to predict \cite{DadoSG} that most of 
the SGRBs associated with Ligo-Virgo detections of NS/NS mergers would 
be beamed far off-axis. Consequently, only a small fraction of 
them would be visible as low-luminosity far off-axis SGRBs \cite{DadoSG}. 
Also \cite{DadoSG}
the early afterglow, powered by the spin down  
of the newly born millisecond pulsar, would generate a characteristic isotropic 
universal light-curve  \cite{Dado2018}, visible from all 
SHBs associated with gravitational wave detections of neutron star mergers.

Shortly after the detection of SHB170817A, several 
authors claimed \cite{Goldstein2017} that it was an ordinary near-axis SGRB, 
even though its $E_{iso}$ was four orders of magnitude smaller
than that of typical SGRBs. Others, who claimed in the past that ordinary 
and low luminosity GRBs belong to two different classes,
suggested that SHB170817, was ordinary, but viewed far off-axis. 
Moreover,  other changes --labeled ``structured jets"-- were introduced. 
Despite the availability of
many adjustable parameters, all such models failed to 
predict correctly the future evolution of the 
AG light curves of SHB170817A, even further readjusting 
multiple parameters to best fitting the entire  
earlier data, as shown later in Figures 
\ref{fig:FB_AG170817A5}, \ref{fig:FB_AG170817A7}.

\subsection{The properties of the SHB170817A ejecta}

\subsubsection{A superluminally moving source}

\begin{figure}[]
\centering
\includegraphics[width=8.5 cm]{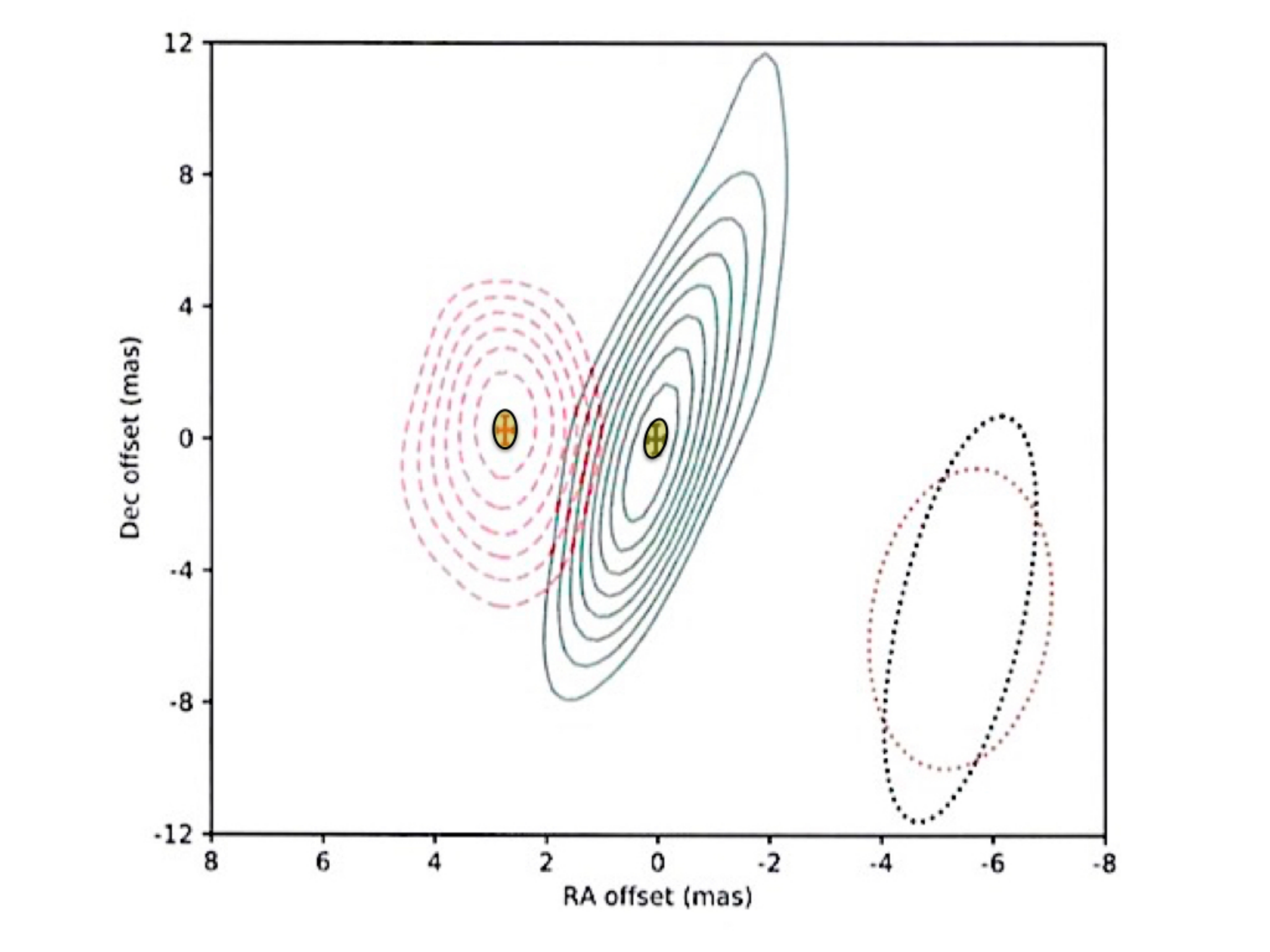}
\caption{Proper motion of the radio counterpart of GW170817. 
Its authors \cite{MooleyJET} explain: {\it The
centroid offset positions (shown by $1\,\sigma$ error bars) and 
$3\,\sigma$-$12\,\sigma$ contours of the radio source detected 75 d 
(black) and 230 d (red) post-merger with VLBI at 4.5 GHz. 
The radio source is consistent with being unresolved at both epochs. 
The shapes of the synthesized beam 
for the images from both epochs are shown as dotted ellipses in the 
lower right corner. The proper motion vector of the radio source has 
a magnitude of $2.7\!\pm\!0.3$ mas}. The $1\,\sigma$
domains have been colored
not to deemphasize the point-like nature of the source (a CB). }
\label{fig:MooleyRadio}
\end{figure}

The VLBI/VLBA observations of the 
radio AG \cite{MooleyJET} of SHB170817A provided images 
of an AG source escaping from the GRB location with
superluminal celerity.  Such a behavior in GRBs was predicted by the CB 
model \cite{DD2004} two decades ago \cite{MooleyJET}. Figure \ref{fig:MooleyRadio}, 
borrowed from \cite{MooleyJET}, shows a displacement with time of a point-like radio 
source --as seen before in micro-quasars and blazars \cite{DD2004}--, rather than 
an unresolved image of a GRB and its AG expanding with a 
superluminal speed, as claimed before in the case of GRB030329
 \cite{Taylor2004}. 
The Figure displays the angular locations of the radio 
AG of SHB170817A moving away in the plane of the
sky from the SHB location by 
$2.68\!\pm\!0.3$ mas \cite{MooleyJET} between day 75 and day 230. 
In \cite{MooleyJET} this image is called ``a jet". 

\subsubsection{Superluminal motion} 

Consider a highly relativistic ($\beta\!\approx\!1$) CB. Its apparent velocity,
Equation \ref{eq:Vapp}, can also be expressed 
in terms of observables:
\begin{equation} 
 V_{app}\!\approx\! 
{c\,\sin\theta\over (1\!+\!z)\,(1\!-\!\cos\theta)}\!\approx\! 
{D_A\,\Delta \theta_s\over (1\!+\!z)\Delta t}\,,
\label{eq:Vapp2}
\end{equation} 
where $\Delta\theta_s$ is the angle by which the source is seen to have moved in a time 
$\Delta t$.
The angular distance to SHB170817A to its host galaxy NGC 4993, 
at $z\!=\! 0.009783$ \cite{Hjorthetal2017}, is $D_A\!=\!39.6$ Mpc, for the local value 
 $H_0\!=\!73.4 \pm 1.62\,{\rm km/s\, Mpc}$ obtained from Type Ia SNe \cite{Riess 2016}.  
The location of the VLBI-observed source
--which moved $\Delta\theta_s\!=\! 2.7\!\pm\!0.3$ mas in a time 
 $\Delta t\!=\!155$ d (between days 75 and 230)-- implies
 $V_{app} \!\approx\! (4.0\pm 0.4)\,c$, which, solving for the
 viewing angle $\theta$
in Equation \ref{eq:Vapp2}, results in $\theta\!\approx\!  27.8 \pm 2.9$ deg.
 This value agrees with $\theta\!=\!25\pm 8$ deg, 
obtained \cite{Mandeletc2017}  from the gravitational wave observations
\cite{Abbott}
 for the same $H_0$ \cite{Riess 2016} and the hypothesis that the CB 
was ejected along the rotation axis of the binary system.

\subsubsection{Initial Lorentz factor}

In the CB model SGRBs, much as LGRBs, can be treated as approximately standard 
candles viewed from different angles. They satisfy similar 
correlations. In particular low luminosity (LL) 
SHBs such as SHB170817A are ordinary (O) SHBs viewed far 
off-axis. Consequently, their $E_{iso}$ and $E_p$ are expected to obey the relations
\begin{equation}
E_{iso}{\rm (LL\,SHB)}\! \approx\! \langle\! E_{iso}{\rm (O\,SHB)}\!\rangle /
[\gamma^2\,(1\!-\!\cos\theta)]^3 \,,
\label{eq:LLSHBEiso}
\end{equation}
\begin{equation}
(1+z)\,E_p{\rm(LL\,SHB)}\!\approx\! \langle (1+z)\,E_p{\rm (O\,SHB)}\rangle /
[\gamma^2\,(1\!-\!\cos\theta)].
\label{eq:LLSHBEp}
\end{equation} 

Given the measured value  $E_{iso}\!\approx\! 5.4\times 10^{46}$ erg of 
SHB170817A \cite{Goldsteinetal2017}, the mean 
value $\langle E_{iso}\rangle\!\approx\!1\times 10^{51}$ erg of 
ordinary SGRBs,  and  the  viewing angle 
$\theta\!\approx\!28$ deg obtained  from the observed superluminal 
velocity of the source of its radio afterglow \cite{MooleyJET}, Equation \ref{eq:LLSHBEp}  
yields  $\gamma_0\!\approx\!14.7$ and $\gamma_0\,\theta\!\approx\! 7.2$.

\subsubsection{Prompt Emission Observables}

The CB model correlations of Equations
  \ref{eq:Eiso} and \ref{eq:Corr1} are well satisfied by each of 
the three major types of 
GRBs: SN-LGRBs, SN-less LGRBs and SN-less SGRBs, as was demonstrated 
in Figures  \ref{fig:epeiso17GRBs}, \ref{fig:epeisoLLGRBs} and \ref{fig:epeisoallshbs}.
Equations \ref{eq:LLSHBEiso} and \ref{eq:LLSHBEp}, for the viewing angle of 
the CB obtained from its apparent superluminal 
motion \cite{MooleyJET}, result in the following additional tests of CB model predictions:

{\bf Peak energy.}
Assuming that SHBs have the same redshift distribution as GRBs
(with a mean value $\langle z\rangle \!\approx\!2$), and given the observed 
$\langle\! E_p\!\rangle\!=\!650$ keV of SHBs \cite{Goldsteinetal2017},
one obtains $\langle(1\!+\!z)\,E_p\rangle\! \approx\! 1950$ keV. 
Consequently, Equation \ref{eq:LLSHBEp} with $\gamma_0\,\theta\!\approx\! 7.2$ and
$z\!\approx\!1$
yields  $E_p\!\approx\!75$ keV for SHB170817A. This is to be
compared with $E_p\!=\!82\!\pm\!23$ keV ($T_{90}$) reported in
\cite{20a},  
$E_p\!=\!185\!\pm\!65$ keV estimated in \cite{20c},
and  $E_p\!\approx\!65\!+\!35(\!-\!14)$ keV estimated in \cite{20d}, 
from the same data, with a mean value $E_p\!=\!86\!\pm\!19$ keV,
agreeing with the expectation.

{\bf Peak time.}
In the CB model the peak time $\Delta t$ after the beginning 
of a GRB or SHB pulse is roughly equal to half of its FWHM.
Assuming again that SHBs are roughly standard candles, the
dependence of their $\Delta t$ values on the viewing angle $\theta$ is 
\begin{equation}
\Delta t{\rm (LL\,SHB)}\,\approx\! \gamma_0^2\,(1\!-\!\cos\theta)
\langle \Delta t{\rm (O\,SHB)}\rangle,
\label{eq:Peakt}
\end{equation}
For  $\theta\!\approx\!28$
deg, obtained from the superluminal motion of the source of 
the radio AG of SHB170817A,     
$\Delta t\!\approx\! 0.58$ s obtained  from the prompt emission pulse 
of  SHB170817A (see Figure \ref{fig:fig02}), and $\langle{\rm FWHM(SHB)}\rangle\!=55$ ms, 
Equation \ref{eq:Vapp2} yields $\gamma_0\!\approx\! 14.7 $. 
 Using Equations 
\ref{eq:LLSHBEiso} and \ref{eq:LLSHBEp}, and $\gamma_0\,\theta\!\simeq \!7.2$ 
one checks that this value of
$\gamma_0$ is  consistent 
with  $E_{iso}\!=\!5.4\times 10^{46}$ estimated in \cite{Goldstein2017}, and 
$\langle E_p\rangle\!=\! 86\!\pm\! 19$ keV  
the mean of the estimates in  \cite{Dado2018}.

Moreover, in the CB model the shape of resolved SHB and GRB  pulses 
satisfies $2\,\Delta t\!\approx\! {\rm FWHM} \!\propto\! 1/E_p$. Using the observed 
$\langle {\rm FWHM(SHB)}\rangle\!\approx\!55$ ms, $\gamma_0\!\sim\!14.7$, 
and  $\theta\!\approx\! 28$ deg, Equation \ref{eq:Peakt} for SHB170817A results in
$\Delta t \!\approx\!0.63$ s, in good agreement with its observed value, 
$0.58\!\pm\! 0.06$ s\footnote{Quite obviously, the replacements of physical 
parameters by their means
may not be completely reliable,
not only because of the spread in their values, but 
also because of detection thresholds and selection effects.}.

\subsection{The single pulse's correlation between energy and time}

\begin{figure}[]
\centering
\includegraphics[width=8.5 cm]{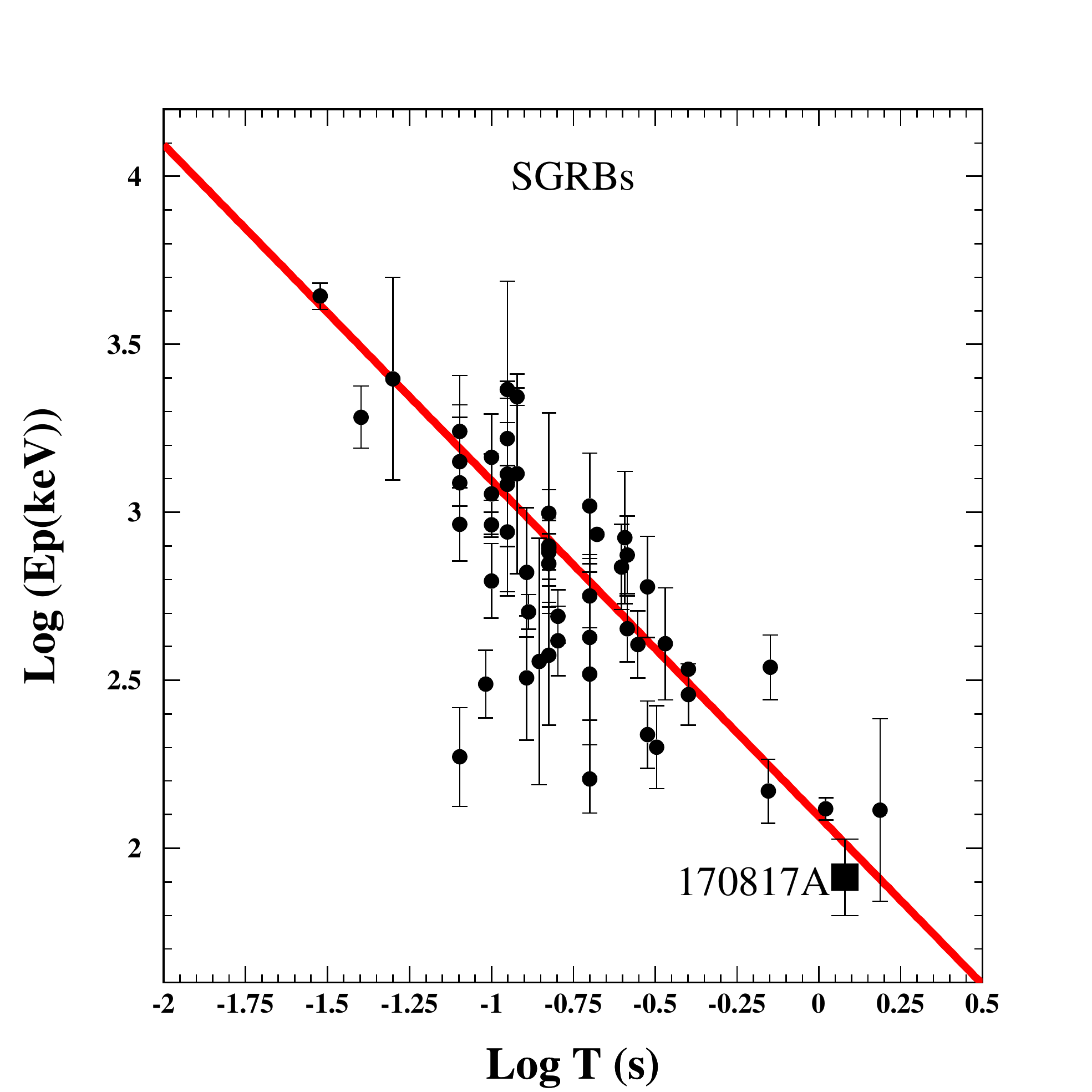}
\caption{
Comparison between the predicted $[E_p,T_p]$ correlation
of Equation \ref{eq:EpTp} and the corresponding data in GCN circulars
for 54 resolved pulses of SGRBs, obtained by the Konus-Wind
and Fermi-GBM collaborations.}
\label{fig:epvt_shb}
\end{figure}

In the CB model the peak-time of a pulse, $T_p\!\propto\!(1\!+\!z)/\gamma_0\, \delta_0$,  
and its peak energy $E_p$ of Equation \ref{eq:Ep0}
are properties of resolved or isolated pulses, 
which occur in a large fraction of SHBs, not GRBs. SHB170817A,
being a one-peak event, is a good case to study this variable.
One of the simplest CB-model predictions is the $[E_p,T_p]$ correlation 
\begin{equation}
E_p\!\propto\! 1/T_p\,.
\label{eq:EpTp} 
\end{equation}
In Figure \ref{fig:epvt_shb} this correlation is compared with the
values of $E_p$ and $T_p$ in the GCN circulars for resolved SGRB pulses 
measured by the Konus-Wind and by the Fermi-GBM collaborations. 
The figure shows that the prediction is satisfied by most of the 
measurements, in particular the ones with small error bars.

\subsubsection{The Early Time Afterglow}

\begin{figure}[] 
\centering
\includegraphics[width=8.5 cm]{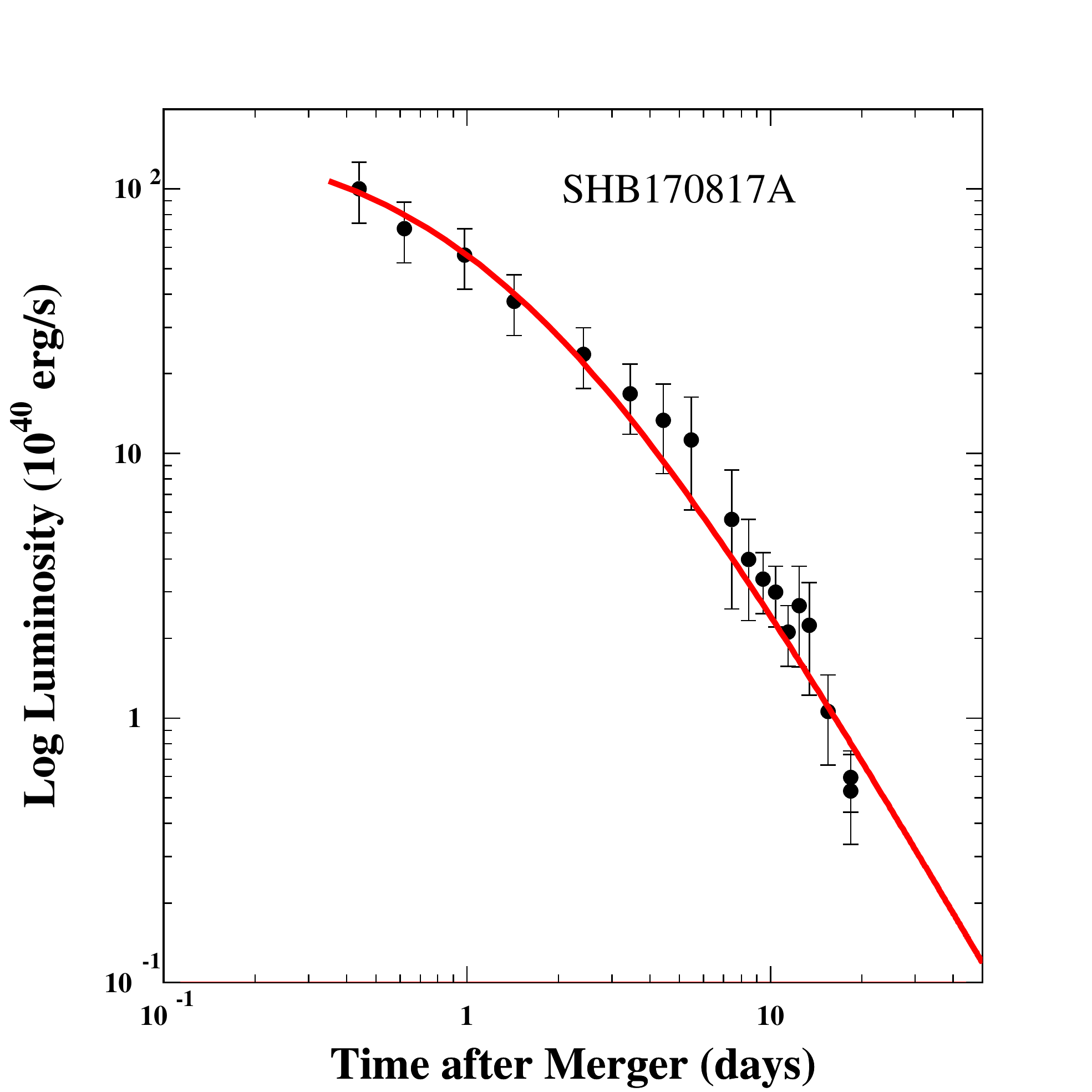} 
\caption{Comparison between the observed \cite{Drout2017} bolometric light curve 
of SHB170817A and the universal light curve of Equation \ref{eq:PWN},
assuming the presence of a milli-second pulsar  
 with $L(0)\!=\!2.27\times 10^{42}$ erg/s and  
$t_b\!=\!1.15$ d. The fit has $\chi^2/{\rm dof}\!=\!1.04$.} 
\label{fig:fig04a}
\end{figure}

The bolometric AG of SHB170817A, 
during the first few days after burst, has the universal shape of 
the early-time X-ray AGs of all SGRBs and SN-less LGRBs well sampled 
 during the first few days after the prompt emission. 
This universal shape is the one already shown for SGRBs in Figure \ref{fig:XAGS12SHBMSP}.
In Figure \ref{fig:fig04a}, it is shown for the bolometric light 
curve \cite{Drout2017} of SHB170817A during the first two weeks after burst.
In the absence of reliable observational information
on the environments of compact NS/NS and NS/BH binaries
shortly before they merge, and glory of optical-energy photons, similar to that
adopted in the CB model of LGRBs, was erroneously hypothesized 
for SHBs as well. It has been replaced  by a soft X-ray
glory once the superluminal motion of the ejected CB in 
SHB170817 was measured \cite{MooleyJET}.

\subsubsection{The late-time afterglow}

As long as the Lorentz factor of a decelerating CB is such that $\gamma^2\!\gg\! 1$,
 $\gamma\,\delta\!\approx\!1/(1\!-\!\cos\theta)$ and
the spectral energy density of its
unabsorbed synchrotron AG --Equation \ref{eq:Fnu}-- can be rewritten as 
\begin{equation}
F_{\nu}(t,\nu)\propto n(t)^{\beta_\nu+1/2}\,[\gamma(t)]^{2 \beta_\nu -4}\,\nu^{-\beta_\nu}\,
\label{eq:Fnu2}
\end{equation} 
with  $n(t)$ the baryon density of the medium encountered
by the CB and $\beta_\nu $ the spectral index of the
emitted synchrotron radiation.

For a constant density, the deceleration of the CB results in a late-time
$\gamma(t)\propto t^{-1/4}$ \cite{Dado 2013}, and as long as $\gamma^2\!\gg\!1$,
\begin{equation}
F_\nu(t,\nu)\!\propto\! t^{0.72 \pm 0.03}\nu^{-0.56\!\pm\!0.06},
\label{eq:Fnu3}
\end{equation}
with use of the observed \cite{MooleySL} $\beta_\nu\!=\!0.56\!\pm\!0.06$,
which extends from the radio (R) band, through the optical (O) band, to
the X-ray band. 

If the CB moved out from within a cloud of constant 
internal density into a  wind-like density
distribution (proportional to $r^{-2}$) its deceleration rate diminished
and $\gamma(t)$  became practically constant. 
Consequently, the time dependence of $F_\nu$ in
Equation \ref{eq:Fnu2} becomes a fast decline described by
\begin{equation}
F_{\nu}(t,\nu) \propto t^{-2.12\!\pm\!0.06}\nu^{-0.56\!\pm\!0.06}.
\label{eq:Fnu4}
\end{equation}

\begin{figure}[]
\centering
\includegraphics[width=8.5 cm]{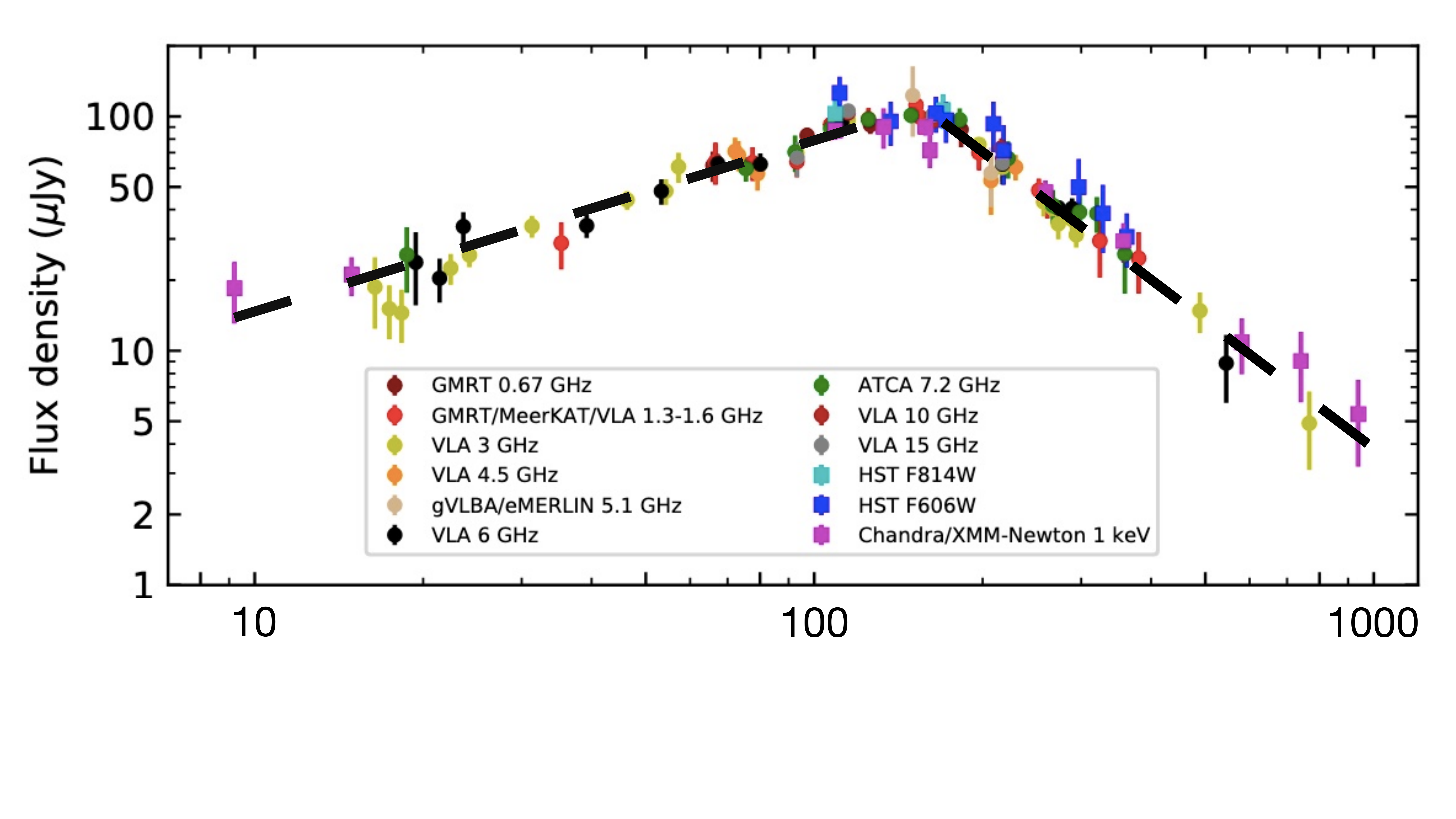} 
\vspace{-1cm}
\caption{Adapted from Figure 2 of \cite{Makha}. The radio light curve of
SHB170817A, measured until 940 days post-merger, spanning multiple
frequencies, and scaled to 3 GHz using the spectral index $-0.584$. 
The early-time trend expected in the CB model is the rising 
black-dashed line. The late-time trend, also black-dashed, is for an assumed 
$1/r^2$ ISM density decline.}
\label{fig:RadioAG}
\end{figure}

These CB-model approximate rise and fall power-law
 time dependences of the light curves of the ROX afterglow
of SHB170817, with temporal indices $0.72\!\pm\!0.03$
and $-2.12\!\pm\!0.06$, respectively, are in good agreement with 
the power-law indices extracted in \cite{MooleyKP8},
$0.78\!\pm\!0.05$  and $-2.41+0.26/-0.42$, respectively, from
a phenomenological parametrization of the measured radio
light curves of SHB170817 \cite{MooleyKP8}
during the first year after burst. They also agree with the indices 
subsequently extracted
in \cite{Makha}, $0.86\!\pm\!0.04$ and $-1.92\!\pm\!0.12$, as shown
in Figure \ref{fig:RadioAG}. 

\section{FB Model Interpretations of SHB170817A} 

Soon after the discovery of the late-time radio, optical and 
X-ray afterglows of SHB170817A, many FB model best fits to these initially
rising light curves were published. They involved many choices and 
best-fit parameters 
(e.g.~\cite{MooleyKP7} and references therein). When new observations were obtained, 
they did not follow the FB models' predictions. They were 
replaced with ``postdictions" and a readjusting of
assumptions and parameters. These repeated failures 
illustrate the flexibility of the FB models, 
rather than their validity. Some examples are summarized 
below.

In November 2017 the numerous authors of \cite{MooleyKP7} concluded that
{\it The off axis jet scenario as a viable explanation of 
the radio afterglow of SHB170817A is ruled out} and that a chocked 
jet cocoon is most likely the origin of the gamma rays and rising
AG of SHB170817A. Their claim was based on a  best fit 
(Figure 2 in \cite{MooleyKP7})
to the 3 GHz radio observations obtained with ATCA 
and VLA before November 2017. 

In April 2018, the observers among the authors of \cite{Dobie2018} reported 
that their 2-9 GHz radio observations of GW170817 covering the 
period 125-200 days post-merger, taken with the Australia 
Telescope Compact Array and the Karl G.~Jansky Very Large Array,
unexpectedly peaked at day $149\!\pm\!2$ post merger and 
declined thereafter. RXO observations of 
the AG were continued until two years 
after burst. They are shown in Figure \ref{fig:RadioAG}
(Figure 2 of \cite{MooleyKP8}). The parametrization is 
 a smoothly broken power-law \cite{Beuermann99} with a 
temporal index $\alpha\!=\!0.84\!\pm\!0.05$ on the rise, peak 
time $149\!\pm\! 2$ day, and a temporal index $1.6\!\pm\!0.2$ on 
the decay. The authors of \cite{MooleyKP8} reached conclusions opposite 
 to their earlier ones  \cite{MooleyKP7,Dobie2018} and to their previous
arXiv versions. To wit, in \cite{MooleyKP8} they have reported a {\it strong 
jet signature in the late-time light-curve of GW170817}. They 
justified their new conclusion by the fact that the post break 
flux density parametrized as $F_\nu(t)\!\propto\! 
t^{\alpha}\nu^{\beta}$  yields 
 $\beta\!=\!-0.54\!\pm\! 0.06$ and $\alpha\!=\!\!-2.17$, 
consistent with the FB model prediction
 $\alpha\!=\!-p$ post break with $p$ the power-law 
 index of the energy distribution of the radiating electrons \cite{Sari98}.

\begin{figure}[]
\centering
\includegraphics[width=7.5 cm]{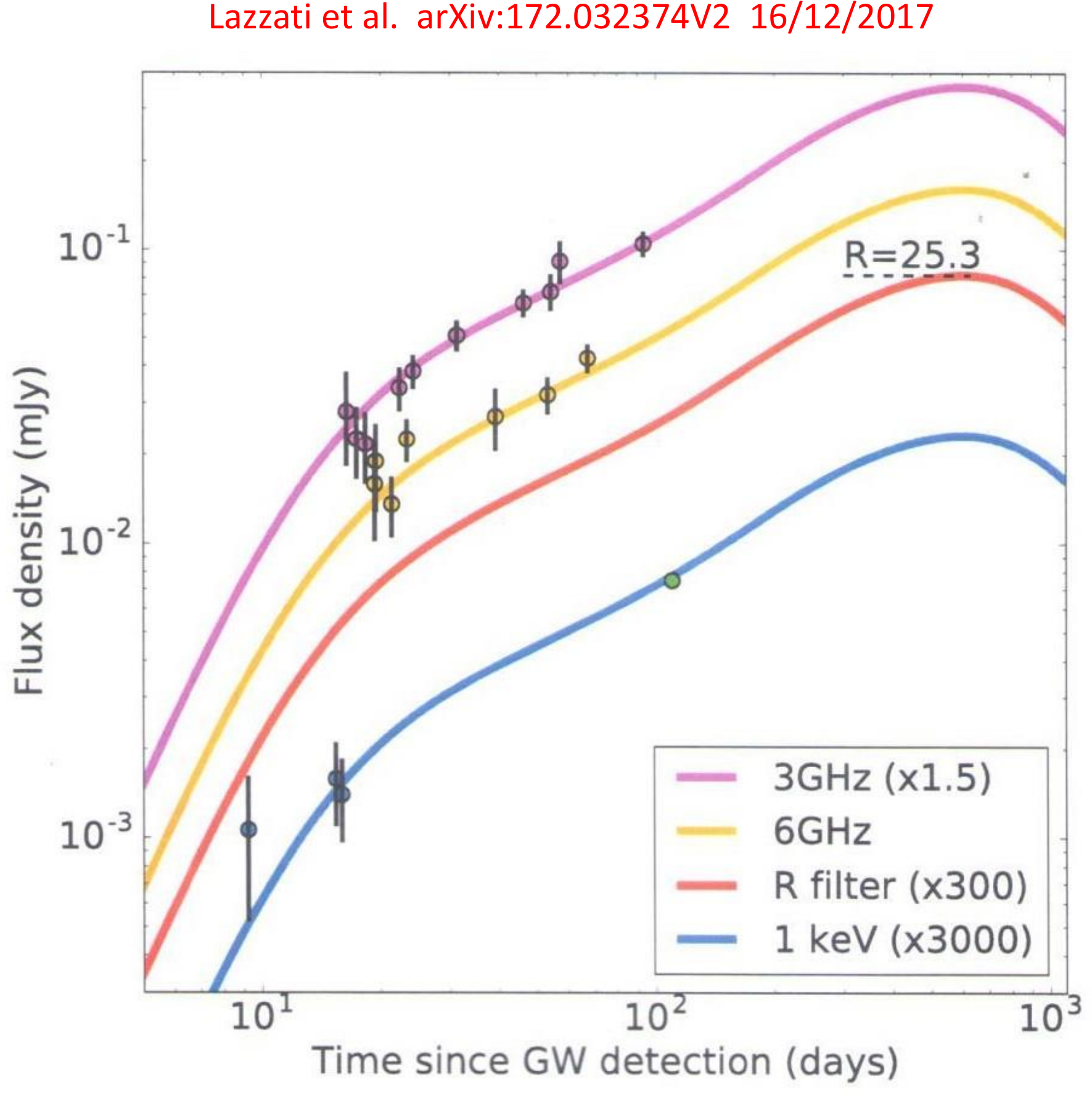} 
\caption{Best fit  light curves 
of an off-axis structured jet model 
\cite{Lazzati1} to the ROX 
AGs of SHB170817A measured before December 2017
(Figure from the first version of \cite{Lazzati1} posted in the arXiv on December 8th, 2017).}  
\label{fig:FB_AG170817A5}
\end{figure}

\begin{figure}[]
\centering
\includegraphics[width=8.5 cm]{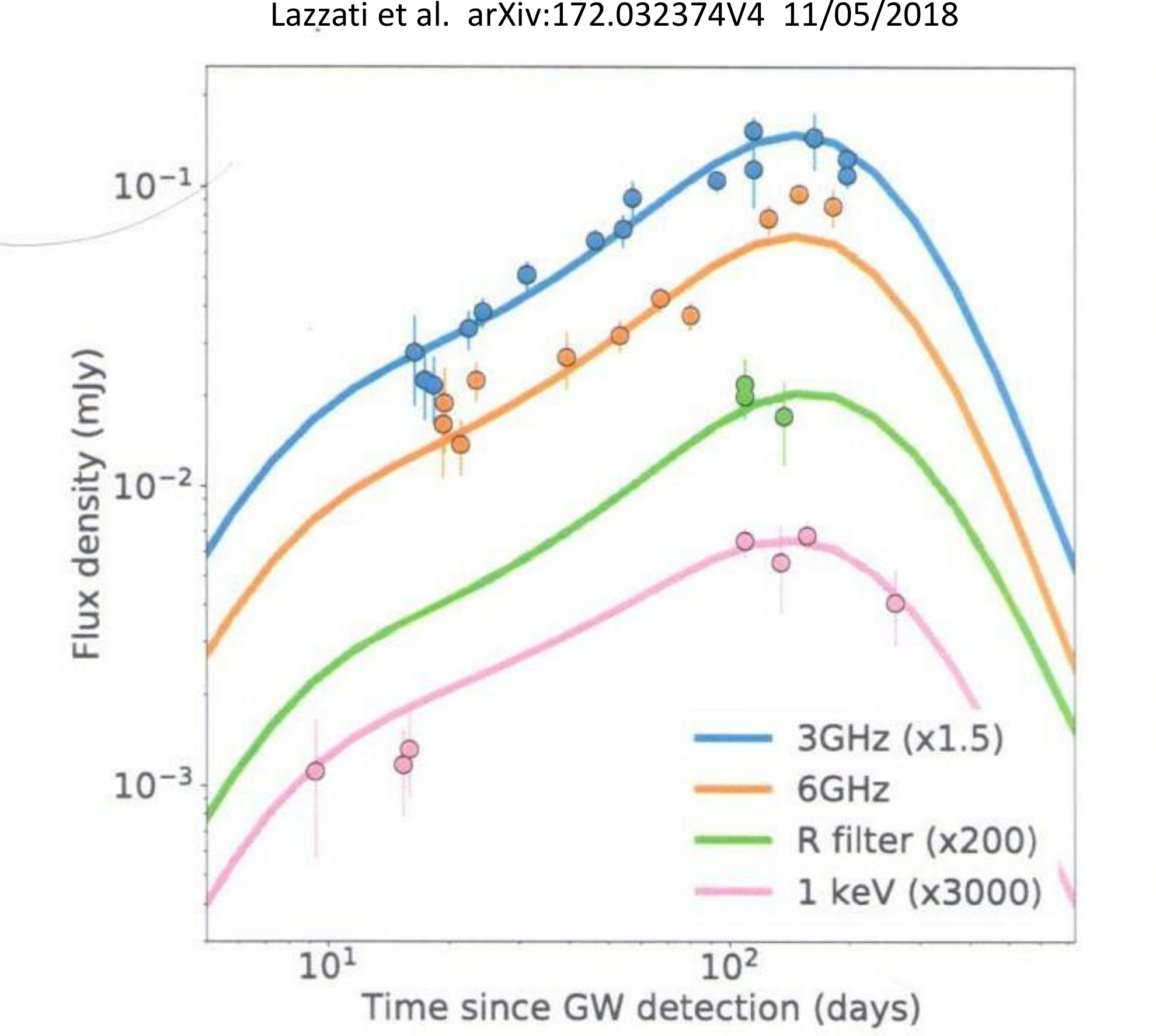} 
\caption{Best fit light curves to
the ROX AG of SHB170817A up to April 2018, 
obtained from a structured jet model \cite{Lazzati2}. Reported in 
version 4 of \cite{Lazzati1} posted in the arXiv on May 11th 2018.}
\label{fig:FB_AG170817A7}
\end{figure}

The cited FB model interpretations are not self-consistent for various reasons: \\
(a) The 
relation $\alpha\!=\! -p$ is only valid for a
conical jet with a fast lateral expansion ($V_\perp\!\approx\!c$) 
that has stopped propagating after the jet break 
time \cite{Sari98}. The fast spreading and the stopped 
propagation of the jet are not supported by the VLBI 
observations of the radio AG of SHB170817A \cite{MooleyJET}, which 
show a superluminal compact source (a CB), rather than 
the cited features.\\
(b) The relation $\alpha\!=\!-p$ post break 
is seldom satisfied by GRB AGs  and often yields $p\!<\! 2$. \\
(c) Due to rather large measurement errors, it is not yet clear that the 
temporal behavior of the AG of SHB170817A after break can be well 
parametrized by a broken power-law satisfying 
$\alpha\!=\! - p$.\\
(d) All types of FB models --with conical or structured jets-- used to fit the multi-band afterglow
of SHB170817A failed to correctly predict the subsequent data.  
This is demonstrated, for example, by the arXiv versions 1-4 of 
\cite{Lazzati1} where the evolution of the AG was 
first incorrectly predicted by a structured jet with a 
relativistic, energetic core surrounded by slower and less energetic 
wings, propagating in a low density ISM, as shown in Figure \ref{fig:FB_AG170817A5}. 
As soon as the AG break around day 150 and its subsequent fast 
decline were observed, the 
structured jet model with its dozen or so adjustable parameters 
had no problem to accommodate this behavior, see Figure \ref{fig:FB_AG170817A5}.

\section{Conclusions}

Table \ref{table3} contains a summary of many observational tests of 
the CB and FB models. Even without insisting on the cases
for which the CB model provided successful {\it predictions} and without reference to Table
\ref{table4} on the evolution of the FB model(s), the conclusions are not new,
but they should be overwhelmingly clear.

\begin{table*}
\caption{Critical Tests of The Cannonball and Fireball  models of GRBs and SHBs}
\label{table3}
\centering
\begin{tabular}{l l l l l l}
\hline
\hline
 ~ Test  &~~~~~~~ Cannonball Model~~~~~~~   &    &~~~~~~~~ Fireball Model~~ &             \\
\hline
Test 1   & Large GRB linear polarization &$\surd$~~~~~~& Small GRB  polarization  & X \\
Test 2   & Prompt emission correlations  &$\surd$~~~~~~& Frail relation          & X \\
Test 3   & Inverse Compton GRB pulses    &$\surd$~~~~~~& Curvature-shaped pulses  &X\\
Test 4   & SN-GRBs: Canonical afterglow  &$\surd$~~~~~~& Canonical AG not expected  &X\\
Test 5   & AG's break correlations         &$\surd$~~~~~~& AG's Break correlations &     X\\
Test 6   & Post-break closure relation   &$\surd$~~~~~~& Post-break closure relation & X\\  
Test 7   & Missing breaks (too early)    &$\surd$~~~~~~& Missing  breaks (too late)  & X\\
Test 8   & Chromatic afterglow           &$\surd$~~~~~~& Achromatic afterglow        & X\\
Test 9   & MSP-powered AG of SN-less GRB &$\surd$~~~~~~& Magnetar jet re-energization  & X\\
Test 10  & GRB rate $\propto$ SFR        &$\surd$~~~~~~& GRB rate not  $\propto$ SFR & X\\
Test 11  & LL GRBs = far off-axis GRBs   &$\surd$~~~~~~& LL GRBs = Different GRB class &X\\
Test 12  & Super-luminal CBs             &$\surd$~~~~~~& Superluminal fireball        &X\\
Test 13  & SHBs optical AG powered by NS &{?}~~~~~~& SHBs + macronova               &{ ?}\\
Test 14  & XRFs = Far off-axis LGRBs     &$\surd$~~~~~~& Different class of LGRBs     &X\\ 
Test 15  & XRTs = NS-powered  AGs        &$\surd$~~~~~& AGs of Far-off-axis GRBs      &X\\   
Test 16  & Radio image of SHB170817A: a CB        &$\surd$~~~~~& A complex structured jet      &X\\   
\hline   
\end{tabular}
\end{table*}

\newpage
\vspace{3cm}
\begin{table*}
\caption{\bf Majority and minority views on GRBs preceding decisive observations}
\label{table4}
\hspace{-.5 cm}
\begin{tabular}{l l l l l l}
\hline  
\hline
~~Key property      & ~~~~Majority  view&   &~~~~~~~Minority view&~~~ \\ 
\hline 
Location:           & Galactic  &X &Extragalactic& $\surd$\\ 
Produced by         & Relativistic $e^+e^-\gamma$ fireball& X& Highly relativistic 
                    plasmoids &$\surd$\\
Production mechanism& Collisions of $e^+e^-$ shells &X&ICS of light by plasmoids (CBs)&$\surd$\\ 
Prompt Emission     & Synchrotron radiation (SR)& X& Inverse Compton scattering &$\surd$ \\    
GRB geometry        & Isotropic  & X            & Very narrowly beamed & $\surd$ \\
LGRBs origin        & Stellar collapse to BH & X   & Stripped-envelope SN &$\surd$\\   
\hline
Afterglows' origin    & SR from shocked ISM & X & Synchrotron from CBs &$\surd$ \\ 
Afterglows' geometry  & Isotropic &X & Narrowly beamed &$\surd$\\
\hline
SN1998bw/GRB980425  & Rare SN/Rare GRB & X & SNIc-GRB viewed far off-axis &$\surd$ \\   
LL GRBs             & Different class of GRBs &X &  Normal GRBs seen far off-axis&$\surd$\\ 
SN-Less LGRBs       & Stellar Collapse to BH   & ? & Phase Transition in HMXRBs&? \\
\hline
AG plateau origin   & Jet re-energization &X &  Early time jet deceleration&$\surd$\\ 
AG break origin     & Deceleration of conical jet &X&  Deceleration of CBs   &$\surd$\\  
Missing jet breaks  & Too late to be seen  &X& Too early to be seen   &$\surd$ \\ 
\hline
Observed rate of GRBs & $\propto$ SFR + evolution &X& $\propto$ SFR, modified by beaming&$\surd$\\ 
\hline
Geometry & Spherical $\to$ Conical shells &X& Succession of cannonballs &$\surd$\\ 
\hline
\end{tabular}
\end{table*}

\acknowledgments{ADR acknowledges that this
project has received funding/support from the European Union's Horizon 2020 research and innovation programme under the Marie Sklodowska-Curie grant agreement No 860881-HIDDeN.}


\end{document}